\documentclass[preprint,authoryear]{elsarticle}
\usepackage[section] {placeins}
\usepackage{amsmath,amssymb,amstext,amscd,bm}
\usepackage{rotating}
\usepackage{pdflscape}
\journal{New Astronomy}

\newcommand\aap{A\&A}
\newcommand\aaps{A\&AS}
\newcommand\mnras{MNRAS}
\newcommand\apj{ApJ}

\newcommand\apjs{ApJS}
\newcommand\aj{AJ}
\newcommand\araa{ARA\&A}

\newcommand\pasp{PASP}

\usepackage{graphics}
\usepackage{graphicx}
\usepackage{epsfig}
\usepackage{amssymb}
\usepackage{amsthm}
\usepackage{lineno}

\journal{New Astronomy}
\begin{document}
\begin{frontmatter}

\title{CCD $UBV(RI)_{C}$ Photometry of Twenty Open Clusters}

 \author[kayseri,a]{{\.I}nci Akkaya Oralhan\corref{cor1}} 
 \author[istanbul]{Y\"uksel Karata\c{s}}
 \author[mexico]{William J. Schuster}
 \author[mexico]{Ra\'ul Michel}
 \author[mexico]{Carlos Chavarr\'{\i}a}
 \fntext[*]{E-mail: iakkaya@erciyes.edu.tr}
 \cortext[cor1]{Corresponding author}

\address[kayseri]{Erciyes University, Science Faculty, Department of Astronomy and Space Sciences, Talas Yolu, 38039, Kayseri, Turkey}
\address[istanbul]{Department of Astronomy and Space Sciences,  Science Faculty, {\.I}stanbul University, 34119, University, {\.I}stanbul, Turkey}
\address[mexico]{Observatorio Astron\'omico Nacional, Universidad Nacional Aut\'onoma de M\'exico, Apartado Postal 877, C.P. 22800, Ensenada, B.C., M\'exico\\}

\begin{abstract}

Fundamental astrophysical parameters have been derived for 20 open clusters (O\!Cs) using   
CCD~$U\!BV\!(RI)_C$ photometric data observed with the 84~cm telescope 
at the San Pedro M\'artir National Astronomical Observatory, M\'exico.  

The interstellar reddenings, metallicities, distances, and ages have been compared
to the literature values.  Significant differences are usually due to the usage of
diverse empirical calibrations and differing assumptions, such as concerning cluster metallicity,
as well as distinct isochrones which correspond to differing element-abundance ratios, internal
stellar physics, and photometric systems.  Different interstellar reddenings, as well as varying
reduction and cluster-membership techniques, are also responsible for these kinds of systematic
differences and errors. 

The morphological ages, which are derived from the morphological indices ($\delta V$ and $\delta 1$) 
in the CM diagrams, are in good agreement with the isochrone ages of 12 O\!Cs, those with good red
clump (RC) and red giant (RG) star candidates.  No metal abundance gradient is detected for
the range $6.82 \leq R_{GC} \leq 15.37$ kpc, nor any correlation between the cluster ages and metal
abundances for these 20 O\!Cs. 

Young, metal-poor O\!Cs, observed here in the third Galactic quadrant, may be associated with stellar
over-densities, such as that in Canis Major (Martin et al.) and the Monoceros Ring (Newberg et al.),
or signatures of past accretion events, as discussed by Yong et al. and Carraro et al.

\end{abstract}

\end{frontmatter}
\newpage
\section{Introduction}

Open clusters (O\!Cs) are valuable for studying stellar evolutionary models, and 
the age-metallicity relation and metal-abundance gradient in the Galactic 
disc \citep[e.g.][]{cam85, car94,fri95}, as well as luminosity and mass functions
\citep{pis08}.
By fitting the photometric observations of open clusters to theoretical isochrones,
the fundamental parameters such as interstellar reddening, metallicity,
distance modulus, and age can be precisely and accurately inferred.  

The aims within the Sierra San Pedro M\'artir National Astronomical Observatory (SPMO,
hereafter) open cluster survey \citep[cf.][]{sch07,tap10,akk10} are the following:  
\begin{enumerate} 
\item a common {\sl UBVRI} photometric scale for open clusters, 
\item  an atlas of colour--colour and colour--magnitude diagrams for these clusters, 
\item  a homogeneous set of cluster reddenings, distances, ages and, if possible, 
metallicities, 
\item an increased number of old, significantly reddened, or distant open clusters, and 
\item a selection of interesting clusters for further study.
\end{enumerate}
The O\!Cs for the present study have been selected from the large (and mostly complete) 
catalogue, ``Optically Visible Open Clusters and Candidates'' \citep{dia12}, which 
is now also available at the CDS (Centre de Donn\'ees Astronomiques de Strasbourg).  This 
work aims to provide the fundamental parameters of reddening, metallicity, distance modulus, 
and age for the 20 O\!Cs.  Our final intention is to publish a set of homogeneous photometric
$U\!BV\!(RI)_{C}$ data for over 300 Galactic OCs \citep{sch07, tap10}.

As emphasized by  \cite{moi10}, thousands of studies of individual O\!Cs 
in the literature have non-concordant results due to the variety of techniques used to 
observe, reduce, and derive the fundamental astrophysical parameters. 
These kinds of biases lead to inhomogeneous compilations.  While deriving the
astrophysical parameters, the choice of the reference lines from the Hyades or from
\citet[hereafter SK82]{schm82}, the differing theoretical and observational ZAMS,
as well as varying cluster membership methods are responsible for these nonuniform
results.  The fundamental astrophysical parameters of open star clusters were compiled  
in the catalogues of O\!Cs by  \citet[the Lund catalogue]{lyn87} and
\citet[][hereafter Dias]{dia02, dia12}.  In this same sense the WEBDA OC database \citep{mer92}
is a  very valuable, but inhomogeneous, on-line catalogue.  

The metal-abundance information has an important influence on the choice and fit of the isochrones 
to the photometric observations. Many authors fit isochrones which correspond to the Sun's heavy-element
abundance, $Z_\odot$, to the cluster members in the two-colour and colour-magnitude diagrams. 
For the determination of $(Z, [Fe/H])$ for a cluster, some authors prefer to apply a fit 
by matching the theoretical ZAMS of the age libraries, such as \cite{ber94}, \cite{gir00}, and
\cite{yi03}, to F- and G-type dwarf stars in the $(U$--$B),(B$--$V)$ two-colour diagram, 
and thus fix the mass fraction and logarithmic metal abundance, $(Z, [Fe/H])$, of the cluster.
\cite{cam85} and \cite{tad03} have used a $\delta(U$--$B)$ photometric technique to determine
photometric abundances of the O\!Cs.  Recently, \citet[][hereafter T10]{tap10} and
\citet[][hereafter A10]{akk10} have also used this $\delta(U$--$B)$ technique with an improved
approximation to F-type dwarf stars in the two-colour diagrams from the CCD~$U\!BV\!(RI)_C$
photometric observations of O\!Cs.  These ultraviolet excesses have thus been converted
to $Z$ heavy element abundances, which are necessary for selecting the isochrones and determining
reliable ages and distances of the O\!Cs.  

For O\!Cs, \cite{pau06} stressed the importance of certain analysis techniques for obtaining
the best results; these were reinforced by A10, and include:
(a) a good technique for determining cluster membership, (b) realizing the importance of the
$(U$--$B)$ colour index in determining the interstellar reddening and the stellar metal abundance,
(c) obtaining a clear visibility of the cluster turn-off and identifying possible Red-Clump
candidates in colour-magnitude diagrams while isochrone fitting, and (d) awareness of the
possible contamination and shifting of the main sequence by as much as 0.75 mag due to the
presence of binary stars, or even more due to multiple stars.

The issues mentioned above affect the interpretation of metal-abundance gradients in the Galaxy, our 
ability to detect and establish a gap, or discontinuity, within an $[Fe/H]$--$R_{GC}$ relation,
and also the possibility to see an age-metallicity relation for the O\!Cs.  Taking into consideration
these facts, O\!Cs need to be uniformly analysed, homogeneously with regards to the instrumentation,
observing techniques, and reduction and calibration methods.  

In this paper, 20 O\!Cs within the SPMO Open Star Cluster Project have been analyzed to determine the
astrophysical parameters, and thus to reveal the nature of possible differences in the astrophysical
parameters found in the literature and to identify those sources most compatible with our results.  
However, in order to study the $[Fe/H]-R_{GC}$ and age--metallicity relations for O\!Cs in general, 
we stress that our sample is relatively small, but uniformly treated, and selection effects 
may also introduce limitations.  Nonetheless, our 20 O\!Cs do have this advantage of being uniformly
analysed, homogeneous regarding instrumentation, observing techniques, reduction methods, photometric
calibrations, and analyses.

This paper is organized as follows:  Section~2 describes the observation and reduction techniques.
The technique for determining cluster membership is presented in Section~3.
In Section~4 the $U\!BV\!(RI)_C$ photometric system is employed to derive interstellar reddenings
and metallicities of the clusters from a two-colour diagram, and distance moduli and ages
from colour-magnitude diagrams.  Comparisons of these parameters with previous results from 
the literature are made in Section~5, and comparisons to the morphological ages are given in Section~6.
The metal-abundance gradient and age-metallicity relation are presented in Section~7.  The spatial
distribution, and the identifications of Red-Clump (RC) and Blue-Straggler (BS) candidates are
presented in Sections~8 and 9, respectively, and the conclusions are given in Section~10.

\section{Observations and Reduction Procedures}

A CCD $U\!BV\!(RI)_C$ survey of northern O\!Cs has been undertaken at SPMO
using always the same instrumental setup (telescope, CCD, filters), observing
and data reduction procedures, and system of standard stars \citep{lan83,lan92,clem13,mich14}.
The CCD $U\!BV\!(RI)_C$ observations of the 20 O\!Cs of this paper have been made exclusively
 with the 0.84-m f/13 Cassegrain telescope of SPMO, during nights 
of June 2001, February 2002 (subdiveded in four contigous nights to check the 
night parameters) and January, May (subdiveded as February 2002 for the same 
reason), September, and November 2003.  The telescope hosted the filter-wheel 
`Mexman' with the SITe~1 (SI003) CCD camera\footnote{no longer in use}, which 
had a 1024$\times$1024 square pixel array, with a pixel size of 
$24\mu$m$\times24\mu$m; this CCD had non-linearities less than 0.45 per cent 
over a wide range, with no evidence for fringing even in the $I$ band, and
Metachrome II and VISAR coverings to increase sensitivity at the blue and
near-ultraviolet wavelengths. 
 
The 0.84-m telescope was re-focused before the observation of each OC, using 
the $V$ filter of our parfocal set of filters.  The O\!Cs have been observed 
with exposure times of $3 \times 240$ seconds for the $U$ filter, $3 \times 
180$ for $B$, $3 \times 100$ for $V$, $3 \times 100$ for $R$, and $3 \times 
120$ for $I$.  For the $U$ band, extra integrations ($\approx 600$ seconds) 
were sometimes made to improve the signal-to-noise ratio.  Also, for some 
clusters exposures as short as 10 seconds were made in the $R$ and $I$ filters 
to avoid saturation of the brightest stars.\\

In Table 1, a general log sheet of the observing runs is presented. Several 
standard-star fields from \cite{lan92,clem13,mich14} were observed to permit the 
determination of the atmospheric extinction coefficients and the 
derivation of the photometric transformations to the Johnson-Cousins photometric
system\footnote{The transformations resulted linear for our purposes in the case of 
the $U\!BV\!(RI)_C$ system of the SPMO.}.  The standard-star fields have been 
observed with exposures of $1 \times 240$ seconds for the $U$ filter, $1 \times 
120$ for B, $1 \times 60$ for V, $1 \times 60$ for R and I. 

Usually one, or more, Landolt fields were re-observed with an air-mass range of 
$\approx 1.1-2.3$, in order to measure the atmospheric extinction coefficients.  
Due to the wide band-passes of the Johnson-Cousins filters, second-order colour 
terms were included in the atmospheric-extinction corrections.  For the large air-
mass observations, the filters were frequently observed with both forward and 
backward sequences (i.e.~$UBVRI-IRVBU$); this was also occasionally done for other 
standard-star fields to increase precision and observing efficiency of 
the photometric observations.

The usual calibration procedures for CCD photometry were carried out during
each of our observing runs:  fifty to a hundred `bias' exposures were made each
night, and fifty or more `dark' images' were made during each run with exposures 
according to the longest of our stellar exposures; these `darks' were usually 
made during the a non-photometric weather spell or nights.  Flat-fields were 
obtained at the beginning and end of the nights by observing a `clear of stars'  
sky-patch in the opposing direction to the sunrise or sunset directions; at 
least five flat-fields per night were obtained for each filter with exposures 
greater than five seconds (keeping shutter errors negligible), and with small 
offsets ($\approx 10"$) on the sky between each flat-field exposure. 

\begin{table*}
\centering
\tiny
\caption{A general log sheet of the observing runs}
~\\
\tiny
\begin{tabular}{lllccccc}
\hline
Run     &$N_{G}$&$N_{std}$&  Landoldt Fields observed       &(B-V)$_{min}$ & (B-V)$_{max}$ & $X_{min}$ & $X_{max}$ \\ 
 \hline
        &    &     &MARK A, PG1633+099, PG1528+062    &        &       &      &       \\  
Jun 2001&  6 & 128 &PG1530+057, PG1525-071, PG1657+078& $-$0.252 & +1.14 & 1.07 & 2.33  \\    
        &    &     &                                  &        &       &      &       \\ 
        &    &     &PG0918+029, PG0942-029, PG1047+003&        &       &      &       \\ 
Feb 2002&  8 &  35 &PG1323-086, PG1528+062, SA 095 112& $-$0.298 & +1.41 & 1.14 & 2.72  \\ 
        &    &     &SA104 336, SA107 600              &        &       &      &       \\ 
        &    &     &                                  &        &       &      &       \\  
        &    &     &PG1323-086, PG1525-071, PG1528+062&        &       &      &       \\ 
May 2003&  9 & 445 &PG1530+037, PG1633+099, PG1657+078& $-$0.252 & +2.33 & 1.10 & 3.94  \\ 
        &    &     &SA 204 334, SA107 599, SA 110 503 &        &       &      &       \\
        &    &     &                                  &        &       &      &       \\ 
        &    &     &SA 092 330, SA 092 498, SA 092 500&        &       &      &       \\
Sep 2003& 10 & 410 &PG0231+051, PG2331+055, PG2336+004& $-$0.320 & +2.53 & 1.11 & 3.56  \\  
        &    &     &SA 092 501, SA 095 139, SA 098 670&        &       &      &       \\  
        &    &     &SA 110 364                        &        &       &      &       \\ 
        &    &     &                                  &        &       &      &       \\ 
        &    &     &PG0231+051, PG0918+020, PG2213-006&        &       &      &       \\ 
Nov 2003&  9 & 423 &SA 092 252, SA 095 ~96, SA 098 653& $-$0.320 & +1.91 & 1.11 & 4.02  \\ 
        &    &     &SA 113 191, SA 113 339, RU 149    &        &       &      &       \\ 
\hline
\end{tabular} 
\begin{list}{Table Notes.}
\item $N_G$ and $N_{std}$ are the number of Landolt fields  and standard star 
measurements done during a run, respectively.  
\item (B-V)$_{min}$, (B-V)$_{max}$, X$_{min}$ and X$_{max}$ are the minimum and 
maximum colour and air-mass values observed during the run, respectively.   
\end{list}
\end{table*}

More details of the principal parameters of the detector and instrument used in 
the allotted observational runs are given in {\tt http:$//$haro.astrossp.unam. \\
mx$/$telescopios}, and concerning the data observations, reductions, and errors in T10 and A10. \\

Standard, absolute photometry outside the earth's atmosphere has usually been 
reduced from a instrumental to a 
standard or reference system with two forms of equations:  one deducing stellar 
\textbf{colours} in a standard or reference system, based on a subset of the 
reference stars measured with the same instrumental set-up one is calibrating, 
and the other one deducing \textbf{magnitudes} in the standard system \citep[c.f.]
{mit60,har62}. The following is an equality relating instrumental colours with 
their equivalent standard values: 

\begin{equation}
(\alpha -\beta)_s =  z_{\alpha \beta} + c_{\alpha \beta} \cdot
\frac{(\alpha-\beta)_i}{1 + p_{\alpha \beta}\cdot <X>_{\alpha \beta}} + 
\frac{K_{\alpha \beta}\: <X>_{\alpha \beta}} {1+p_{\alpha \beta}\: <X>_{\alpha \beta},} 
\end{equation} 

\noindent 
where the $(\alpha-\beta)$ refer to the colour defined by the filters or passbands 
$\alpha$ and $\beta$, and the subscripts ``s'' and ``i'' stand for standard or 
instrumental colours, respectively. The second-order atmospheric extinction 
coefficient $p_{\alpha \beta}$ $( = p_{\alpha\alpha\beta} - p_{\beta\alpha\beta}$, 
from the equation~2) is assumed known, but one can determine its value 
from the observations in a simple way , as will be shown below. $X_{\alpha\beta}$ 
is the mean of the air masses measuring filters ``$\alpha$" and ``$\beta$", and  
except for the instrumental colour, all other quantities are usually computable to 
an accuracy equal or better than that of the reference system that has usually a 
pair of hundreths of a magnitude uncertainties, see  \citet[references therein]{john66}.
 $<X>_{\alpha \beta}$ is the mean air-mass of passbands 
$\alpha$ and $\beta$, and $z_{\alpha \beta}$, $c_{\alpha \beta}$ and $K_{\alpha 
\beta}$ the unknowns for the least-square reduction\footnote{When a subscript is 
conformed by two letters, it refers to the two passbands defining the colour with 
which we are dealing.}. One recovers  the first order extinction $k_{\alpha \beta}$ 
from the relation \newline 
\begin{equation}
K_{\alpha \beta} = c_{\alpha \beta}\cdot k_{\alpha \beta}
\end{equation}  

In general we preferred and used only this transformation of \textbf{colours} 
to the standard system. If the same equipment is used, there is no physical reason 
for the second order extinction coefficient to change significantly during an observing run 
and its mean value will do the job (see table 2, $p_u$).\\

The other form of equation consists in deriving the standard stellar magnitude of a 
filter $\alpha$ $(= -2.5\;log\: ADU_\alpha$, were $ADU_\alpha$ is the net stellar 
signal after doing the usual cosmetic corrections to the image $\alpha$), with the 
addition of a corrective colour term that compensates  for systematic shifts of the 
effective wavelength (i.e. $\lambda_{e}$) due to deviations of the instrumental 
sensitivity curve from the original curve of the defining (standard or reference) 
instrument, mainly  becuse to the differences in the transmission curve of the 
optics (i.e. mirrors, filters, windows, etc...) and the quantum-efficiency function 
of the detector. \\

The magnitude transformation equation can be written as follows

\begin{equation} 
\alpha_s = z_{\alpha}  + \alpha_{i} + (k_{\alpha \alpha \beta} + 
p_{\alpha \alpha \beta}\: (\alpha - \beta)_s)\cdot X_\alpha + 
A_{\alpha \gamma \delta} \: (\gamma-\delta)_s
\end{equation}

\noindent
Two colours, $(\alpha - \beta)_s$ and $(\gamma - \delta)_s$, are intentionally 
written down in the above equality to stress that they do not have to be the same, 
but they usually are. Again subscripts ``s'' and ``i'' denote quantities in the 
standard and instrumental systems, respectively. The $z_{\alpha},\: k_{\alpha \alpha \beta}$ 
and $A_{\alpha \gamma \delta}$ are the solutions of the reduction 
by least mean squares. 

$p_{\alpha \alpha \beta}$ is the second-order atmospheric extinction coefficient of 
magnitude $\alpha$ for the colour $(\alpha-\beta)$.\footnote{The first of a three 
letter subscript refers to the magnitude at that pass-band, the following two 
letters refer to the colour in the standard system being used to make the second-
order correction.}  The second-order atmospheric extinction coefficient is 
necessary in the reduction because it corrects for shifts in the effective 
wavelength of a (broad-band filter) magnitude with respect to a (mean) AOV star defining the 
pass-band, due to the convolution of the incoming stellar flux (SED) with the 
atmospheric transmission: Within the observational errors it is the same for all 
$(\gamma - \delta)_s$ colours except for Johnson's filter "U" which is affected by
the effective temperature and the gravity of the stars. In the case of Johnson's 
"V" magnitude, this second-order term should be close to zero, since the extinction curve at SPMO is 
fairly flat at this wavelength range \citep[and references therein]{sp01}. 
For a better understanding of the reduction procedure discussed 
here, the reader is referred to the early work by \citet[][]{kin52a,kin52b}.

\subsection{On the determination of the second-order coefficients} 

Among the Landolt standard-star fields, in several of them one finds blue and red standard 
stars within the ($\approx 7'\times 7'$) field of view of the CCD used for the observations 
(mainly those fields with prefixes "PG", "Mark" or "Rubin"). They enable us to determine 
the second-order cefficients $p_{\alpha \beta \gamma}$ with help of Equation~2 by a least 
square reduction, assuming $A_{\alpha \gamma \delta} 
= 0$. This assumption introduces a small but systematic error in the least-mean square 
solution since $A_{\alpha \gamma \delta}$ is small but not zero (see Table 
2). (In a perfect match between the instrumental and the standard system $A_{\alpha \gamma \delta} = 0$). 
Since their standard magnitudes and colours are known, by a least square reduction of 
the data of the observed standard stars, one obtains $z_\alpha,\; k_{\alpha \alpha \beta}$ and $p_{\alpha \alpha \beta}$.  
(Contiguous colours to magnitude $\alpha$ should be preferred).\\

The atmospheric extinction-coefficient will depend on the stellar spectral type 
as a result of the convolution of the instrumental response curve of the photometer 
with the extinction curve of the atmosphere \citep[see extinction of SPMO in][and references therein]{sp01}. 
At this point, one expects $p_{\alpha \gamma \delta}$ 
to be fairly constant for an observing run or even more (cf. Table 2), and the 
nightly estimate for the first order extinction coefficient $k_{\alpha \alpha 
\beta}$, and the zero-point $z_\alpha$ may vary 10 \% or less from night to night in a 
given observing run (see February 2002a, b, c and May2003a through May2003d). 

On the other hand, due to the convolution of the instrumental response curve with 
the spectral energy distribution of the star that results in a shift of $\lambda_{e}$, 
we need to correct the magnitude with a colour term.\\

A three parameter reduction, i.e. zero-point correction for the magnitude, 
$z_{\alpha}$, the first- and second-order atmospheric extinction coefficients, 
$k_{\alpha\alpha\beta}$ and $p_{\alpha\alpha\beta}$, respectively, and a colour correction 
coefficient due to shifts in the effective wavelength $A_{\alpha\delta\gamma}$, {\bf have sense  only} if the standard fields were 
observed with enough air-mass and colour difference between measurements, i.e. $\Delta X \gtrsim 0.7$ and $\Delta (B-V) \gtrsim 0.7$ 
suffice in most of the cases for the prevailing SPMO sky conditions. If any one condition does not fulfil the 
respective range, it is better to use the mean respective coefficient. It is also recommendable to 
determine zero-points and atmospheric extinction-coefficients 
nightly, but the data of the whole run can be used to determine  
$A_{\alpha\delta\gamma}$ and $p_{\alpha\alpha\beta}$. \\

Finally, the stability of the system has been gratifying, as seen by checking 
zero-points, atmospheric extinction, and transformation coefficients for the 
more problematic filter, Johnson's "U", resulting from the five observing runs reported in this work. 
The system is robust for the observer and, at least during 
an observing run, the reduction parameters repeat well, and one can measure good 
colours of the program stars in the standard system with help of Eq.~(1), 
i.e. see Table~2. Significant variations in the zero point, $z_u$, were due to the 
aluminization of the telescope mirrors, extreme dry or hot weather (see \citet[]
{sp01}).

\begin{table}
\centering
\tiny
\caption{The four principal reduction parameters for the U filter}
\tiny
\begin{tabular}{lcccc}
\hline
Run     & $z_u$            & $k_{U}$            & $p{_u}$             &   $c{_u}$      \\  
\hline 
June~2001 & $4.535\pm 0.007$&  $0.520\pm 0.001$ & $-0.054\pm 0.003$ & $-0.004\pm 0.003$ \\ 
Feb~2002a& $4.710\pm 0.002$ &  $0.426\pm 0.008$ & $-0.051\pm 0.002$ & $-0.006\pm 0.003$ \\   
Feb~2002b& $4.757\pm 0.002$ &  $0.411\pm 0.001$ & $-0.052\pm 0.001$ & $-0.007\pm 0.002$ \\   
Feb~2002c& $4.621\pm 0.015$ &  $0.500\pm 0.005$ & $-0.068\pm 0.002$ & $+0.001\pm 0.002$ \\  
May~2003a& $4.260\pm 0.001$ &  $0.518\pm 0.013$ & $-0.067\pm 0.002$ & $+0.021\pm 0.004$ \\  
May~2003b& $4.250\pm 0.002$ &  $0.474\pm 0.002$ & $-0.021\pm 0.001$ & $-0.032\pm 0.002$ \\  
May~2003c& $4.259\pm 0.002$ &  $0.509\pm 0.003$ & $-0.069\pm 0.002$ & $+0.022\pm 0.002$ \\  
May~2003d& $4.257\pm 0.002$ &  $0.529\pm 0.017$ & $-0.049\pm 0.004$ & $-0.009\pm 0.014$ \\  
Sep~2003 & $4.704\pm 0.003$ &  $0.438\pm 0.002$ & $-0.021\pm 0.004$ & $-0.013\pm 0.020$ \\ 
Nov~2003 & $4.213\pm 0.006$ &  $0.454\pm 0.005$ & $-0.031\pm 0.001$ & $-0.022\pm 0.003$ \\ 
\hline
\end{tabular} 
\end{table}

Two additional remarks: i) if a proper match between the instrumental and the 
reference systems has been achieved, the $A_{\alpha\delta\gamma}$ of Eq.\ (2) 
should be small ($\le 0.10$, in a perfect match, =0.00). ii) When the filter's bandwidth is large, 
one needs to determine the second-order extinction coefficient $p_{\alpha\alpha\beta}$ only 
once per run (in the five runs spread over the almost 
2.5 years of photometric data discussed here, $p_{\alpha\alpha\beta}$, which was measured almost nightly, 
changed less than about 12$\%$, except for about five 
nights of extremely good weather in May, September, and November of 2003, 
when it was about 50$\%$ under its mean value of $-$0.0778 for the filter U.  A 
good guess for the standard magnitude of a problem star would be to assume  
$A_{\alpha\delta\gamma} = 0.0$ and $p_{\alpha\alpha\beta}$ equal to its previous 
determination (for an overall error less than about 2\%). The data reductions and 
transformations of this CCD photometry have been carried out using the usual 
techniques and packages of IRAF\footnote[6]{IRAF is distributed by NOAO, which is 
operated by the Association of Universities for Research in Astronomy, Inc., 
under cooperative agreement with the NSF.}.  Aperture and PSF photometry techniques were 
used for handling and combining the standard-star and cluster-star observations, 
respectively \citep{how89,how90,ste87, ste90}.  The reduced CCD~$U\!BV\!(RI)_C$ 
standard photometric data for these 20 O\!Cs will be provided upon request to Ra\'ul Michel.

\section{Determination of Cluster Members}

For the possible members of the 20 O\!Cs, a java-based computer program, `SAFE',
\citep[]{mc10} has been utilized for the visualization and analysis
of the photometric data of O\!Cs \citep{sch07}.  This program is capable of displaying
each cluster's data simultaneously in different colour-colour (hereafter CC) and
colour-magnitude (CM) diagrams and has an interactive way to identify a star, or
group of stars, in one diagram and to see where it falls in the other diagrams, thus
facilitating the elimination of field stars and the apperception of cluster features.
Since some O\!Cs in our sample are rather close to the Galactic bulge and/or disc, the
field-star contamination can become significant.  In this sense, the field-star
decontamination technique used by \cite{bon07} has the advantage with respect to the
SAFE program for removing field stars, and has been successfully applied for large-sized
fields around O\!Cs.  However, since the O\!Cs within the SPMO survey have been selected
to be small or comparable to the size of the CCD, 6.9$\times$6.9 arc minutes, the
central part of each cluster has been isolated, using SAFE, to increase the contrast
of the cluster with respect to field stars in the various CC and CM diagrams.
High-mass stars are transferred to the cores of the clusters as a result of mass
segregation, and so the SAFE algorithm has the advantage for identifying various groups
of stars in CC and CM plots, such as RC/RG stars and possible BS stars.
Note that RC/RG stars are also quite important for determining the ages and especially
the distances of the clusters.

\begin{figure}
\centering
\epsfig{file=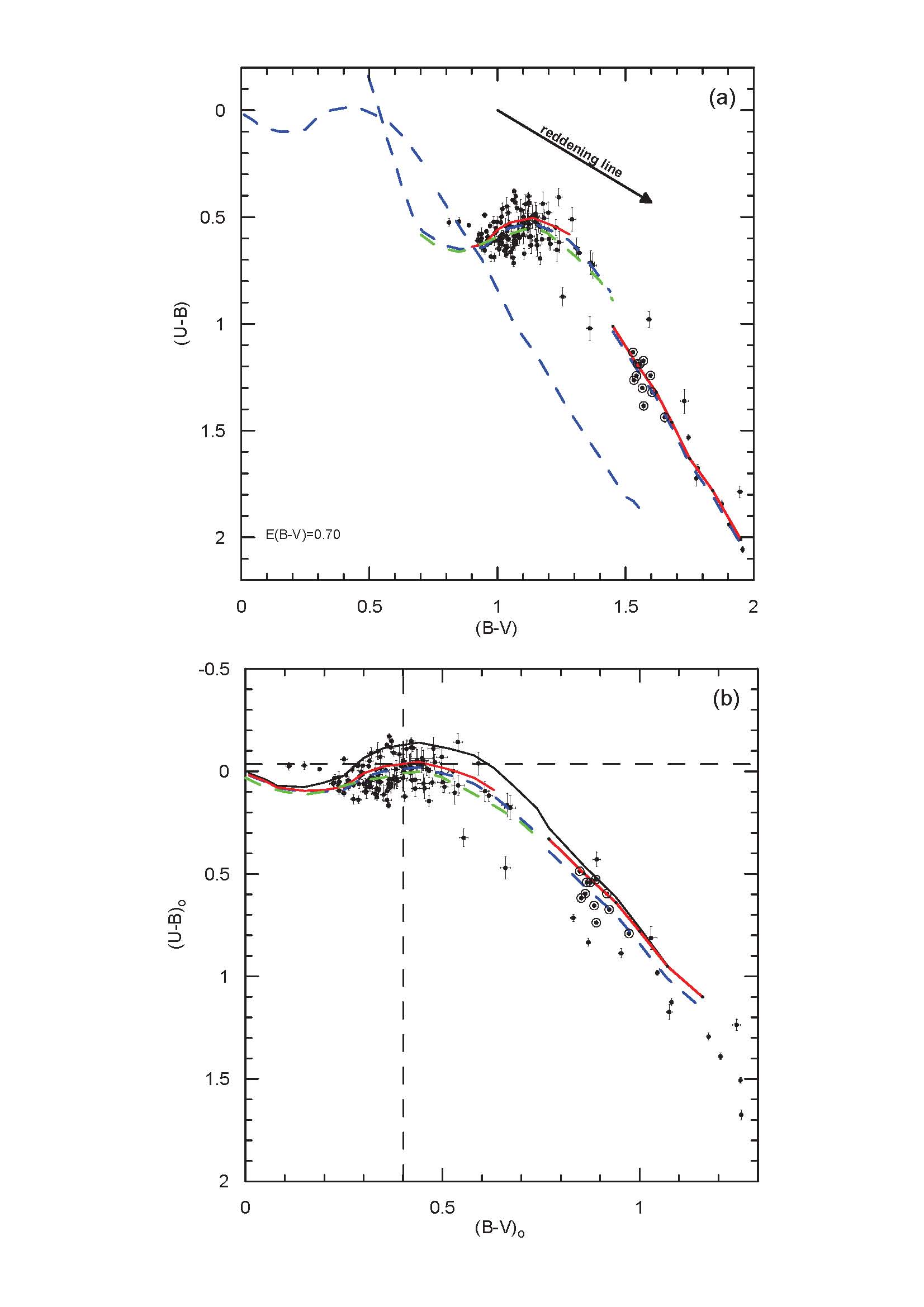, width=9cm, height=13.4cm}
\caption {Shows the reddened, $(U$--$B, B$--$V)$, and the de-reddened,  
$((U$--$B)_0, (B$--$V)_0)$, diagrams for Ki~05.  Panel~(a):  Blue dashed lines
show the reddened and de-reddened SK82 relations for MS (upper section) and RG (lower
section) stars.  Panel~(b):  Green dashed and the upper solid black lines denote the
Hyades main-sequence and the metal-free upper envelope of Melbourne (1959) and Sandage
(1969), respectively.  The red solid curve shows the estimated iso-metallicity line
for [Fe/H]$=$$-$0.17 (Z =+0.013), as measured for this cluster; the vertical and
horizontal lines show the mean values of $\langle(U$--$B)_0\rangle$ and
$\langle(B$--$V)_0\rangle$ for F-type stars on the Ki~05 main sequence.  A reddening
vector is also shown as an arrow in panel~(a).  Big open circles mark the RC candidates.}
\end{figure}

\begin{figure}
\epsfig{file=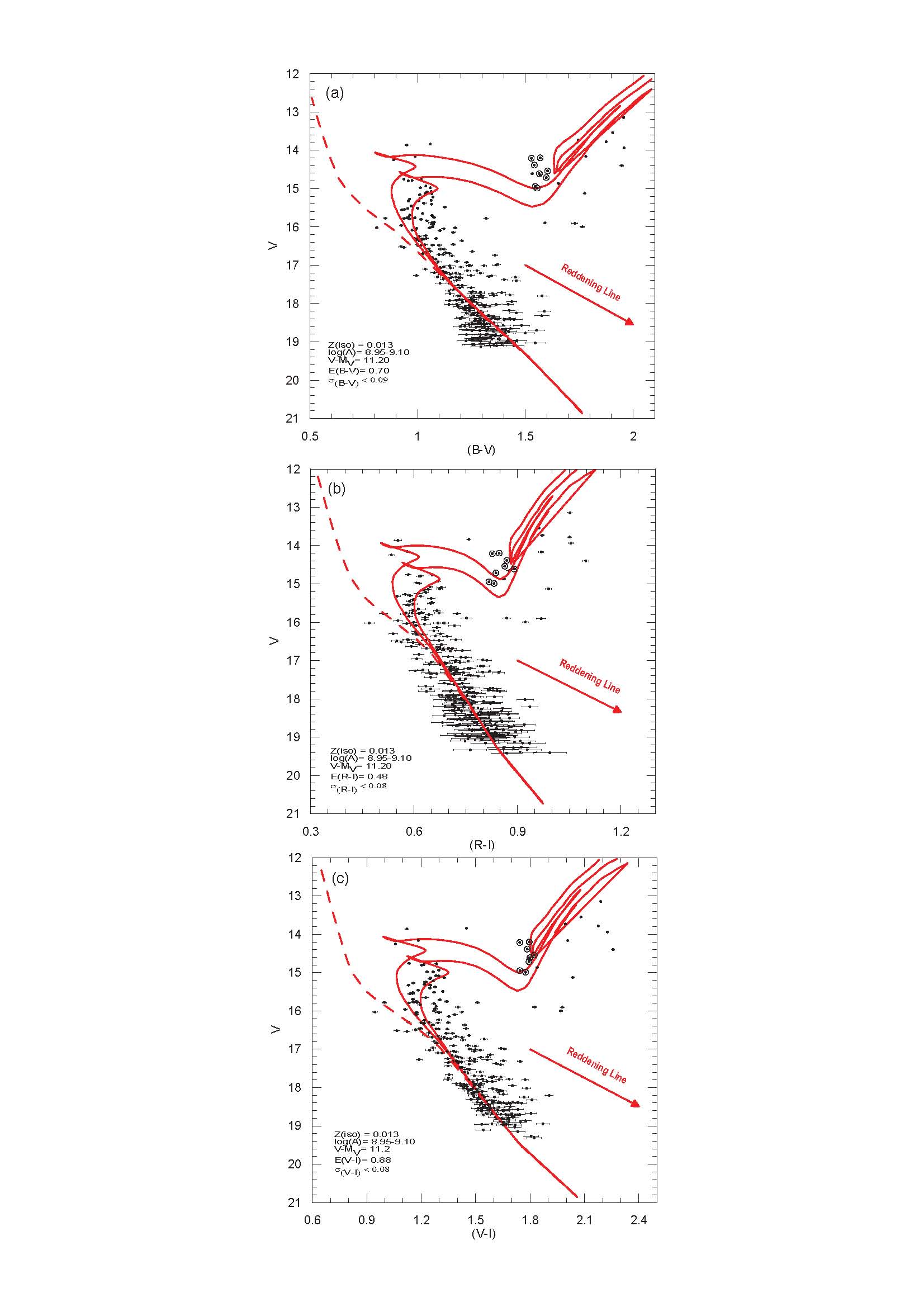, width=9cm, height=11cm}
\caption {Panels~(a)--(c): the CM diagrams of $(V,B$--$V)$, $(V,R$--$I)$, and  $(V,V$--$I)$,
respectively, for Ki~05.  Solid lines show the M08 isochrones interpolated to $Z=+0.013$.
Big open circles mark the RC candidates.}
\end{figure}

\begin{figure}
\centering
\epsfig{file=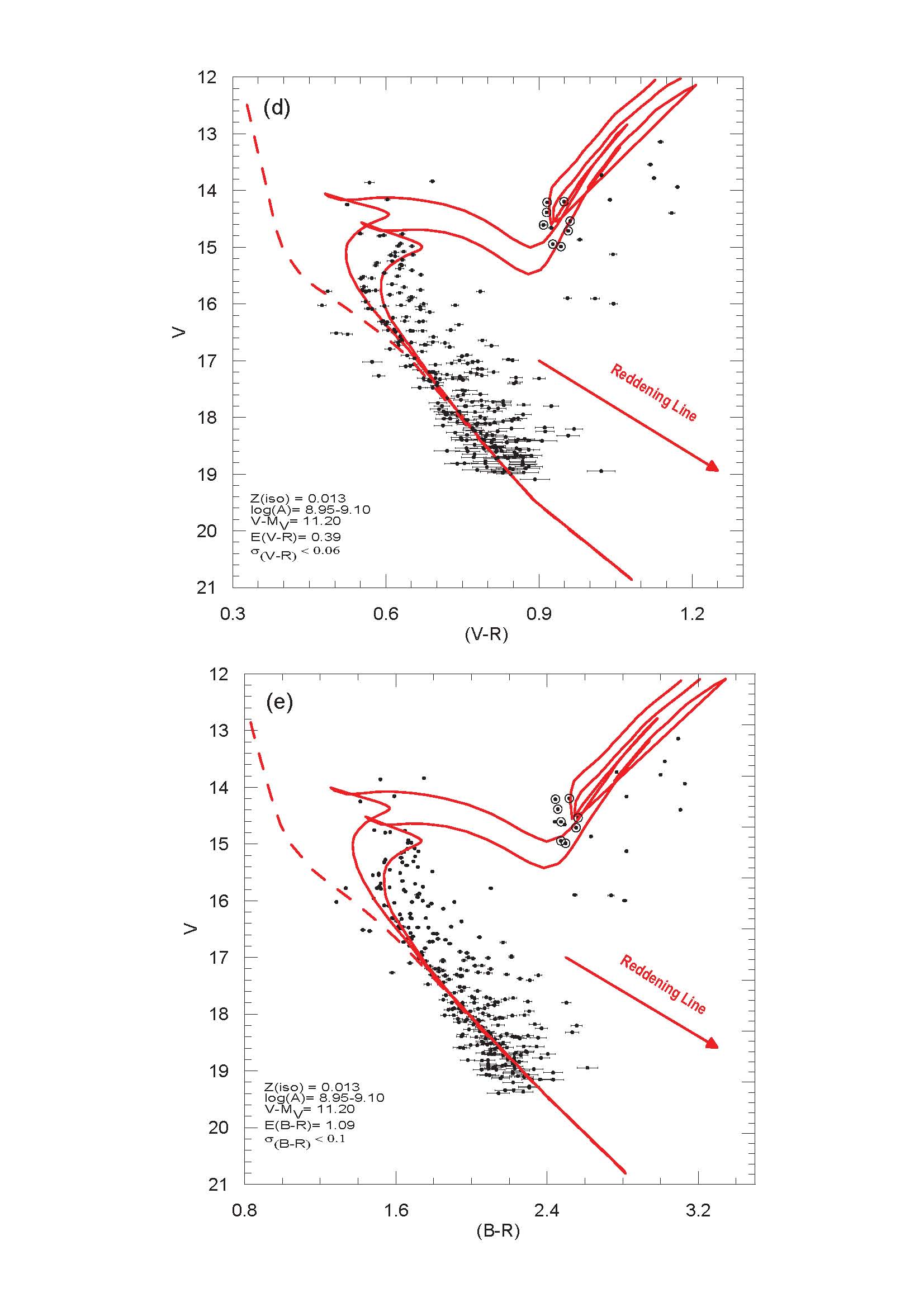, width=10cm, height=12cm}
\caption {Panels (d)--(e): the CM diagrams of $(V,V$--$R)$ and $(V,B$--$R)$ for Ki~05. 
The symbols are the same as in Fig.~2.}
\end{figure}

\begin{table}
\tiny
\centering
\tiny
\caption{The values of $(B$--$V)$, $(U$--$B)_{H}$, $(U$--$B)_{M}$ for the Hyades main sequence
(``H'' suffix) and for approximately metal-free stars (``M'').  The parts of the data of
$(B$--$V)< 0.35$ and $(B$--$V)>0.35$ are taken from Melbourne (1959) and Sandage (1969),
respectively.  $\delta(U$--$B)_{M}$ and $\delta_{0.6}/\delta_{M}$ in Columns~4 and 5 represent 
this maximum ultraviolet excess and the value needed to normalize this to $(B$--$V)=0.60$,
respectively.  Other normalization factors are available in table 1A of Sandage (1969)
for lesser UV excesses.} 
\begin{tabular}{ccccc}
\hline
$(B$--$V)$&$(U$--$B)_{H}$ &$(U$--$B)_{M}$ &$\delta(U$--$B)_{M}$&$\delta_{0.6}/\delta_{M}$ \\
\hline
0.00 &0.03 &$+$0.01    &0.02 &15.5 \\
0.05 &0.08 &$+$0.05    &0.03 &12.4 \\
0.10 &0.10 &$+$0.08    &0.03 &12.4 \\
0.15 &0.11 &$+$0.06    &0.05 &5.96 \\
0.20 &0.10 &$+$0.00    &0.10 &3.10 \\
0.25 &0.07 &$-$0.08    &0.15 &2.01 \\
0.30 &0.04 &$-$0.17    &0.21 &1.46 \\
0.35 &0.03 &$-$0.22    &0.25 &1.24 \\
0.40 &0.01 &$-$0.25    &0.26 &1.19 \\
0.45 &0.00 &$-$0.27    &0.27 &1.15 \\
0.50 &0.03 &$-$0.25    &0.28 &1.11 \\
0.55 &0.08 &$-$0.22    &0.30 &1.03 \\
0.60 &0.13 &$-$0.18    &0.31 &1.00 \\
0.65 &0.19 &$-$0.11    &0.30 &1.03 \\
0.70 &0.25 &$-$0.03    &0.28 &1.10 \\
0.75 &0.34 &$+$0.08    &0.26 &1.19 \\
0.80 &0.43 &$+$0.19    &0.24 &1.29 \\
0.85 &0.54 &$+$0.32    &0.22 &1.41 \\
0.90 &0.64 &$+$0.44    &0.20 &1.55 \\
0.95 &0.74 &$+$0.55    &0.19 &1.63 \\
1.00 &0.84 &$+$0.67    &0.17 &1.82 \\
1.05 &0.94 &$+$0.79    &0.15 &2.06 \\
1.10 &0.99 &$+$0.87    &0.12 &2.58 \\
\hline
\end{tabular}  
\end{table}

\section{Analyses of 20 O\!Cs}

Twenty O\!Cs which contain F-type stars have been selected for the scope of
this paper. Their selection has been based mostly on this presence of F-type stars, plus
the fact that they have had limited previous attention in the literature; only about half of
them have had previous metallicity determinations by photometric or spectroscopic methods.
These O\!Cs with F-type stars are quite valuable for photometric metal-abundance
determination from the {\sl U--B, B--V} plane in addition to providing interstellar
reddenings, distances and photometric ages.   However, our sample is rather small and may contain 
selection biases affecting the results of this paper.  Moreover, our sample includes only
a few O\!Cs in each of the quadrants I, II, and III of the Galactic disc, the largest
number in quadrant III, none in quadrant IV, and metal-poor old O\!Cs are not well
represented, as can be seen from Fig.~14.  The majority of our cluster sample is
located within $50^{o}$ of the Galactic anticentre direction.

\begin{table*}
\centering
{\scriptsize
\caption{The mean values of $\langle(U$--$B)_{H}\rangle$ of the Hyades reference line, plus the mean
$\langle(U$--$B)_0\rangle$ values, which correspond to the given $\langle(B$--$V)_0\rangle$ as set
by the iso-abundance lines for the distributions in the (B--V) range (Col.~2) of F-type stars.
The resulting ultraviolet excesses, $\delta(U$--$B)$, and their normalized
values, $\delta_{0.6}$, plus their uncertainties (A10), are given in Columns~6 and 7, respectively.
The number of dwarf stars in the F-star region is listed in the last column.}
\begin{tabular}{lccccccc}
\hline
Cluster &(B-V) range &$\langle(B$--$V)_0\rangle$ & $\langle(U$--$B)_{H}\rangle$ &$\langle(U$--$B)_0\rangle$&$\delta(U$--$B)$& $\delta_{0.6}$&N \\
\hline
NGC 6694 &(0.50$-$1.00)&0.39 &0.010 &$-$0.025 &0.035 &0.042$\pm0.025$&58 \\
NGC 6802 &(0.90$-$1.20)&0.30 &0.044 &$-$0.010 &0.054 &0.079$\pm0.020$&71 \\
NGC 6866 &(0.30$-$0.60)&0.44 &0.000 &$-$0.040 &0.040 &0.044$\pm0.010$&36 \\
NGC 7062 &(0.70$-$1.10)&0.37 &0.022 &$-$0.044 &0.066 &0.080$\pm0.015$&31 \\
Ki 05    &(0.90$-$1.30)&0.40 &0.010 &$-$0.036 &0.046 &0.058$\pm0.040$&42 \\
NGC 436  &(0.50$-$1.10)&0.40 &0.010 &$-$0.090 &0.100 &0.119$\pm0.050$&46 \\
NGC 1798 &(0.70$-$0.90)&0.31 &0.041 &$-$0.046 &0.087 &0.111$\pm0.040$&41 \\
NGC 1857 &(0.60$-$0.90)&0.33 &0.036 &$-$0.034 &0.070 &0.088$\pm0.030$&14 \\
NGC 7142 &(0.60$-$1.00)&0.50 &0.030 &$-$0.020 &0.050 &0.056$\pm0.023$&40 \\
Be 73    &(0.50$-$0.70)&0.32 &0.038 &$-$0.012 &0.050 &0.064$\pm0.010$&18 \\
Haf 04   &(0.50$-$0.90)&0.28 &0.054 &$+$0.004 &0.050 &0.084$\pm0.030$&14 \\
NGC 2215 &(0.50$-$1.00)&0.37 &0.022 &$-$0.056 &0.078 &0.095$\pm0.040$&37 \\
Rup 01   &(0.40$-$0.80)&0.38 &0.018 &$-$0.042 &0.060 &0.070$\pm0.030$&14 \\
Be 35    &(0.40$-$0.80)&0.49 &0.024 &$-$0.025 &0.049 &0.052$\pm0.030$&35 \\
Be 37    &(0.40$-$0.70)&0.40 &0.010 &$-$0.020 &0.030 &0.036$\pm0.020$&52 \\
Haf 08   &(0.60$-$0.90)&0.44 &0.000 &$-$0.080 &0.080 &0.093$\pm0.040$&34 \\
Ki 23    &(0.40$-$0.60)&0.44 &0.000 &$-$0.040 &0.040 &0.046$\pm0.020$&18 \\
NGC 2186 &(0.50$-$1.00)&0.44 &0.000 &$-$0.080 &0.080 &0.093$\pm0.030$&34 \\
NGC 2304 &(0.30$-$0.60)&0.37 &0.002 &$-$0.050 &0.052 &0.062$\pm0.030$&23 \\
NGC 2360 &(0.30$-$0.60)&0.44 &0.000 &$-$0.040 &0.040 &0.046$\pm0.020$&70 \\
\hline
\end{tabular} 
}
\end{table*}

\subsection{Interstellar reddenings and photometric metallicities}

By following the analytic methods presented in detail in the works of T10 and A10,
20 O\!Cs have been analysed in the two-colour ({\sl U--B, B--V}) diagram and five
CM diagrams together with the ZAMS intrinsic-colour calibrations of SK82, the
Hyades main sequence colours of \citet[]{san69} (Table 1A) and \citet[]{mel59, mel60},
and the Padova isochrones, \citet[][M08]{mar08},
to obtain reddenings, metallicities, distance moduli, and ages for these O\!Cs. 

As is presented in the works of T10 and  A10, interstellar reddenings of the 20 O\!Cs
have been estimated from displacements of the intrinsic-colour sequences (dwarfs plus red giants)
of SK82 in the CC diagram, as shown for Ki~05 in Fig.~1(a), until the best fit to the
data of the clusters with an {\sl U--B} shift of 0.72E{\sl (B--V)}+0.05{\sl E(B--V)}$^2$
and a {\sl (B--V)} shift of {\sl E(B--V)}. 

Photometric metal abundances [Fe/H] have been measured for F-type cluster stars in the CC diagram
with respect to the Hyades mean main-sequence line, which is shown as dashed green lines in Figs.~1(a)
and (b).  Once the interstellar reddening shift has been made, the $\langle(U$--$B)_0\rangle$ and
$\langle(B$--$V)_0\rangle$ colours have been fixed as mean values from the distribution of the F-type
stars in each cluster.  By using these $\langle(U$--$B)_0\rangle$ and $\langle(B$--$V)_0\rangle$ values,
the iso-metallicity line (solid red line, Figs.~1(a) and (b)) as representative of the mean metal abundances
of the cluster has been estimated, and these average values are given in Table~4 for the 20 O\!Cs together
with the data for the mean Hyades main sequence.  From these $\langle(U$--$B)_0\rangle$ and
$\langle(B$--$V)_0\rangle$ values of the 20 O\!Cs, the values of
$\delta(U$--$B) = (U$--$B)_{H}-\langle(U$--$B)_0\rangle$ have been measured, and normalized to
$(B$--$V)_0=0.6$ via the data of Table~1A given by \citet[]{san69}. Then, the metallicity values, [Fe/H],
for the 20 O\!Cs have been derived from the empirical calibration, [Fe/H]--$\delta(U$--$B)_{0.6}$, as
given in equation~(6) of \cite{ks06}.  These final [Fe/H] values are given for each cluster in Table~6.

Heavy-element abundances, $Z$, of the 20 O\!Cs have been converted from the photometric metal abundances,
[Fe/H], via the expression $Z = Z_\odot \cdot 10^{[Fe/H]}$.  The solar abundance value has been taken as
$Z_\odot = +0.019$, which is that adopted by M08 for their isochrones.  However, \cite{hou11},
\cite{caf09}, and  \cite{asp09} have published the values of $Z_\odot = +0.0142$, $+0.0156$, and
$+0.0134$, respectively, based on helioseismology methods or spectroscopic chemical-composition
analyses with 3D hydrodynamical solar models.  A review of these and other solar values can be found in
\citet[][Table 4]{asp09}.  These lower solar metal abundances would have systematic effects on our results;
for example, the isochrone ages would be systematically increased, and distances decreased.

Values of $\langle(B$--$V)_0\rangle$, $\langle(U$--$B)\rangle_{H}$, $\langle(U$--$B)_0\rangle$,
$\delta(U$--$B)$, and $\delta_{0.6}$ have been given in Table 4 for the 20 O\!Cs, and
their [Fe/H] and $Z$ values are listed in Table 6.  To be able to appreciate the photometric metal-abundance
determination and the iso-metallicity line (the red curve in Fig.~1),  Ki~05 has been taken
as an example.  The interstellar reddening value for this cluster has been measured as E(B--V)=0.70 mag by an
appropriate shift in the CC diagram.  The stars above the Hyades mean relation in  the CC diagrams of Figs.~1(a)
and (b) occupy the regions of $(B$--$V) \approx 0.9-1.3$, or $(B$--$V)_{0} \approx 0.20-0.60$, i.e. mostly
F types.  The mean values of the distribution correspond to $\langle (B$--$V)_0\rangle = 0.40$ and
$\langle(U$--$B)_0\rangle = -0.036$ for the F-type members of Ki~05.  Then, 
$\delta(U$--$B)=(U$--$B)_{H}-\langle(U$--$B)_0\rangle = +0.01 -(-0.036)=+0.046$, and
$\delta_{0.6}/\delta(U$--$B)=1.19$ has been obtained via Table 1A.  Finally,
$\delta_{0.6}=+0.058$ is converted into $[Fe/H] = -0.17$ $(Z=+0.013)$ from the calibration of
[Fe/H]--$\delta(U$--$B)_{0.6}$ of \cite{ks06}, who also have used the Hyades mean colours as reference.
Thus, from the estimated values of $\langle (U$--$B)_0 \rangle$ for $(B$--$V)_{0} \approx 0.24-0.44$,
the iso-metallicity line which corresponds to $Z=+0.013$ has been drawn in Fig.~1 (the red curve).

In addition, as is seen from the plots of the CM diagrams (Figs.~2-3 and Figs.~S1-S19), O\!Cs with RC/RG 
candidate stars are as follows:  Ki~05, NGC~6802, NGC~6866, NGC~7062, NGC~1798, 
NGC~7142, Ru~01, Be~35, Be~37, Ki~23, NGC~2304, and NGC~2360.  For these twelve O\!Cs, 
to determine photometric metal abundances, the F-type stars
in CC plots have been fit above the ZAMS colours of Hyades main sequence and simultaneously the RC/RG
stars above the red-giant colours of SK82 with consistent ultraviolet excesses according to the
normalizations of \cite{san69}. The best fit, the solid curve in
CC diagrams, has the brightest bluer stars ($0.20<(B$--$V)_{0}<0.60$) slightly above the F-star hump
of the Hyades main sequence, while the RC/RG stars ($0.90<(B$--$V)_{0}<1.30$) are slightly above the
red-giant colours of SK82.  The mean $(B$--$V)_{0}$ values and the numbers of F-type stars falling
in their observed (B--V) range (Col.~2; Table 4) are given in
Cols.~3 and 8 of Table~4, respectively.  For the other eight O\!Cs, for which RC/RG candidates are not clearly
observed, only the F-type dwarf stars have been used for the metal abundance.

The O\!Cs:  Ki~05, NGC~6802, NGC~7062, NGC~1798, NGC~7142, Be~35, Be~37, NGC~2304 and NGC~2360 show a
clumpy distribution over $0.80<(B$--$V)_{0}<1.05$ in the CM diagrams, showing convincingly the
presence of RC stars.  On the other hand, Ki~23 shows a rather clear red-giant (RG) sequence in the
CM diagrams (panels (b)--(f) of Figs.~S15) with a few stars blueward of this RG
branch which are also good RC candidates.  NGC~6866 and Ru~01 (Figs.~S3 and S12) show 
RC/RG stars which are probable good candidates for an RC star due to their position with respect to
the isochrone.  All of these RC stars have been emphasized with big open circles in the CC and CM
diagrams due to their considerable usefulness as distance indicators.

\subsection{Distances and Ages}

The CM plots from the analysis of Ki~05 have been displayed in Figs.~2--3.  The astrophysical
parameters of Ki~05 are given in Table 5, and then summarized in Table~6 together with the
fundamental astrophysical parameters for the other 19 O\!Cs.  The CC and CM diagrams, and the
corresponding tables for the other 19 O\!Cs have been presented as Tables S1--S4, and Figs.~S1--S19
in the supplementary electronic section.

As is seen from Figs.~2(a)--(c) and 3(d)--(e) for Ki~05, the M08 isochrones, corresponding to
the $Z$ value given in Table~4, have been over-plotted in five CM diagrams:  $V,(B$--$V)$;
$V,(R$--$I)$; $V,(V$--$I)$; $V,(V$--$R)$; and $V,(B$--$R)$, after reddening the isochrones along the
colour axis with a colour excess corresponding to the $E(B$--$V)$ value given in Table~6,
converted with help of the interstellar extinction ratios given in Table~6 of A10, and adding a
visual extinction of $A_{\rm V} = 3.1\times E(B$--$V)$ to the absolute magnitudes of the isochrones.
The isochrones have then been shifted vertically to obtain the best fit to the observed intermediate
section of the main sequence (MS), as well as the RC sequence.  This vertical shift is the (true)
distance modulus, $DM =(V_{0}$--$M_{\rm V})$.  The average distance moduli and distances from
the five CM diagrams for the 20 O\!Cs have also been given in Table~6.  To derive an age estimate for the
clusters, the M08 isochrones, for the appropriate $Z$ values, have been shifted in the CM planes as
above, i.e.\ $ M_{V} + 3.1E(B$--$V) + DM$ and $C_0(\lambda_1-\lambda_2) + E[C(\lambda_1-\lambda_2)]$,
respectively, and then the isochrone ages have been varied until a satisfactory
fit to the data has been obtained through the observed upper MS, TO, and RC sequences.
The resulting average inferred mean ages are presented in Table~6, Col.~8.

For most of these CM diagrams, two isochrones have been plotted to provide a means for
appreciating the uncertainties of the derived distances and ages.  In Column~4 of Table~5 and of
Tables S1-S4 (see the supplementary section), the range in ages provided by these isochrone pairs
is given.  The final mean values for the distances and ages from the five CM diagrams are given in
the final line for each cluster in these tables; the error estimates and the calculation of the
weighted mean values of {\sl E(B--V)}, [Fe/H], $(V_{0}$--$M_{\rm V})$, $d$ (kpc), log(A), and A
have been calculated as described in Sect.~3.4 of A10.  For the parameters, $(V_{0}$--$M_{\rm V})$,
$d$ (kpc), and log(A), the mean values and the associated mean uncertainty have been calculated from
the individual uncertainties of the five CM diagrams, weighted with their respective precisions,
using the expressions (8)--(9) given in A10. These errors take into account an attempt to estimate
external errors, as in A10, since the photometric indices and the five resulting CM diagrams are
not independent.

\begin{table}
\centering
\tiny
\caption{The derived fundamental astrophysical parameters of Ki 05.} 
\begin{tabular}{lccccc}
\hline
Colour &$(V_{0}$--$M_{V})$ &  d~(kpc) &log(A)-range &log(A) & A~(Gyr) \\
\hline
\multicolumn{ 5}{l}{$E(B$--$V)=0.70\pm0.08$, $[Fe/H]=-0.17\pm0.25$, $Z=0.013\pm0.007$}  \\
\hline
$(B$--$V)$ &11.20$\pm$0.12 &1.74$\pm$0.09 &8.95$-$9.10 &9.10$\pm$0.10 &1.26$\pm$0.33 \\
$(R$--$I)$ &11.20$\pm$0.20 &1.74$\pm$0.16 &8.95$-$9.10 &9.10$\pm$0.15 &1.26$\pm$0.52 \\
$(V$--$I)$ &11.20$\pm$0.10 &1.74$\pm$0.08 &8.95$-$9.10 &9.10$\pm$0.10 &1.26$\pm$0.33 \\
$(V$--$R)$ &11.20$\pm$0.10 &1.74$\pm$0.08 &8.95$-$9.10 &9.10$\pm$0.10 &1.26$\pm$0.33 \\
$(B$--$R)$ &11.20$\pm$0.10 &1.74$\pm$0.08 &8.95$-$9.10 &9.10$\pm$0.10 &1.26$\pm$0.33 \\
\hline
Mean  & 11.20$\pm$0.05 &  1.74$\pm$0.04 &            &  9.10$\pm$0.05 &  1.26$\pm$0.16\\
&\multicolumn{1}{l}{}&\multicolumn{ 1}{l}{}&\multicolumn{ 1}{r}{}&\multicolumn{ 1}{l}{}&\multicolumn{ 1}{l}{} \\
\hline
\end{tabular} 
\end{table}

\begin{table*}
\centering
\tiny
\caption{The fundamental astrophysical parameters of the 20 O\!Cs.
Galactic coordinates have been taken from the WEBDA data-base.}
\begin{tabular}{lcccccccc}
\hline
Cluster & $\textit l^{\circ}$  &$b^{\circ}$ &E(B--V) & [Fe/H]  &Z  &$(V_{0}$--$M_{V})$ &A(Gyr) &d~(kpc) \\
\hline
NGC6694 &23.88 &$-$2.91 & 0.51$\pm$0.06&$-0.09\pm0.14$ &0.016$\pm$ 0.005 &11.10$\pm$0.04 &0.18$\pm$0.01 &1.66$\pm$0.03 \\
NGC6802 &55.34 &$+$0.92 & 0.80$\pm$0.07&$-0.30\pm0.13$ &0.009$\pm$ 0.003 &11.19$\pm$0.05 &1.12$\pm$0.08 &1.73$\pm$0.04 \\
NGC6866 &79.56 &$+$6.84 & 0.06$\pm$0.05&$-0.10\pm0.05$ &0.015$\pm$ 0.002 &10.61$\pm$0.02 &0.75$\pm$0.04 &1.32$\pm$0.01 \\
NGC7062 &89.96 &$-$2.75 & 0.43$\pm$0.08&$-0.31\pm0.09$ &0.010$\pm$ 0.002 &11.40$\pm$0.02 &0.71$\pm$0.04 &1.91$\pm$0.02 \\
Ki05    &143.78&$-$4.29 & 0.70$\pm$0.08&$-0.17\pm0.25$ &0.013$\pm$ 0.007 &11.20$\pm$0.05 &1.26$\pm$0.16 &1.74$\pm$0.04 \\
NGC436  &126.11&$-$3.91 & 0.40$\pm$0.07&$-0.55\pm0.33$ &0.005$\pm$ 0.004 &11.90$\pm$0.05 &0.18$\pm$0.03 &2.40$\pm$0.05 \\
NGC1798 &160.70&$+$4.85 & 0.47$\pm$0.07&$-0.50\pm0.28$ &0.006$\pm$ 0.004 &12.70$\pm$0.04 &1.78$\pm$0.22 &3.47$\pm$0.06 \\
NGC1857 &168.40&$+$1.26 & 0.47$\pm$0.08&$-0.36\pm0.19$ &0.008$\pm$ 0.003 &11.98$\pm$0.04 &0.32$\pm$0.04 &2.49$\pm$0.05 \\
NGC7142 &105.35&$+$9.48 & 0.35$\pm$0.08&$-0.16\pm0.12$ &0.013$\pm$ 0.004 &11.60$\pm$0.05 &3.55$\pm$0.57 &2.10$\pm$0.05 \\
Be 73   &215.28&$-$9.42 & 0.28$\pm$0.06&$-0.21\pm0.06$ &0.012$\pm$ 0.002 &14.50$\pm$0.03 &1.41$\pm$0.08 &7.93$\pm$0.11 \\
Haf 04  &227.94&$-$3.59 & 0.47$\pm$0.09&$-0.33\pm0.19$ &0.009$\pm$ 0.008 &13.22$\pm$0.05 &0.42$\pm$0.05 &4.39$\pm$0.10 \\
NGC 2215&215.99&$-$10.10& 0.23$\pm$0.07&$-0.40\pm0.27$ &0.008$\pm$ 0.005 & 9.60$\pm$0.03 &0.64$\pm$0.05 &0.83$\pm$0.01 \\
Rup 01  &223.99&$-$9.69 & 0.17$\pm$0.06&$-0.25\pm0.18$ &0.011$\pm$ 0.005 &10.85$\pm$0.04 &0.48$\pm$0.04 &1.48$\pm$0.03 \\
Be 35   &212.60&$+$5.35 & 0.11$\pm$0.07&$-0.13\pm0.18$ &0.014$\pm$ 0.006 &13.50$\pm$0.04 &0.89$\pm$0.06 &5.01$\pm$0.10 \\
Be 37   &217.23&$+$5.94 & 0.05$\pm$0.05&$-0.05\pm0.08$ &0.017$\pm$ 0.003 &13.60$\pm$0.02 &0.63$\pm$0.06 &5.25$\pm$0.06 \\
Haf 08  &227.53&$+$1.34 & 0.32$\pm$0.07&$-0.39\pm0.26$ &0.008$\pm$ 0.005 &11.88$\pm$0.04 &0.56$\pm$0.07 &2.38$\pm$0.04 \\
Ki 23   &215.53&$+$7.20 & 0.03$\pm$0.02&$-0.11\pm0.11$ &0.015$\pm$ 0.004 &12.40$\pm$0.02 &1.78$\pm$0.07 &3.02$\pm$0.03 \\
NGC 2186&203.54&$-$6.19 & 0.26$\pm$0.07&$-0.39\pm0.26$ &0.008$\pm$ 0.005 &11.40$\pm$0.03 &0.32$\pm$0.04 &1.91$\pm$0.03 \\
NGC 2304&197.21&$+$8.90 & 0.03$\pm$0.03&$-0.20\pm0.18$ &0.012$\pm$ 0.005 &12.79$\pm$0.02 &0.93$\pm$0.03 &3.61$\pm$0.03 \\
NGC 2360&229.81&$-$1.42 & 0.01$\pm$0.07&$-0.11\pm0.11$ &0.015$\pm$ 0.004 &10.25$\pm$0.02 &1.12$\pm$0.07 &1.12$\pm$0.01 \\
\hline
\end{tabular} 
\end{table*}

\begin{figure}
\epsfig{file=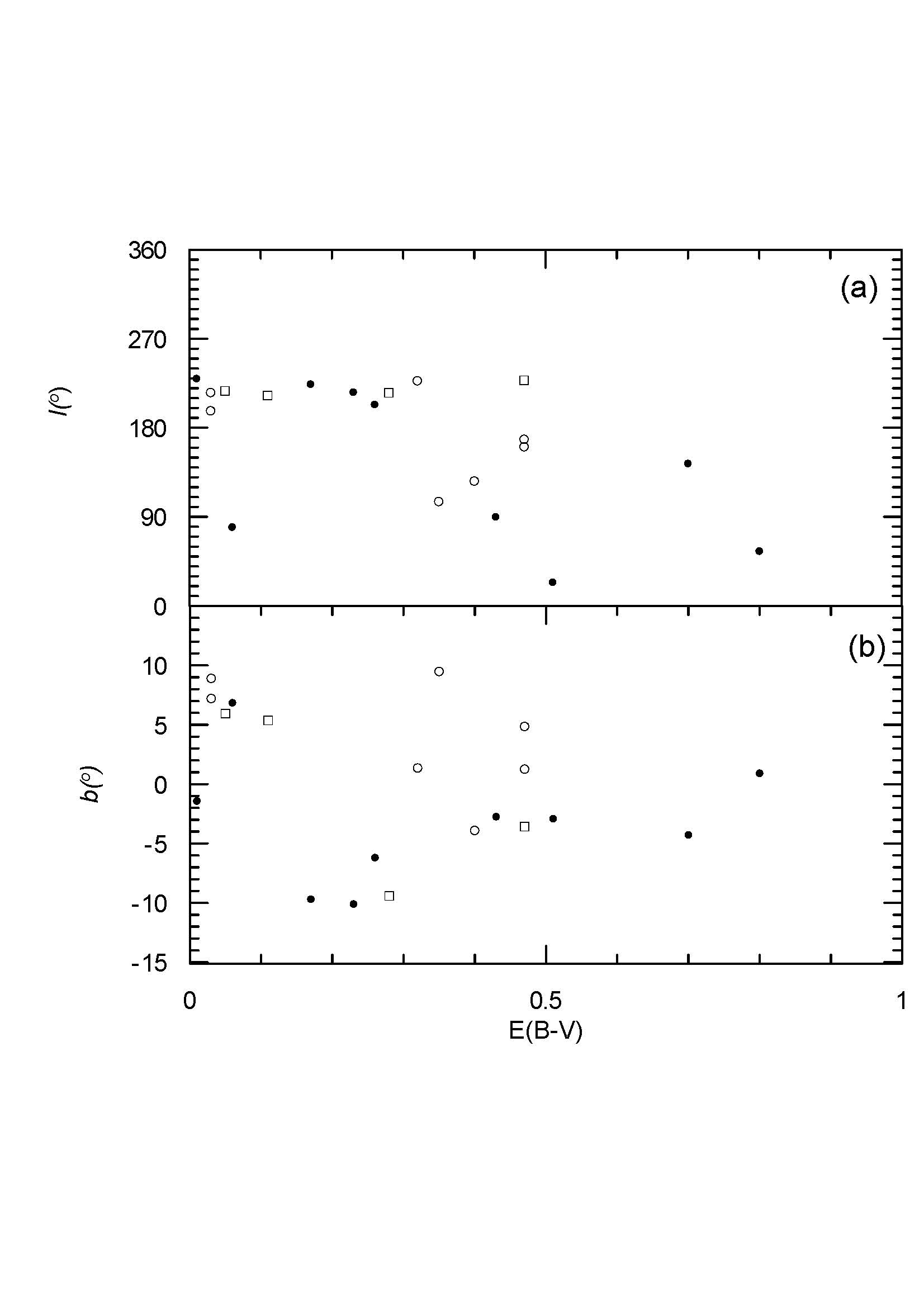, width=7cm, height=8.5cm}
\caption {(a)  $E(B$-$V)$ versus  $\textit l^{\circ}$ and (b) versus  $\textit b^{\circ}$.
 Filled and open circles show the O\!Cs with d = [0, 2] kpc and d = [2, 4]
kpc, respectively, while open squares represent those with d $>$ 4 kpc}
\end{figure}

\begin{figure}
\epsfig{file=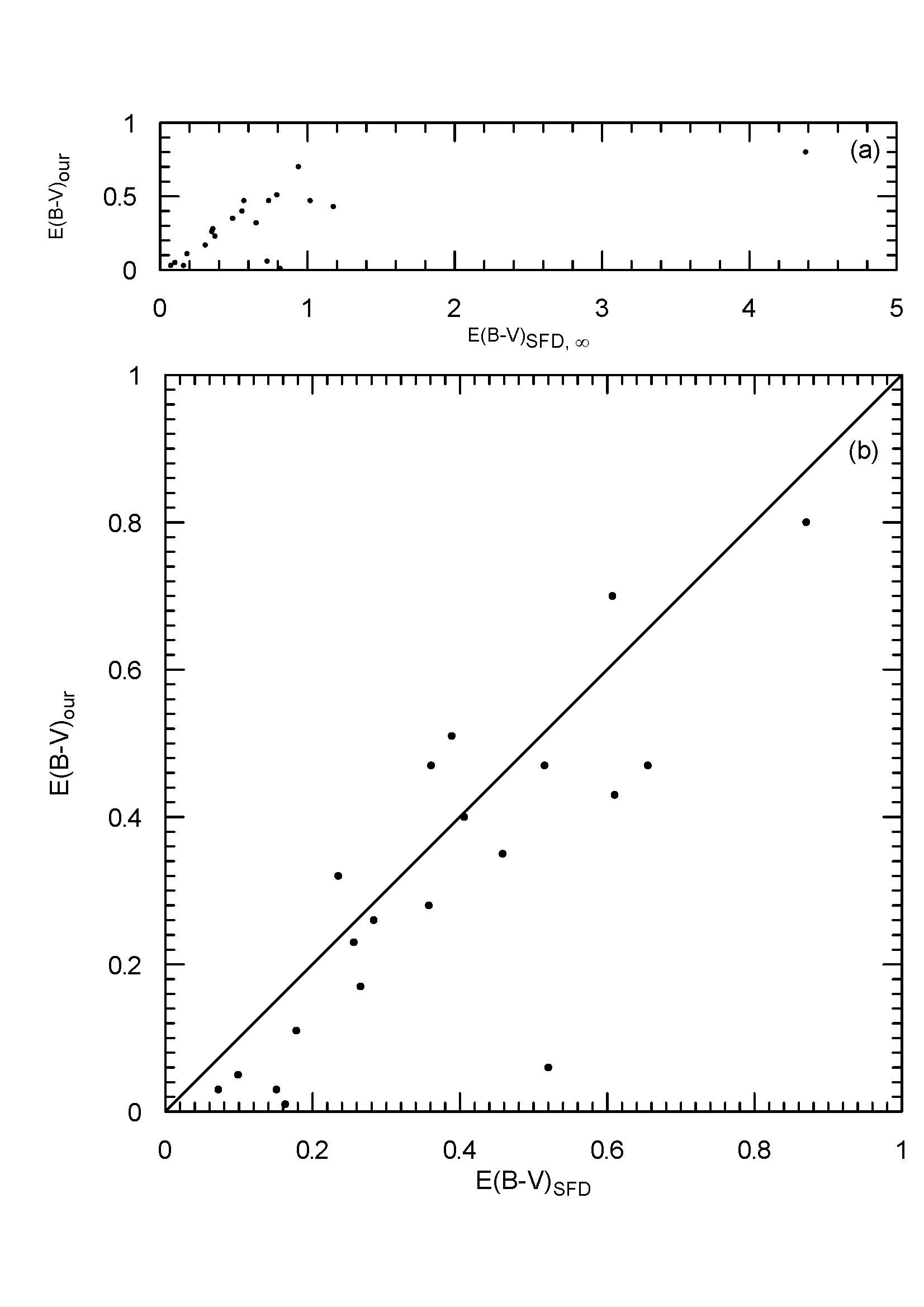, width=7cm, height=8cm}
\caption {(a) $E(B$--$V)_{\rm SFD,\infty}$ versus $E(B$--$V)$, 
and (b) $E(B$--$V)_{\rm SFD}$ versus $E(B$--$V)$.}
\end{figure}

\subsection{Correlations of the Interstellar Reddenings}

The $E(B$--$V)$ versus Galactic longitude ($\textit l^{\circ}$) and latitude ($\textit b^{\circ}$) plots,
as a function of the cluster distances, have been given in Figs.~4(a) and (b), respectively, where
filled and open circles show the $d = [0,~2]$ kpc and  $d = [2,~4]$ kpc subsets, respectively; open
squares represent those with $d > 4$ kpc.  
As is seen from panel~(a), except for Ki~05 with  E(B$-$V)=0.70 and l=143$^{\circ}.78$, the reddenings
of the O\!Cs in the anticentre directions have $E(B$--$V) < 0.50$; the distances of these O\!Cs fall
in the range $0 < d < 8$ kpc.
There are two O\!Cs with $E(B$--$V) > 0.50$ in the Galactic centre
direction, and the distances of clusters in quadrant I are less than 2 kpc.  The distribution of
$E(B$--$V)$--$\textit l^{\circ}$ in Fig.~4(a) is in quite good agreement with the one of
\citet[][fig.~6]{josh05}.  It is seen from panel~(b) that the O\!Cs fall in the range of
$-10^{\circ}$ $\le$ b $\le$  $+10^{\circ}$.  O\!Cs with $|b|$ $>$ $5^{\circ}$ have
$E(B$--$V)$ $<$ $0.40$, whereas the O\!Cs inside $|b|$$<$$5^{\circ}$ fall in the range
$0.30<E(B$--$V)<0.80$. 

The $E(B$--$V)$ values from extinction maps given by \citet[][hereafter SFD]{sch98} (based on the 
IRAS 100-micrometer surface brightness converted to extinction) have been compared with our values.
The relations of $E(B$--$V)_{\rm SFD,\infty}$ versus $E(B$--$V)$, and $E(B$--$V)_{\rm SFD}$
versus $E(B$--$V)$ of the 20 O\!Cs are displayed in Figs.~5(a) and (b), respectively.  It is seen from
Fig.~5(a) that there is a large discrepancy between the $E(B$--$V)$ values for five O\!Cs, with the
$E(B$--$V)_{\rm SFD,\infty}$ values being much larger than ours.
For a correction of the SFD reddening estimates, the equations of \cite{bon} and  
\cite{sch04} have been adopted.  Then the final reddening, $E(B$--$V)_{\rm SFD}$, for a
given star is reduced compared to the total reddening $E(B$--$V)(\ell, b)_\infty$
by a factor $\lbrace1-\exp[-d \sin |b|/H]\rbrace$, given by \cite{bs80}, where $b$, $d$, and $H$ are
the Galactic latitude (Column~3 of Table~6), the distance from the observer to
the object (Column~9 of Table~6), and the scale height of the dust layer in the
Galaxy, respectively; here we have assumed $H = 125$ pc \citep{bon}.  Note that Galactic latitudes
of our O\!Cs are less than $10^\circ$.  These reduced final reddenings have been compared with 
our measured ones in Fig.~5(b).  The differences in $\Delta E(B$--$V)$ in Fig.~5(b) are at the level of 0.11--0.46  for eight O\!Cs
between corresponding $E(B$--$V)$ values. For the rest, the $E(B$--$V)$ values of the
clusters are in fairly good concordance with the ones of SFD.  However, as discussed by \cite{chen99},
from samples of open and globular clusters at $|b|>2.5^{\circ}$, the SFD reddening values tend to overestimate
$E(B$--$V)$ by a factor of up to 1.16--1.18.  \cite{arc99}, \cite{camb05}, and \cite{josh05} have also
confirmed that SFD maps overestimate the extinction in several parts of the sky.  (The corrections of
\cite{bon} and  \cite{sch04} have been derived from such studies.)  According to \cite{camb05}, 
the reason is due to the presence of fluffy/composite grains leading to an enhanced far-infrared
emissivity.

\begin{figure}
\center
\epsfig{file=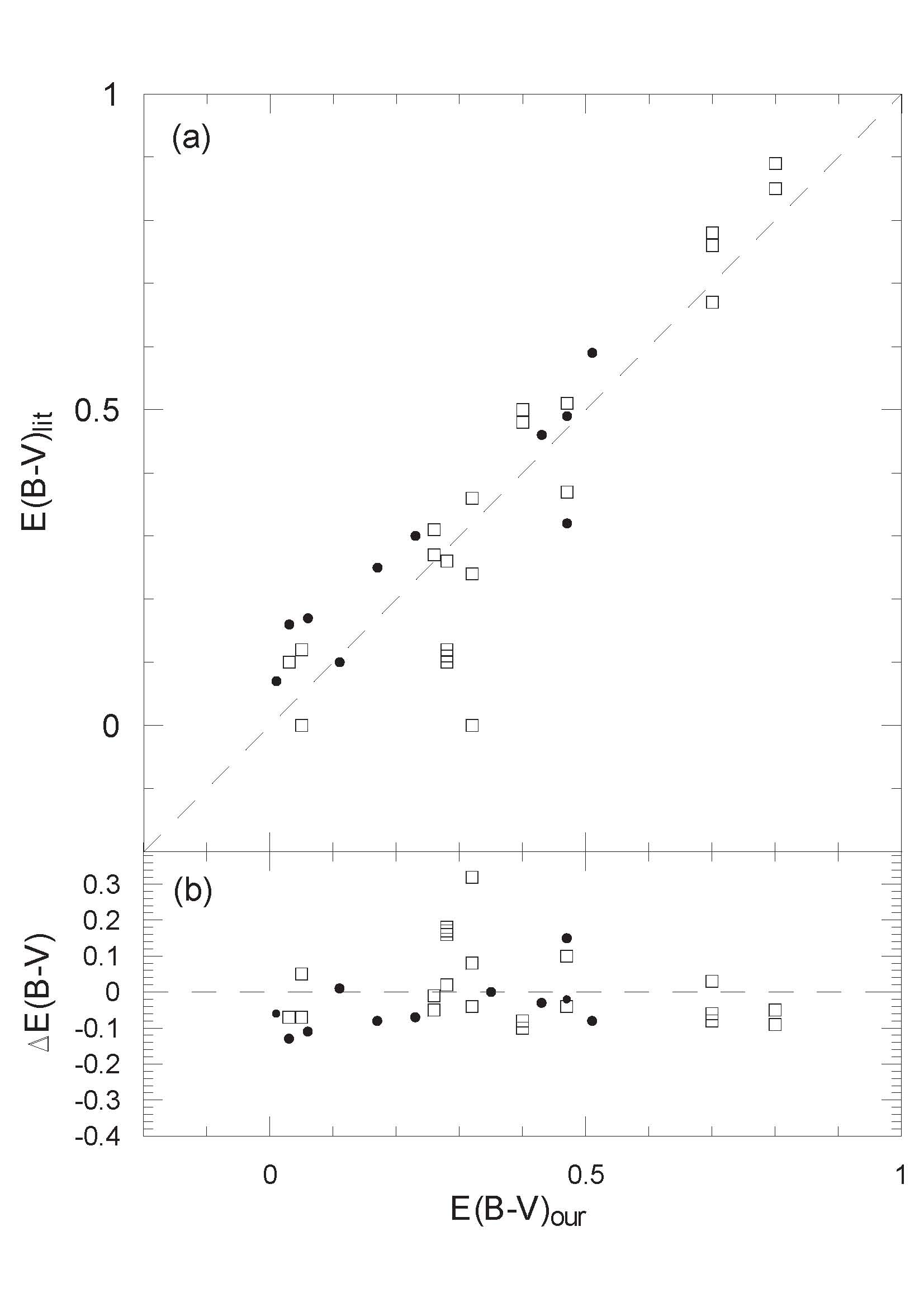, width=7 cm, height=7.5cm}
\caption {Comparison of our $E(B$--$V)$'s with ones from the
literature.  Filled circles represent O\!Cs with only one literature value, 
whereas open squares show ones with more than one literature value.  Panel
(b) shows differences between our values and those of the literature,
$\Delta E(B$--$V) = E(B$--$V)_{\rm our} - E(B$--$V)_{\rm lit}$.}
\end{figure}

\begin{figure}
\center
\epsfig{file=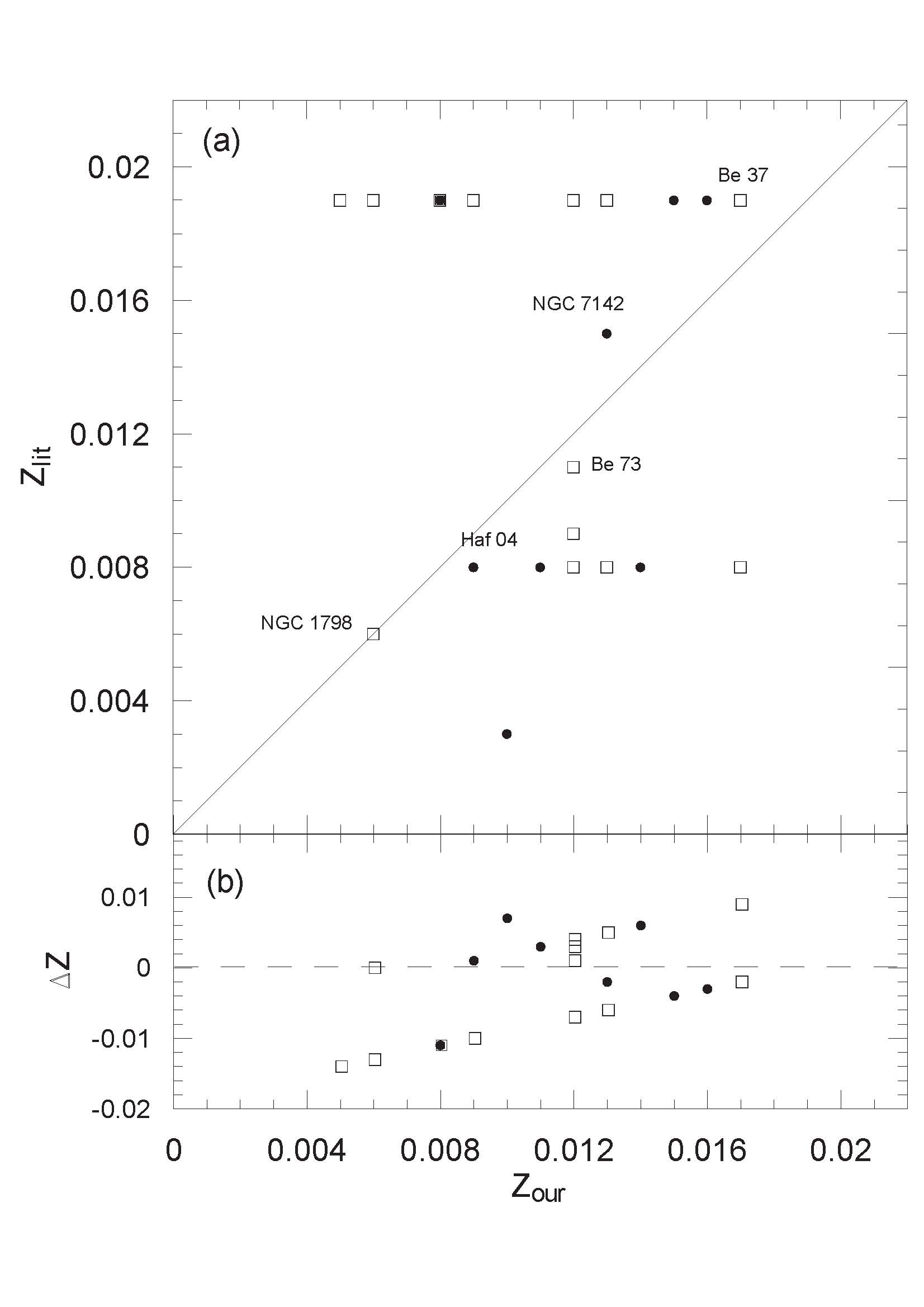, width=7 cm, height=7.5cm}
\caption {Comparison of the mass-fraction heavy-element abundances, $Z$,
with values from the literature.  Again, panel (b) shows our differences with
respect to the literature as a function of $Z_{\rm our}$,
$\Delta Z = Z_{\rm our} - Z_{\rm lit}$.
The symbols are the same as in Fig.~6.}
\end{figure}

\begin{figure}
\center
\epsfig{file=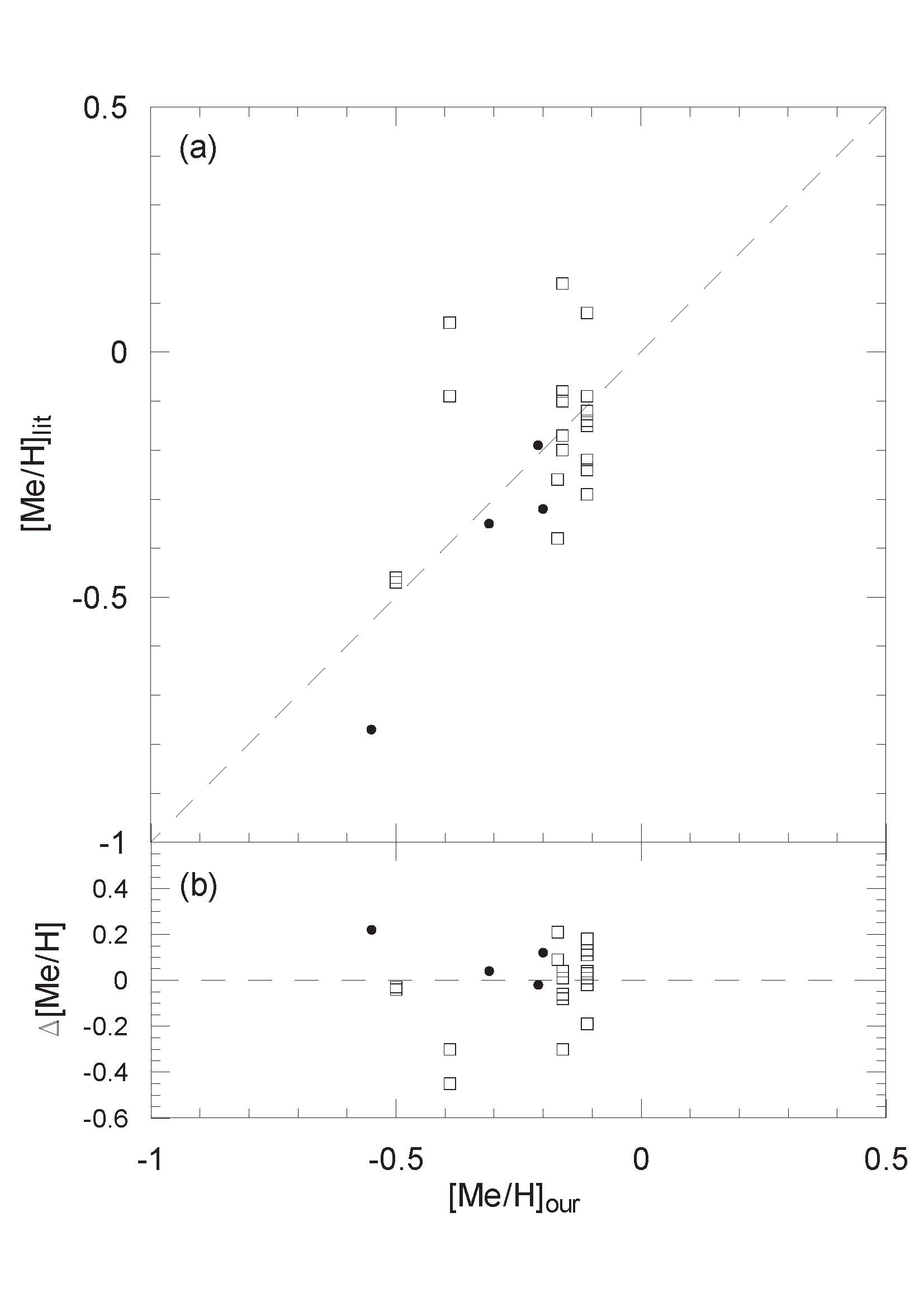, width=7 cm, height=7.5cm}
\caption {Comparison of the logarithmic metal abundances, [Me/H], of nine O\!Cs
with values from the literature.  The symbols and panels are the same as in Fig.~6.}
\end{figure}

\begin{figure}
\center
\epsfig{file=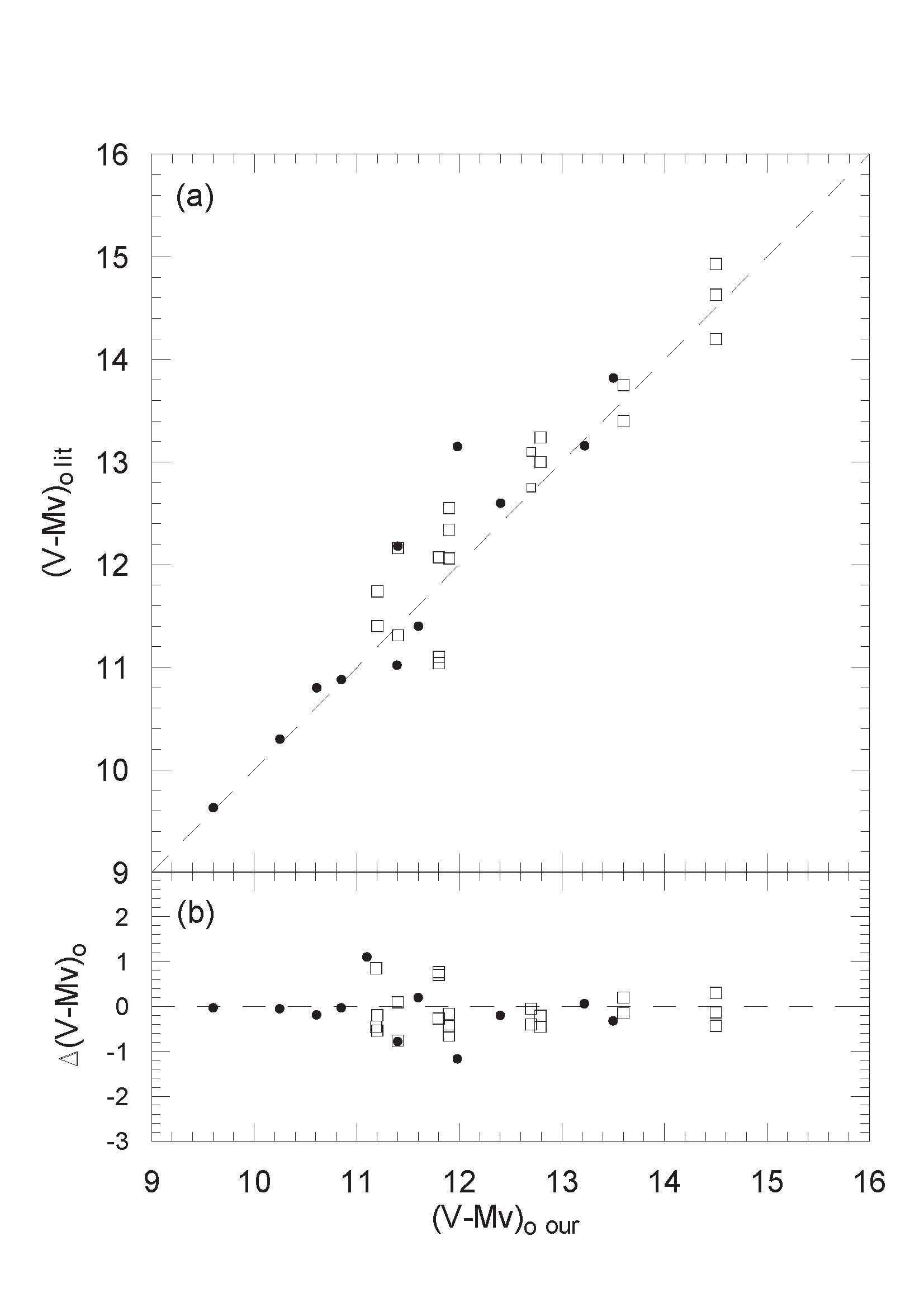, width=7.3 cm, height=7.8cm}
\caption {Comparison of our distance moduli with ones
from the literature.  The symbols and panels as in Fig.~6.}
\end{figure}

\begin{figure}
\center
\epsfig{file=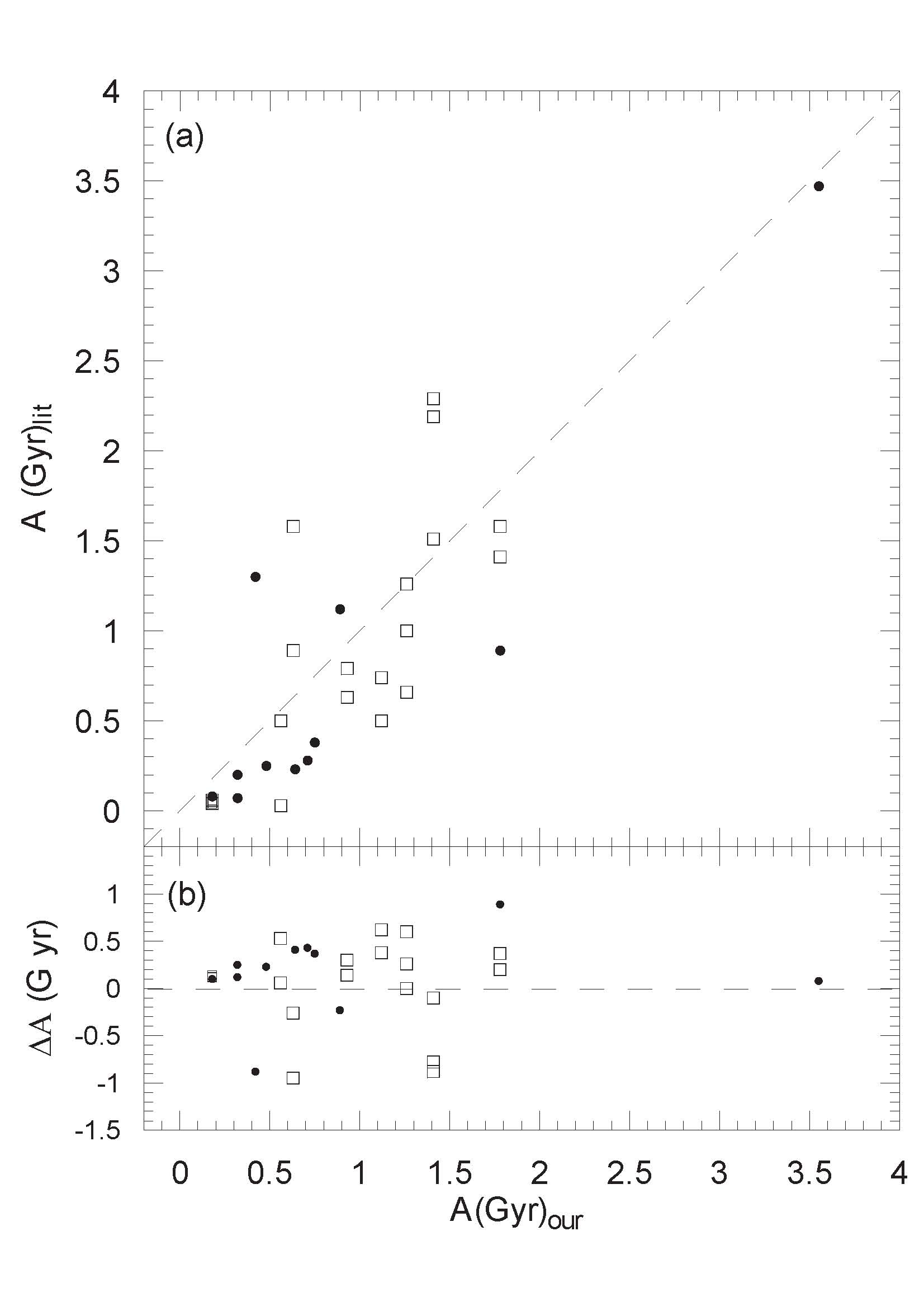, width=7.3 cm, height=7.8cm}
\caption {Comparison of our ages with ones from the
literature.  The symbols and panels as in Fig.~6.}
\end{figure}
\newpage

\begin{sidewaystable}
\centering
\tiny
\caption{The comparisons of reddenings, heavy-element abundances, distance moduli/distances, and ages of the 20 O\!Cs.  
Columns$~$(2)--(6) show the values of this work, and Columns$~$(7)--(12) the values from the literature.  In Column$~$(9)
ZAMS/OMS indicates whether a ZAMS from either isochrones or from an observational main sequences (OMS) has been used in
the literature.  References for the data in Columns~7--12 are listed in the last column.}
\begin{tabular}{lcccccccccccc}
\hline
Cluster  &  $E(B$--$V)$ & Z &$(V_{0}$--$M_{V})$ & d~(kpc) &  A~(Gyr) & $E(B$-$V)$ & $Z$ & ZAMS/OMS &$(V_{0}$--$M_{V})$&    d~(kpc) & A(Gyr) & References \\
\hline
NGC 6694 &0.51 &0.016 &11.10 &1.66 &0.18 &0.59 &0.019 &OMS                        &10.00 &1.00 &0.08 &1 \\
NGC 6802 &0.80 &0.009 &11.19 &1.73 &1.12 &0.85 &0.019 &OMS                        &10.34 &1.17 &0.74 &2  \\
         &     &      &      &     &     &0.89 &0.019 &Claret et al. 2003; ZAMS   &11.64 &2.13 &0.50 &3  \\
NGC 6866 &0.06 &0.015 &10.61 &1.32 &0.75 &0.17 &0.019 &OMS                        &10.80 &1.45 &0.38 &2  \\
NGC 7062 &0.43 &0.010 &11.40 &1.91 &0.71 &0.46 &0.003 &Vandenberg 1985; ZAMS      &12.18 &2.73 &0.28 &4  \\
Ki 05    &0.70 &0.013 &11.20 &1.74 &1.26 &0.67 &0.019 &Bertelli et al. 1994; ZAMS &11.74 &2.23 &1.26 &5  \\
         &     &      &      &     &     &0.78 &0.008 &Vandenberg 1985; ZAMS      &11.40 &1.90 &1.00 &6  \\
         &     &      &      &     &     &0.76 &0.008 &Bertelli et al. 1994; ZAMS &11.40 &1.90 &0.66 &7  \\
NGC 436  &0.40 &0.005 &11.90 &2.40 &0.18 &0.48 &0.019 &Bertelli et al. 1994; ZAMS &12.34 &2.94 &0.06 &8  \\
         &     &      &      &     &     &0.50 &0.019 &Maeder and Meynet 1991; ZAMS&12.55 &3.24 &0.04&9 \\
         &     &      &      &     &     &0.48 &0.019 &Maeder and Meynet 1989; ZAMS&12.06 &2.58 &0.06&10 \\
NGC 1798 &0.47 &0.006 &12.70 &3.47 &1.78 &0.37 &0.019 &Bertelli et al. 1994; ZAMS &12.75 &3.55 &1.58 &11 \\
         &     &      &      &     &     &0.51 &0.006 &Bertelli et al. 1994; ZAMS &13.10 &4.17 &1.41 &12 \\
NGC 1857 &0.47 &0.008 &11.98 &2.49 &0.32 &0.49 &0.019 &Barbaro et al. 1969; ZAMS  &13.15 &4.27 &0.07 &13 \\
NGC 7142 &0.35 &0.013 &11.60 &2.10 &3.55 &0.35 &0.015 &Mermilliod 1981; ZAMS      &11.40 &1.91 &3.47 &14 \\
Be 73    &0.28 &0.012 &14.50 &7.93 &1.41 &0.10 &0.008 &Bertelli et al. 1994; ZAMS &14.20 &6.92 &2.29 &15 \\
         &     &      &      &     &     &0.12 &0.008 &Girardi et al. 2000; ZAMS  &14.93 &9.68 &1.51 &16 \\
         &     &      &      &     &     &0.11 &0.011 &Girardi et al. 2000; ZAMS  &14.93 &9.68 &1.51 &17 \\
         &     &      &      &     &     &0.26 &0.008 &Bertelli et al. 1994; ZAMS &14.63 &8.43 &2.19 &18 \\
Haf 04   &0.47 &0.009 &13.22 &4.39 &0.42 &0.32 &0.008 &Bertelli et al. 1994; ZAMS &13.16 &4.29 &1.30 &18 \\
NGC 2215 &0.23 &0.008 &9.60  &0.83 &0.64 &0.30 &0.019 &OMS                        & 9.63 &0.84 &0.23 &19 \\
Rup 01   &0.17 &0.011 &10.85 &1.48 &0.48 &0.25 &0.008 &Girardi et al. 2002; ZAMS  &10.88 &1.50 &0.25 &20 \\
Be 35    &0.11 &0.014 &13.50 &5.01 &0.89 &0.10 &0.008 &Bertelli et al. 1994; ZAMS &13.82 &5.81 &1.12 &18  \\
Be 37    &0.05 &0.017 &13.60 &5.25 &0.63 &0.00 &0.008 &Bertelli et al. 1994; ZAMS &13.75 &5.62 &1.58 &18  \\
         &     &      &      &     &     &0.12 &0.019 &Bonatto et al. 2004; ZAMS  &13.40 &4.79 &0.89 &21  \\
Haf 08   &0.32 &0.008 &11.88 &2.38 &0.56 &0.00 &0.019 &SK65; OMS                  &11.10 &1.66 &0.03 &22  \\
         &     &      &      &     &     &0.36 &0.019 &SK65; OMS                  &11.04 &1.61 &   - &23 \\
         &     &      &      &     &     &0.24 &0.019 &Girardi et al. 2000; ZAMS  &12.07 &2.59 &0.50 &17  \\
Ki 23    &0.03 &0.015 &12.40 &3.02 &1.78 &0.16 &0.019 &Bonatto et al. 2004; ZAMS  &12.60 &3.31 &0.89 &21  \\
NGC 2186 &0.26 &0.008 &11.40 &1.91 &0.32 &0.31 &0.019 &SK65; OMS                  &11.31 &1.83 &  -  &23  \\
         &     &      &      &     &     &0.27 &0.019 &Girardi et al. 2002; ZAMS  &12.16 &2.70 &0.20 &24  \\
NGC 2304 &0.03 &0.012 &12.79 &3.61 &0.93 &0.10 &0.009 &Bertelli et al. 1994; ZAMS &13.00 &3.98 &0.79 &25  \\
         &     &      &      &     &     &0.10 &0.019 &Girardi et al. 2002; ZAMS  &13.24 &4.45 &0.63 &24  \\
NGC 2360 &0.01 &0.015 &10.25 &1.12 &1.12 &0.07 &0.019 &Eggen 1968; ZAMS           &10.30 &1.15 &   - &26  \\
\hline
\end{tabular} 
\\
(1) \cite{cuf40}, (2) \cite{hoa61}, (3) \cite{net07}, (4) \cite{pen90}, (5) \cite{mac07},
(6) \cite{dur01}, (7) \cite{car00}, (8) \cite{pan03}, (9) \cite{phel94b}, (10) \cite{hue91},
(11) \cite{mac07}, (12) \cite{par99}, (13) \cite{babu89}, (14) \cite{cri91}, (15) \cite{ort05}, 
(16) \cite{car05}, (17) \cite{car07}, (18) \cite{has08}, (19)  \cite{bec76}, (20) \cite{pia08}, 
(21) \cite{tad08}, (22) \cite{fen72}, (23) \cite{mof75}, (24) \cite{lat10}, (25) \cite{ann02}, 
(26) \cite{egg68}.
\\ 
\end{sidewaystable}

\begin{table*}
\centering
\tiny
\caption{The comparisons of the metal abundances for nine O\!Cs.  The uncertainties are shown
as $\sigma$ in Columns~3 and 6.  In Column~4 the spectroscopic $[Me/H]_{\rm lit}$ values have been
converted from the spectroscopic $[Fe/H]_{\rm lit}$'s in Column~5 via the equation of Zwitter et al.\ (2008).
The details for the abundance determinations from the literature have been indicated in Column~8.}
\begin{tabular}{lcccccccc}
\hline
 Cluster&[Fe/H] & $\sigma$ &$[Me/H]_{\rm lit}$ &$[Fe/H]_{\rm lit}$ &$\sigma$ & References & Remarks&    N \\
\hline
NGC 7062 &$-$0.31 &0.09 &$-$0.35 &$-$0.35 &     & 1 & Str\"omgren photometry             &         \\
Ki 05    &$-$0.17 &0.25 &$-$0.38 &$-$0.38 &     & 2 & Bertelli et al. (1994) ZAMS     &            \\
         &      &       &$-$0.26 &$-$0.30 &0.17 & 3 & Spectroscopy$-$Giants           &            \\
NGC 436  &$-$0.55 &0.33 &$-$0.77 &$-$0.77 &     & 4 & $\delta(U$--$B)$$-$UBV photometry  &        \\
NGC 1798 &$-$0.50 &0.28 &$-$0.47 &$-$0.47 &     & 5 & Bertelli et al.(1994)   ZAMS    &            \\
         &      &       &$-$0.46 &$-$0.46 &     & 4 & $\delta(U$--$B)$$-$UBV photometry  &        \\
NGC 7142 &$-$0.16 &0.12 &$-$0.10 &$-$0.10 &0.10 & 6 & Washington photometry           &            \\
         &      &       &$-$0.17 &$-$0.17 &     & 7 & Washington photometry           &            \\
         &      &       &$-$0.20 &$-$0.23 &0.13 & 8 & Spectroscopy                    &  11 Giants \\
         &      &       &$-$0.08 &$-$0.10 &0.10 & 3 & Spectroscopy                    &  12 Giants \\
         &      &       &+0.14   &+0.14   &0.01 & 9 & Spectroscopy                    &  4 Giants  \\
Haf 08   &$-$0.39 &0.26 &$-$0.09 &$-$0.09 &0.10 &10 & DDO-Washington photometry       &            \\
         &      &       &+0.06   &+0.06   &0.06 &11 & DDO photometry                  &            \\
         &      &       &+0.06   &+0.06   &0.04 &12 & DDO photometry                  &            \\
Be 73    &$-$0.21 &0.06 &$-$0.19 &$-$0.22 &     &13 & Spectroscopy                    & 2 Giants   \\
NGC 2304 &$-$0.20 &0.18 &$-$0.32 &$-$0.32 &     &14 & Bertelli et al (1994) ZAMS      &            \\
NGC 2360 &$-$0.11 &0.11 &$-$0.09 &$-$0.09 &     &15 & $\delta(U$--$B)$$-$UBV photometry  &        \\
         &      &       &$-$0.14 &$-$0.14 &0.07 &11,16 & DDO photometry               &            \\
         &      &       &$-$0.12 &$-$0.12 &0.03 &17 & DDO photometry                  &            \\
         &      &       &$-$0.15 &$-$0.15 &0.11 &12 & DDO photometry                  &            \\
         &      &       &$-$0.29 &$-$0.29 &0.04 & 7 & Washington photometry           &            \\
         &      &       &$-$0.24 &$-$0.28 &0.05 & 8 & Spectroscopy - Giants           &            \\
         &      &       &$-$0.22 &$-$0.26 &0.02 & 3 & Spectroscopy - Giants           &            \\
         &      &       &+0.08   &+0.07   &0.07 &18 & Spectroscopy                    &  7 Giants  \\
\hline
\end{tabular}  
\\
(1) \cite{pen90}, (2) \cite{car00}, (3) \cite{fri02}, (4) \cite{tad03},
(5) \cite{par99},  (6) \cite{can86}, (7) \cite{gei91, gei92}, (8) \cite{fri93}, 
(9) \cite{jac08}, (10) \cite{cla89}, (11) \cite{pia95}, (12) \cite{twa97},
(13) \cite{car07}, (14)\cite{ann02}, (15) \cite{cam85}, (16) \cite{cla99}, 
(17) \cite{cla08}, (18) \cite{ham00}\\ 
\end{table*}

\section{Comparisons of Astrophysical Parameters}

The comparison of the astrophysical parameters {\sl E(B--V)}, [Fe/H], $Z$, $(V_{0}$--$M_{V})$,
$d~(kpc)$, and $A~(Gyr)$ of the 20 O\!Cs, and their differences with the literature have been
given in Figs.~6--10, where the filled circles represent the O\!Cs with only one literature value, 
and open squares show those with more than one literature value.  These literature
values and their sources have been listed in Table~7.  If we are to do a large survey with
between 300 and 500 OCs, such comparisons are necessary and useful to evaluate the
consistency and quality of the results in the literature, and to provide the offsets between
all these different studies, ours and those in the literature.  Our large data set can
provide a standard for many such comparisons, and so lead to a final, more homogeneous, set of
cluster parameters.  As is seen from Figs.~6(a)--(b), there is good consistency,
$\Delta E(B$--$V) \approx \pm0.10$, between $E(B$--$V)_{\rm lit}$ and $E(B$--$V)_{\rm our}$.
For Be~73, there is a discrepancy with the three literature values reaching a level of
$\Delta E(B$--$V) = 0.18$, but with the value given by \cite{has08} agreeing very well with
our value (Table~7).  For Haf~08, the largest discrepancy is due to the $E(B$--$V)=0.00$
value given by \cite{fen72}; two other literature values agree to within 1-$\sigma$.

As can be seen from the comparison of the mass-fraction heavy-element abundances in Figs.~7(a)--(b),
and Table~7, for the O\!Cs:  NGC~1798, Be~73, NGC~7142, Haf~04, and Be~37, there is fairly
good agreement between $Z_{\rm lit}$ and $Z_{\rm our}$ for many of these literature values.  Often
when the agreement is not so good, it is because the authors have assumed the solar value (+0.019)
for the cluster metal mass fraction as a conservative approximation, as can be seen in Column~8 of
Table~7.  (See Sect.~3 for a discussion of other possible solar values.)  As is given in Column~9 of
Table~7, sometimes observational main-sequences and theoretical isochrones with $Z_{\odot}$ have
been adopted for comparison in the literature when the authors have no ultraviolet measure
to determine the line-blanketing effect in the stellar atmospheres.  As is evident from Fig.~7(a),
these authors have adopted the solar mass fraction for nine clusters, in nearly all cases quite
different from our measured values. 

In our case, the photometric $Z$ values have been converted from the photometric [Fe/H] abundances. 
The comparison of the metal abundances for nine O\!Cs with the literature values is given 
in Figs.~8(a)--(b), and Table 8. Photometric abundances from $(U$--$B)$ excesses measure an average
metal abundance, [Me/H], whereas spectroscopic methods measure actual iron abundances, [Fe/H].
For this comparison, the iron abundances of the spectroscopic works have been converted to 
[Me/H] values via $[Me/H]=[Fe/H]+0.11(1-(1-exp(-3.6|[Fe/H]+0.55~|)$ from the relation of \cite{zwi08}.
(However, we prefer to continue using the [Fe/H] notation for the photometric abundances of these O\!Cs.)
The converted values have been listed in Column~4 of Table 8. There is good agreement between the
abundances of this work and \cite{cam85} and \cite{tad03}, which are based on the photometric
$\delta(U$--$B)$ technique for NGC~436, NGC~1798, and NGC~2360.  For the clusters NGC~7142 and Be~73,
our photometric abundances agree very well with the spectroscopic values; for NGC~7142, this
does not seem to be the case for the last spectroscopic value from reference (9),
\cite{jac08}. There seems to be some discrepancy between the abundances of our paper and the
spectroscopic results for Ki~05 and NGC~2360, but these differences are within a combined 1-$\sigma$.
Our photometric abundances in Table~8 are in good agreement with the ones from the DDO and Washington
photometries for the giants of NGC~7142 and NGC~2360, but there is a discrepancy between the
photometric abundances for Haf~08.  The abundance value for NGC~7062, based on the F-type dwarfs
from Str\"omgren photometry, is in very good agreement with our value.  For the clusters Ki~05,
NGC~1798, and NGC~2304, our abundance values are in reasonable agreement with the theoretical ones of
\cite{ber94} from ZAMS fitting.  The literature abundance values in Table 8 are mostly based on
spectroscopy and photometry of giants stars, while our values depend most heavily on F-type dwarf
stars. 

>From the comparison of $(V_{0}$--$M_{\rm V})$ in Figs.~9(a)--(b), the differences with the literature are
mostly at the level of $\Delta(V_{0}$--$M_{\rm V})< 0.50$ mag.  However, for the O\!Cs, NGC~6694, NGC~7062,
NGC~1857, Haf~08, NGC~2186, NGC~6802, and NGC~436, these differences are sometimes larger, at the
level of $\Delta(V_{0}$--$M_{\rm V})= [-1.17,~+1.10]$, greater than we would like due to various systematic
problems in the analyses.  For instance, although the reddenings are nearly
identical for NGC~1857, the theoretical solar-abundance ZAMS of \cite{barb69} has been used by
\cite{babu89} for deriving their distance modulus, whereas the M08 isochrone with $Z = +0.008$ has been
utilised in our work.  For Be~73, the heavy-element abundances of the literature and of this work are
almost identical, but for this cluster our $E(B$--$V)=0.28$ does not agree well with the $E(B$--$V)=0.11$ of
\cite{car07}.  These kinds of differences with the literature lead to the discrepancies in the distance
moduli shown in Fig.~9. 

>From the comparison of the ages in Figs.~10(a)$-$(b), the differences of $\Delta A$~(Gyr) 
between this work and the literature are smaller than $\pm0.25$ for eight O\!Cs, while for the O\!Cs, 
NGC~6802, NGC~6866, NGC~7062, Ki~05, NGC~1798, Be~73, NGC~2215, Be~37, Haf~04, Haf~08, Ki~23, and NGC~2304,
the differences are sometimes larger, in the range of $\Delta A~$ $= [-0.95, +0.89]$ Gyr.

As emphasized by  \cite{pau06}, \cite{pau10}, and  \cite{moi10}, these large discrepancies of the
distance moduli, distances, and ages stem from the differences between the heavy-element abundances and
the reddenings of the literature as compared to this work, as seen in Table~7.  Moreover, the variety of
theoretical and observational ZAMSs used in the literature has also been responsible in part for incurring
these differences.  Effectively using the CC diagram for separating the effects of $E(B$--$V)$ and [Fe/H], and
for determining their values independent of the CM diagrams, as in this paper, will help to overcome 
such differences and reduce systematic errors.

\begin{table*}
\centering
\tiny
\caption {Morphological-age indices.  Their meanings are discussed in the text. 
The morphological and isochrone ages have been given in Columns~8 and 9, respectively. The
symbols RC or RG, in Column~10, indicate the presence of the Red Clump or
the base of the Red Giant branch, respectively.}
\begin{tabular}{cccccccccc}
\hline
\tiny
Cluster  &$V_{TO}$&$V_{RC/RG}$ &$(B$--$V)_{TO}$ &$(B$--$V)_{RC/RG}$ &$\delta V$&$\delta 1$&$logA_{mi}$&$logA_{iso}$&RC/RG \\
\hline
NGC~6802 &15.20 &14.88 &1.00 &1.73 &0.32 &0.73 &8.87$\pm$0.03 &9.05$\pm$0.03 &RC \\
NGC~6866 &12.07 &11.07 &0.18 &1.14 &0.18 &0.96 &8.82$\pm$0.03 &8.89$\pm$0.02 &RG \\
NGC~7062 &13.85 &13.18 &0.51 &1.39 &0.66 &0.88 &9.00$\pm$0.04 &8.85$\pm$0.02 &RC \\
Ki~05    &15.63 &14.71 &0.09 &1.60 &0.92 &1.50 &9.11$\pm$0.04 &9.10$\pm$0.05 &RC \\
NGC~1798 &16.56 &15.75 &0.72 &1.35 &0.81 &0.63 &9.06$\pm$0.03 &9.25$\pm$0.03 &RC \\
NGC~7142 &16.19 &15.19 &0.81 &1.31 &1.89 &0.50 &9.55$\pm$0.04 &9.55$\pm$0.06 &RG,RC \\
Ru~01    &12.14 &12.02 &0.18 &1.12 &0.12 &0.94 &8.78$\pm$0.04 &8.68$\pm$0.03 &RG \\
Be~35    &15.32 &14.83 &0.30 &0.96 &0.49 &0.67 &8.94$\pm$0.01 &8.95$\pm$0.03 &RC \\
Be~37    &14.90 &14.43 &0.16 &0.89 &0.47 &0.73 &8.93$\pm$0.07 &8.75$\pm$0.05 &RC \\
Ki~23    &15.08 &14.08 &0.41 &0.63 &1.40 &0.23 &9.31$\pm$0.08 &9.25$\pm$0.02 &RC \\
NGC~2304 &14.37 &13.37 &0.19 &1.06 &0.54 &0.86 &8.96$\pm$0.06 &8.97$\pm$0.02 &RG \\
NGC~2360 &12.48 &11.35 &0.35 &0.97 &1.13 &0.62 &9.19$\pm$0.03 &9.05$\pm$0.02 &RC \\
\hline
\end{tabular}  
\end{table*}

\section{Comparisons of Morphological and Isochrone Ages}

The definitions of the morphological indices $\delta V$ and $\delta 1$, suggested 
by \citet[][fig.~1]{phel94a}, are the following:  $\delta V$ is the magnitude difference 
between the TO stars and the RC stars, $\delta V = V_{TO} - V_{RC}$, and $\delta 1$ is
the difference in the colour indices between the bluest point on the main sequence at the
luminosity of the TO and the colour at the base of the red giant (RG) branch one magnitude
brighter than the TO luminosity,  $\delta 1=(B$--$V)_{\rm TO}-(B$--$V)_{\rm RG}$.  These
measured quantities are given in Columns~2--7 of Table~9 for 12 of our clusters.
\cite{phel94a} note that $\delta 1$ is especially useful because it can be measured in
O\!Cs without any RC candidates.  Having inspected the CM plots of the 20 O\!Cs, 12 O\!Cs which
exhibit noticeable TO, RG, and/or RC sequences on the CM diagrams are available for
assigning both these morphological indices, $\delta V$ and $\delta 1$.  Depending on the
presence, or no, of the RC and the base of RG sequences, the O\!Cs are labeled with ``RC'' or
``RG'' in Column~10 of Table~9.  Since 3 out of these 12 O\!Cs do not exhibit any RC
candidates, only $\delta 1$ values are measured directly.  These values have been
transformed into $\delta V$ via the equation of $\delta V=3.77-3.75\delta 1$, given by
\cite{phel94a}.

The morphological ages are estimated from the equation, 
$log A=0.04\delta V^2+0.34\delta V+0.07[Fe/H]+8.76$ of \cite{sal04}, applying the metal
abundances, [Fe/H], of the O\!Cs in Table~6.  These ages of 12 O\!Cs together with their
uncertainties have been listed in Column~8 of Table~9.  The comparison of the isochrone
and morphological ages has been done in Figs.~11(a) and (b).  When a linear regression is
applied to the comparison in Fig.~11(a), the resulting equation is
$logA_{\rm mi}=(0.796\pm0.144)logA_{\rm iso}+(1.855\pm1.305)$ with a correlation
coefficient of 0.86 and a dispersion of 0.14 dex, where $A_{\rm mi}$ indicates the
morphological ages and $A_{\rm iso}$ the isochrones ages from the CM diagrams, as given
in Table~6.  The correlation coefficient of 0.86 indicates that the morphological ages
are in fairly good consistency with the isochrone ones of these O\!Cs, for the ones with the
presence of good RC and/or RG candidates.  For four O\!Cs, Ki~05, NGC~7142, Be~35,
and NGC~2304, the differences of $\Delta log A$ are quite small, at the level of
$\pm0.01$ dex, as is seen in Fig.~11(b).  For the other eight O\!Cs, age differences fall in
the range of $\Delta log A =[-0.18,~+0.19]$.

\begin{figure}
\center
\epsfig{file=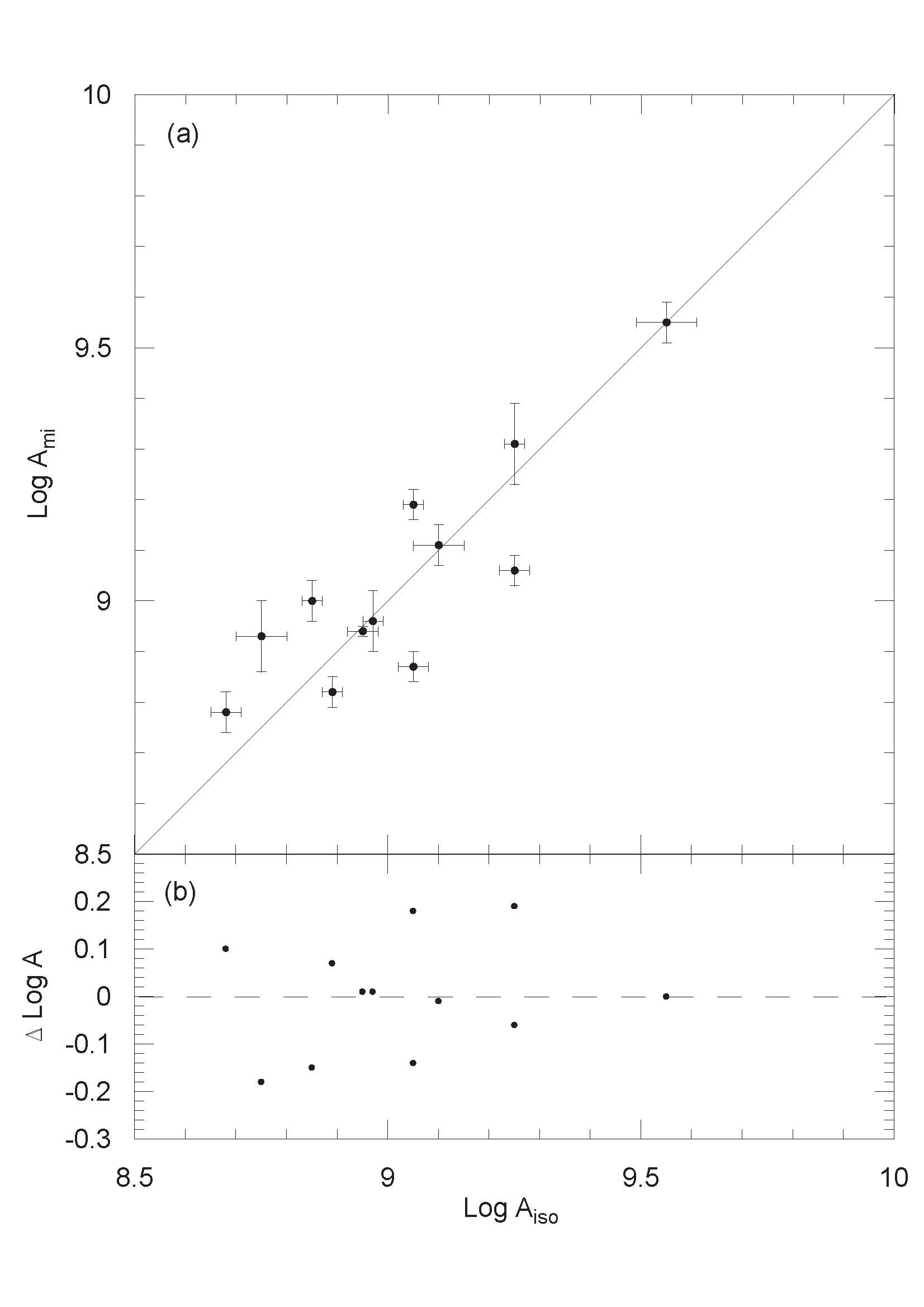, width=8cm, height=9.5cm}
\caption {Panel~(a):  comparison of the isochrone ($logA_{\rm iso}$) and the morphological
($logA_{\rm mi}$) ages for 12 of our 20 O\!Cs.  Panel~(b):  $\Delta log{A}$ versus
$logA_{\rm iso}$.  (All ages in Gyr).  The diagonal solid line shows the one-to-one
relationship.}
\end{figure}

\begin{figure}
\center
\epsfig{file=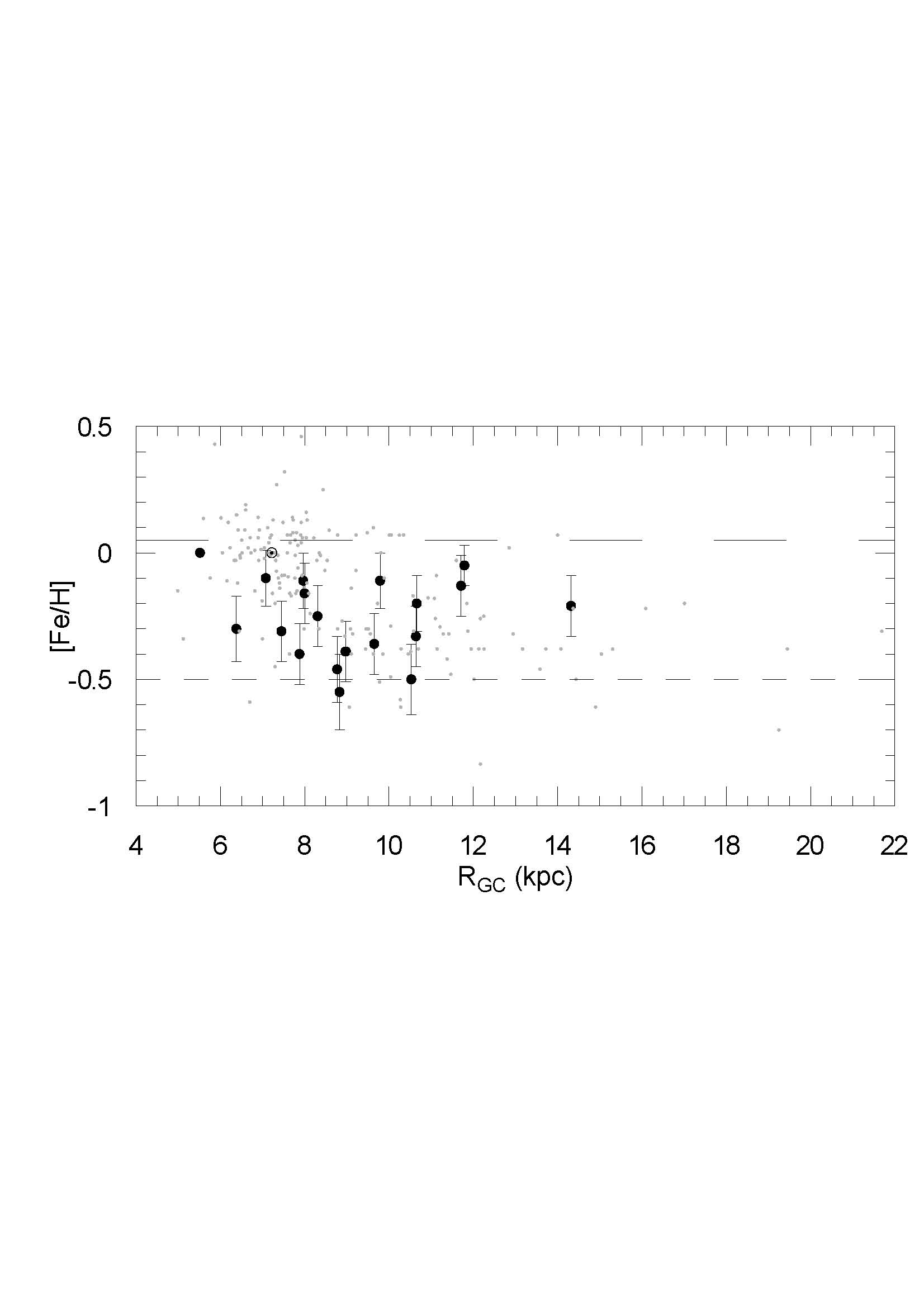, width= 8cm, height=13cm}
\caption {[Fe/H] versus $R_{GC}$ for the 20 O\!Cs.  Filled and grey dots
represent the values of this work and the Dias catalogue, respectively.  
Horizontal lines denote the range of $[Fe/H]=[+0.05,~-0.50 ]$ of Friel
et al.\ (2002).  At $([Fe/H],~R_{\odot}) = [0,~8.3]$ the big circle shows
the solar position.}
\end{figure}

\section{Metal-abundance gradient and Age-Metallicity Relation}

By assuming $R_{\odot}=8.3\pm0.23$ kpc, given by \cite{bru11}, the Galactocentric distances of 20 O\!Cs
have been calculated from the heliocentric distances, d, in Column~9 of Table~6. 
The plot of [Fe/H] versus R$_{GC}$ (kpc) for the 20 O\!Cs is given in Fig.~12,
where filled dots show [Fe/H] and R$_{GC}$ for this sample, while grey dots represent 
the values of 178 O\!Cs in the Dias catalogue for which metal-abundance determinations
are available, but which represent a more inhomogeneous sample with probable systematic
differences.  The Sun is located at $([Fe/H],~R_{\odot}) = [0,~8.3]$.  However, our
sample is rather small and may contain selection biases.  Moreover, our sample
includes only a few clusters in each of the quadrants I, II, III of the Galactic disc,
and  metal-poor old O\!Cs, are not well represented.  These kinds of biases may lead to
possible misinterpretations for mean [Fe/H], R$_{GC}$, and age-metallicity relations.
However, the 20 O\!Cs, in our sample do have the advantage of being uniformly analysed,
and  homogeneous regarding the instrumentation, observing techniques, reduction
methods, and photometric calibrations. 

In Fig.~12 horizontal lines denote the limits of [Fe/H]$=[+0.05, -0.50]$ suggested by
\cite{fri02} for the Galactic O\!Cs.  However, recent spectroscopic studies of O\!Cs indicate
that there are, at least, more metal-poor O\!Cs as compared to these limits.  For example,
\cite{yong05} have measured the spectroscopic abundance of Be~31 as $[Fe/H]=-0.57\pm0.23$,
which is somewhat more metal-poor.  The metal abundances of 178 O\!Cs in Dias catalogue fall
in the range $-0.84 \leq[Fe/H]\leq+0.46$, and NGC~436 in our sample is the most metal-poor
cluster with [Fe/H]$ = -0.55\pm0.33$ dex, while Be~37 the most metal-rich with
[Fe/H]$ = -0.05\pm0.08$ dex.   

When a fit is applied for the 20 O\!Cs, in the range of $6.82 \leq R_{GC} \leq 15.37$ kpc in Fig.~12,
a global abundance gradient of $+0.002\pm0.022$ dex/kpc is calculated indicating that there is no
correlation between [Fe/H] and R$_{GC}$ (kpc) over this wide range.  However, the works of
\cite{twa97}, \cite{mag09}, \cite{lep11}, and \cite{sl13} have
shown that abundance gradients correlating [Fe/H] with R$_{GC}$ for O\!Cs should take into account
a possible discontinuity at $R_{GC} \approx 8$--$10$ kpc.  

Figure 13 shows an age-metallicity plot from the corresponding values of our 20 O\!Cs plus
those from Dias catalogue. It is clear from Fig.~13 that there is not any significant
correlation between the ages and metal abundances for the 20 O\!Cs.  However, it is quite
difficult to draw any firm conclusions due to the small sample and to the deficiency of
old O\!Cs in the present sample, but the Dias sample of Fig.~13 also shows little correlation.
\cite{car98}, \cite{chen03}, and \cite{sal04} also did not find any relation between the
ages and metallicities for their O\!Cs.  The significant scatter in the ages and metallicities
dominates and hides any possible small correlation.  The reason for this scatter is due to
orbital diffusion, radial mixing, and radial migration \citep{sch09}, as well as
inhomogeneous chemical evolution in the Galactic disc \citep{hay08}.

\begin{figure}
\center
\epsfig{file=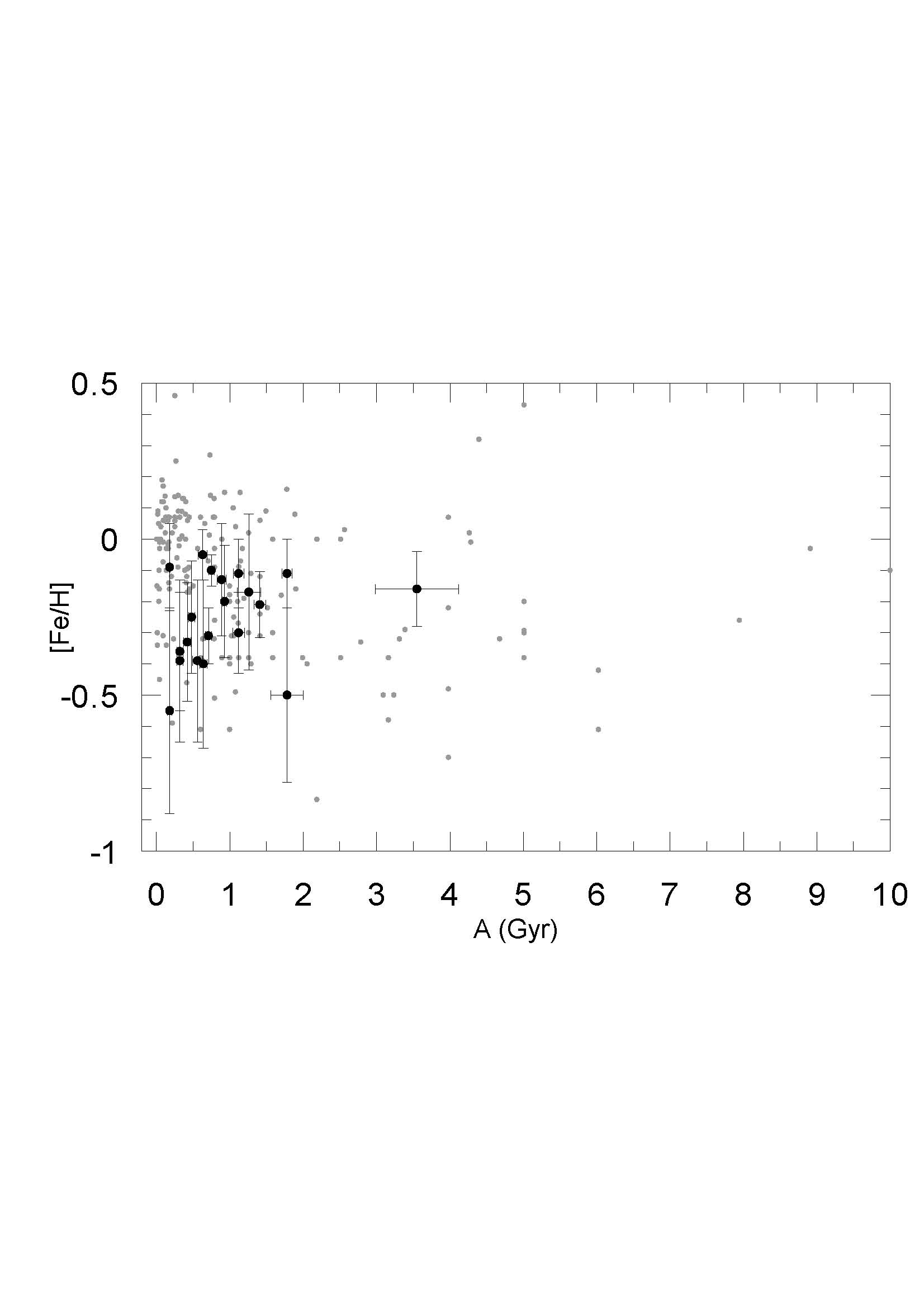, width=8cm, height=12cm}
\caption {[Fe/H] versus A~(Gyr) for the 20 O\!Cs (filled dots). 
The grey dots represent the values in the Dias catalogue.}
\end{figure}

\section{Spatial distribution of 20 O\!Cs}

Fig.~14 shows the spatial distribution of the 20 O\!Cs (filled-circles and open-diamond
symbols) in terms of the Galactic longitude with five spiral arms overlaid.  This schematic
projection of the Galaxy is seen from the north pole.  As is evident from Fig.~14, the
number of old O\!Cs is low in the Galactic center direction.  To first order this is
probably due to our small sample plus our selection procedure,  and the interstellar
extinction, which is very strong mainly in these directions.  But also, this is partly
due to the high frequency of encounters with Giant Molecular Clouds (GMCs) inside the
Solar circle, as discussed by \cite{gie06} and \cite{cam09}, which tends to reduce the
relative number of O\!Cs.

\begin{figure}
\center
\epsfig{file=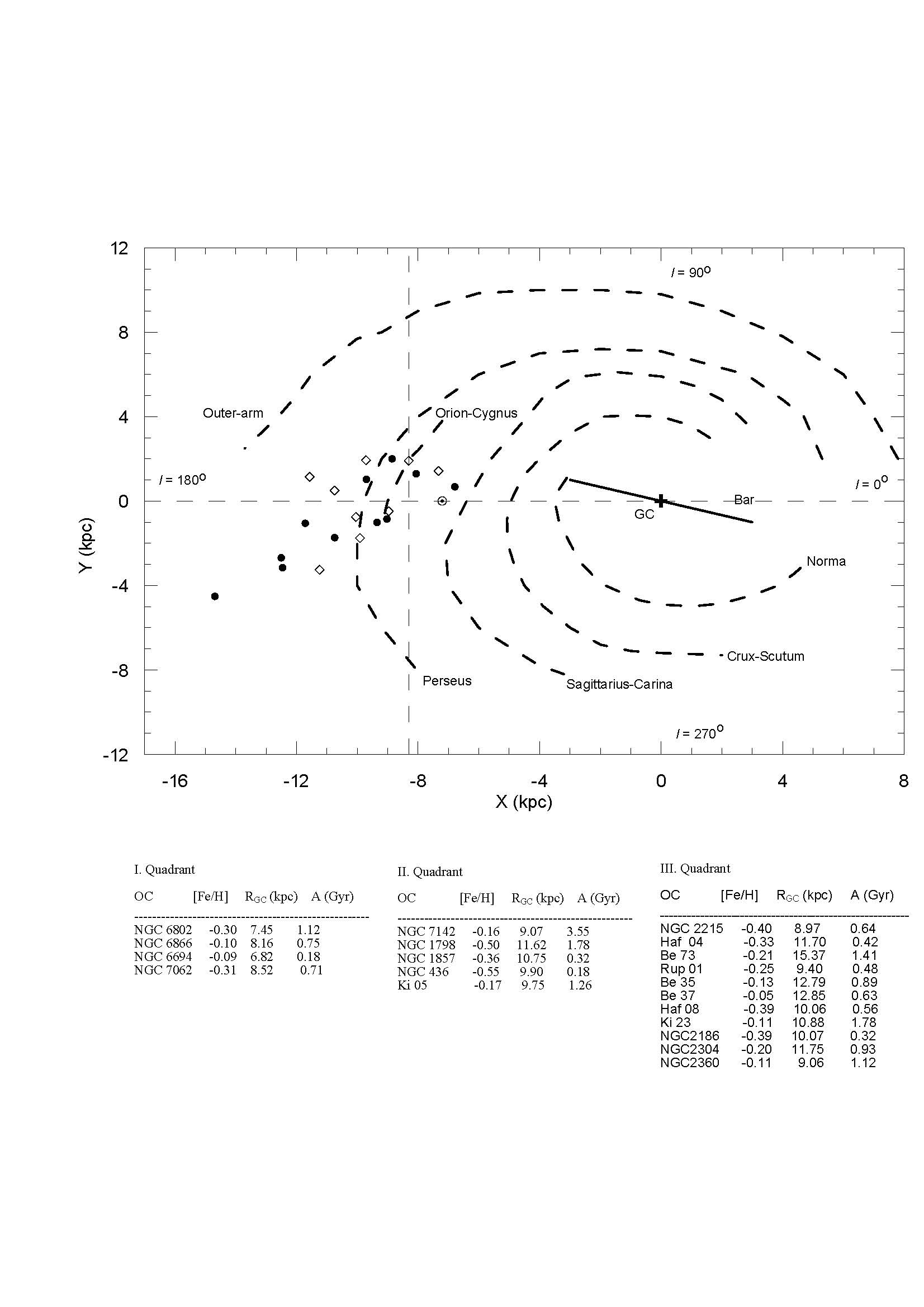, width=7cm, height=12cm}
\caption {The spatial distribution of the 20 O\!Cs with five spiral arms overlaid and in
terms of Galactic longitude.  This schematic projection of the Galaxy is seen from the
North Galactic Pole.  Filled circles and open diamonds show the clusters with $[Fe/H]\ge-0.25$
and $[Fe/H]<-0.25$, respectively.  The cluster names, metallicities, Galactocentric distances,
and ages are listed below the figure for their respective quadrants.}
\end{figure}

Fig.~14 represents an (X, Y) projection in terms of both the metallicity
and the Galactic longitude.  Filled circles and open diamonds show the O\!Cs with
$[Fe/H]\ge-0.25$ and $[Fe/H]<-0.25$, respectively.  Metal abundances, Galactocentric
distances, and ages of these O\!Cs of Fig.~14 are also presented in terms of the Galactic
quadrant. There is not any indication that the distribution of the metal abundance of
the O\!Cs is a function of the quadrant in the (X, Y) plane of Fig.~14.  Young clusters,
$A<1$ Gyr, with $[Fe/H]<-0.30$ are mostly located in the Galactic anticentre
directions, except for NGC~7062, which is located on the border of quadrants I and II. 

A $|z|$ versus $R_{GC}$ plot is displayed in Fig.~15.
Diamonds and filled circles show $A \leq 0.5$ Gyr and $A > 0.5$ Gyr, respectively.
The sun is located at $(R_{GC}, z) = (8.3,~0.0148)$ kpc; the $z$ distance of the Sun
has been taken from \cite{chen99}.  The O\!Cs in Fig.~15 are mostly distributed at the
distances, $|z| < 0.3$~kpc and $R_{GC} < 10$~kpc.  However, for six O\!Cs with
$R_{GC} > 10$~kpc, three reach vertical Galactic distances $|z| > 0.5$~kpc.  These O\!Cs
are located in the Galactic anticentre direction, $160^{\circ} < $ l $ < 228^{\circ}$,  as is
evident from Fig.~14 and Table~6. 

The presence of this kind of cluster, moderately young and metal-poor, may have
originated from the Galactic anticentre over-densities, such as that of Canis Major
\citep{mar04,bel04} and of the Monoceros ring \citep{new02}.  Having inspected the
Galactic coordinates of these O\!Cs, given in Table~6, it appears that they may not be
associated with the Canis Major dwarf galaxy ($\it{l},~\it{b}$)=($240^{\circ}$,$-8^{\circ}$).
However, since these accreting dwarf galaxies have long tidal tails both preceding and
following them, a good agreement of the coordinates is not an exact test.

Similar to this paper, \citet{yong05, yong12} find from their O\!Cs that the disc metallicity
gradient becomes flat beyond $R_{GC}=10$--$12$ kpc, and they discuss the possibility that
these O\!Cs may be associated with an accreted dwarf galaxy, or may have formed as a result
of star formation triggered by merger events in the outer disc. On the other hand, from
observational photometric evidence for a group of young B5--A0 stars along the line of
sight to the Canis Major over-density and also the presence of an older metal-poor
population probably belonging to the Galactic thick disc or halo, \cite{car08} interpret
these O\!Cs as due to the effects of the structure of the Local and Outer spiral arms as
well as the distortion produced by the Galactic warp.

\begin{figure}
\center
\epsfig{file=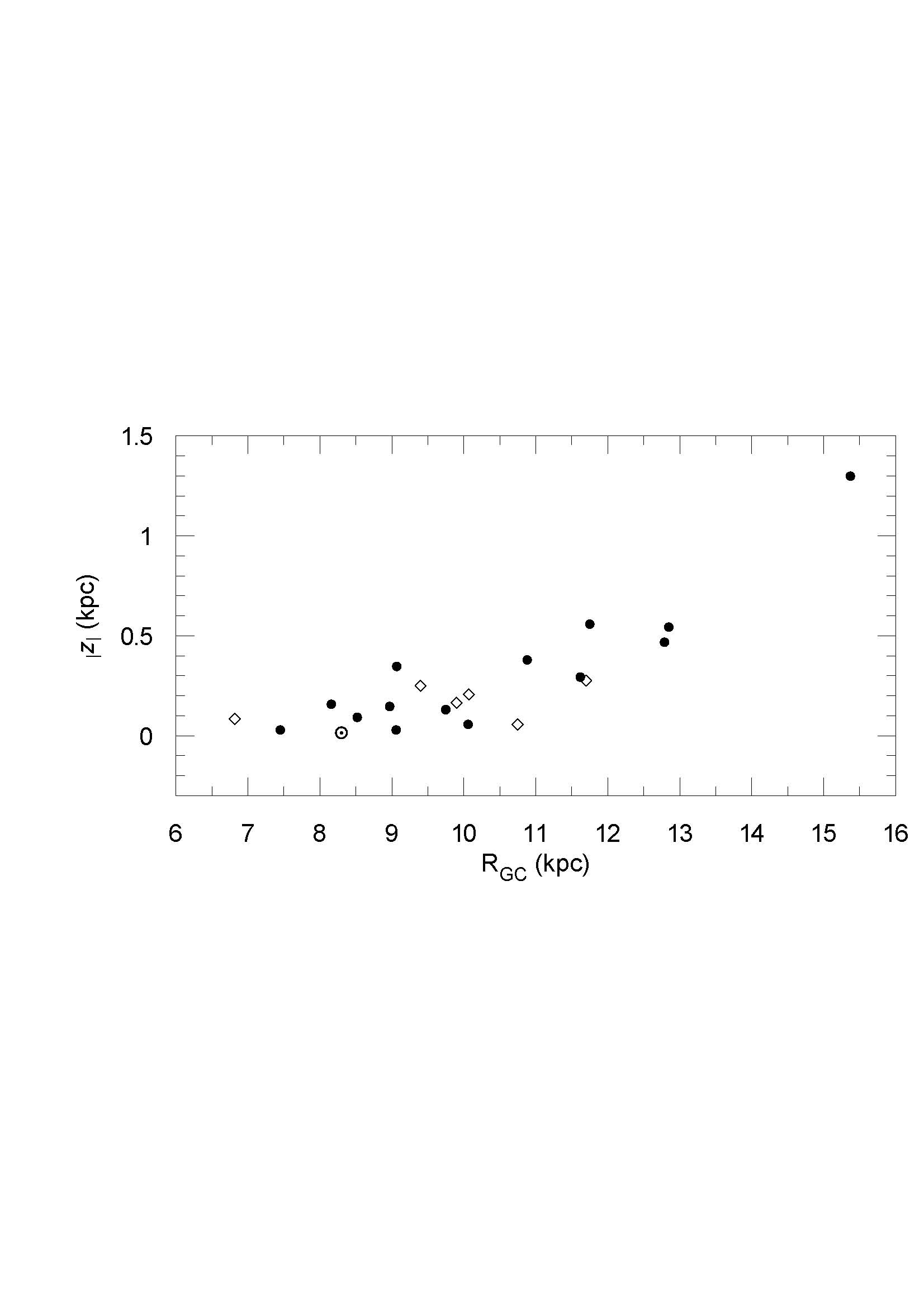, width=8cm, height=10cm}
\caption {$|z|$ versus $R_{GC}$ for the 20 O\!Cs.
Diamonds and filled circles show $A \leq 0.5$ Gyr and $A > 0.5$ Gyr, respectively.
The Sun is located at $(R_{GC}, z) = (8.3, 0.0148)$ kpc.}
\end{figure}

\section{RC and BS candidates}

Ten possible RC groups have been identified in the CM diagrams for the following O\!Cs: 
NGC~6802, NGC~7062, Ki~05, NGC~1798, NGC~7142, Be~35, Be~37, Ki~23, NGC~2304, and NGC~2360,
with single RC stars probable in NGC~6866 and Ru~01.  High mass stars (and their evolved
products, RC/RG stars) are transferred to the cores of the O\!Cs as a consequence of mass
segregation.  The SAFE algorithm is capable of emphasizing this kind of star by focusing on
the central parts of the 20 O\!Cs, which are small or comparable to the size of the CCD
within the SPMO survey.  One can expect to detect these possible RC stars in these O\!Cs
due to their intermediate ages (Table 6).  The other O\!Cs are younger, so it is reasonable
not to detect RC candidates.  Surprisingly, no obvious RC stars have been identified in Be~73
with an intermediate age of $A=1.41$ Gyr.
  
The possible RC candidates have been listed in Table~S5 in the supplementary section.
RC stars are useful as standard candles for estimating the distances to these O\!Cs,
or conversely their distances can be utilised as an initial test of their membership in a
given cluster once its distance has been determined by main-sequence fitting in a CM
diagram.  The mean absolute magnitudes of $\langle M_{\rm V} \rangle=+0.60\pm0.10$ 
\citep{twa97} and  $\langle M_{\rm I} \rangle=-0.22\pm0.03$ \citep{gro08} allow us
to assign mean distances and their errors for RC candidates from their observed magnitudes, 
using the relations of $d = 10^{(V- M_{V} + 5 - A_{\rm V})/5}$ and 
$\sigma_{\rm d}$ = $1\over5$d$ \times \ln(10) \times \sqrt{\sigma_{\rm V}^2+\sigma_{M_{\rm V}}^2+\sigma_{A_{\rm V}}^2}$
for the visual magnitude, with similar equations for $I$.  Here, the total absorption relations
of ${A_{\rm V}} = 3.1~E(B$--$V)$  and $A_{\rm I} = 1.98~E(B$--$V)$ \citep{gim98}, are used,
with the $E(B$--$V)$ values given in Table~6 of this paper.  The magnitudes of $I$ and $V$ in
Columns~3--4 of Table~S5 are used for estimating the $d_{\rm I}$ and $d_{\rm V}$ distances
(in kpc) of these RC candidates, which are listed in Columns~5--6 of this same table, while
the $(\alpha,~\delta)$ coordinates are listed in Columns~1--2. 

The distances in Table~S5 for each RC candidate have been compared to the distances from the 
$V,(B$--$V)$ and $V,(V$--$I)$ CM diagrams, which are also presented in Table~S5 in the header
for each cluster together with its name.  Those stars which show combined 1-$\sigma$ agreement
within the uncertainties of these two distances in Table~S5 have been assigned as members (M),
otherwise as non-members (NM), as shown at the end of Columns~5 and 6.  However, this
membership test has been done considering only distance criteria.  For the confirmation of their
cluster memberships, spectroscopic (radial velocity) observations and proper motion data are
needed for these RC candidates, as well as for other members of these O\!Cs. 

From close inspection of the CC and CM diagrams of the 20 O\!Cs, 12 B\!S candidates with
$V {\lesssim}  15.60$ $(V_{0} {\lesssim} 14.50)$
and $(B$--$V) {\lesssim} 0.80$ $((B$--$V)_{0} {\lesssim} 0.45)$   
in the CM diagram of NGC~7142 have been classified (see Fig.~S8 in the supplementary section,
where these B\!S candidates have been marked with open squares).  
These limits for $V_{0}$ and $(B$--$V)_{0}$, which B\!Ss occupy in CM diagrams, are similar to
those given by \citet[][fig.~19]{carn01} and \citet[][fig.~10]{car10b}.  However, the early-type,
bluer, and brighter B\!S candidates with $0.05\leq(B$--$V)_{0}\leq0.41$ in the CM diagram of
NGC~7142 may also be field stars between us and the cluster, as noticed by \cite{car10b}.  These
are the three additional stars in the CM diagram of NGC~7142, in the same region as the BS
candidates, but which are not marked as such.  The coordinates $(\alpha,~\delta)$ and $BV\!RI$
photometry of these candidates are given in Table~S6 (see the supplementary section).

\section{Conclusions}

Our main conclusions are as follows:

\begin{enumerate}

\item Twenty O\!Cs, observed with the 84 cm telescope at SPMO, provide the advantage of
having been uniformly and homogeneously processed regarding instrumentation, observing
techniques, reduction methods, and analyses.  The fundamental astrophysical parameters of
these 20 O\!Cs:  interstellar reddenings, $E(B$--$V)$; photometric metallicities and mass-fraction
heavy-element abundances ([Fe/H], Z); distance moduli, $(V_{0}$--$M_{\rm V})$, and distances, $d$
(kpc); plus ages, $A$ (Gyr), have been derived, and should be internally quite precise.

\item Differences in interstellar reddenings, metallicities, distance moduli, and ages, as
compared to the literature, have been studied.  Systematic differences, when found, are mainly
due to the usage of distinct mean-colour relations versus isochrones, which correspond 
to differing element abundances, internal stellar physics, and colour-temperature relations. 
Different interstellar reddenings, as well as different assumptions, or measurements, for the
stellar metallicities contribute greatly to these kinds of systematic offsets for the
distances and ages. 

\item The correlation coefficient of 0.86 indicates that the morphological ages, which are derived from  
the morphological indices $\delta V$ and $\delta 1$ in the CM diagrams, are fairly consistent
with the isochrone ages of 12 O\!Cs, for the ones with the presence of good RC and RG candidates.  For
four O\!Cs, Ki~05, NGC~7142, Be~35, and NGC~2304, the differences of $\Delta log A$ are quite
small, at the level of $\pm0.01$ dex, as is seen in Fig.~11(b).  For the other eight O\!Cs, age differences
fall in the range $\Delta log A =[-0.18,~+0.19]$.

\item For our sample of 20 O\!Cs, their interstellar reddenings extend over
$0.01 \pm 0.07 \le E(B$--$V) \le 0.80 \pm 0.07$ [NGC~2360; NGC~6802]; their distances
over $0.83 \pm 0.01 \le d \le 7.93 \pm 0.11$ kpc [NGC~2215; Be~73]; and their ages over
$0.18 \pm 0.01 \le A \le 3.55 \pm 0.57$ Gyr [NGC~6694; NGC~7142].  (See Table~6.)  All 20 O\!Cs
fall in Galactic quadrants I, II, and III, and 15 fall close to the anticentre direction,
$125^{\circ}$ $<$  $l$  $<$ $235^{\circ}$.

\item  For these 20 O\!Cs, which have $6.82 \leq R_{GC} \leq 15.37$ kpc, there is no correlation
between [Fe/H] and $R_{GC}$ over this wide extent (Fig.~12).  Our 20 O\!Cs range over
$-0.55 \pm 0.33 \le [Fe/H] \le -0.05 \pm 0.08$ dex [NGC~436; Be~37].

\item Also, there is not any correlation between the cluster ages and metal abundances 
for these 20 O\!Cs despite a 0.50 dex range in [Fe/H], as shown in Fig.~13.  As discussed in the
literature, this is caused by inhomogeneous chemical enhancements in the Galactic disc
\citep{hay08,jac11}, orbital diffusion, and radial migration and mixing of O\!Cs \citep{sch09}.
These O\!Cs have probably originated from differing Galactic radii and/or from various star forming
regions \citep{lep11}.

\item The presence of young, metal-poor O\!Cs in the second and third Galactic quadrants (Fig.~14)
suggests that they may have originated from the Galactic anticentre over-densities, such as
that in Canis Major \citep{mar04,bel04} or in the Monoceros ring \citep{new02}, as discussed
by \citep{car10a}.  Six of such O\!Cs from our sample are located in the outer Galactic disc,
R$_{GC}$ $>$ 8.3 kpc (Fig.~14)).  According to \cite{yong05}, these O\!Cs may be associated
with an accreted dwarf galaxy or may have formed as a result of star formation triggered by
merger events in the outer disc.  

\item In addition, Fig.~15 shows that of the six O\!Cs with R$_{GC}$ $>$ 11 kpc, three have $|z|> 0.5$
kpc, and all have $|z| \gtrsim 0.3$ kpc.  As discussed by \cite{car08}, their positions may be
distorted by the Galactic warp and the existence of the local arms, the Perseus or Outer-arm.  \cite{mom06}
stress that the Canis Major properties point to an over-density reflecting a normal warped disc
population rather than an accreted dwarf galaxy.  These O\!Cs have mostly intermediate ages, A $>$ 0.5 Gyr,
supporting such an interpretation, i.e.\ a natural consequence of the Galactic disc's warp rather
than an accretion event.

\item Some RC and BS candidates are given in Tables~S5 and S6, respectively, having been classified 
from the CM diagram for a subset of the 20 O\!Cs; 12 O\!Cs show RC candidates, and NGC~7142,
BS candidates.  These are proposed for future radial-velocity and proper-motion observations to
confirm their membership in these O\!Cs.

\end{enumerate}

\section*{Acknowledgments}

We thank the referee, A.~L.~Tadross, for his useful suggestions.  The observations of this publication were
made at the National Astronomical Observatory, San Pedro M\'artir, Baja California, M\'exico, and
we wish to thank the staff of this observatory for their help with acquiring the CCD data.  Also we
greatly appreciated the participation  of C.A.~Santos, and J.~McFarland, who assisted with the
data reductions and programming.  We thank J.~Lepine and C.~Bonatto for the spiral-arm data in
Fig.~14 and for useful comments about the manuscript.  This research made use of the WEBDA open-cluster
database of J.-C. Mermilliod.  This work has been supported by the CONACyT (M\'exico) projects 33940,
45014, and 49434, and PAPIIT-UNAM (M\'exico) projects IN111500 and IN103014.

\clearpage

\section{Supplementary Materials}

Fundamental astrophysical parameters, plus CC and CM diagrams, for 19 O\!Cs.
The tables of fundamental astrophysical parameters for 19 O\!Cs have been given in Tables S1--S4,
and the CC and CM diagrams as Figs.~S1--S19.  ``log($A$)-interval'' in the tables refers to the
derived range in ages possible from the isochrone pairs plotted in Figs.~S1--S19.
Possible RC candidates, which are classified in the CM diagrams of 12 O\!Cs, are listed in Table~S5. 
The data for possible BS candidates in NGC~7142 have been presented in Table~S6.

\clearpage

\begin{table}
\renewcommand\thetable{S1}
\tiny
\centering
\caption {The derived astrophysical parameters of the O\!Cs, NGC~6694, NGC~6802, NGC~6866, NGC~7062, and NGC~436.}
\begin{tabular}{lllclc}
\hline
Colour & $(V_{0}$--$M_{V})$ & $d$ (kpc) & log($A$)-interval & log($A$) & $A$ (Gyr) \\
\hline
 \multicolumn{ 5}{l}{NGC6694:~$E(B$--$V)=0.51\pm0.06$, ~~ $[Fe/H]=-0.09\pm0.14$, ~~ $Z=0.016\pm0.005$} \\
\hline
$(B$--$V)$ &11.10$\pm$0.08 &1.66$\pm$0.06 &  - &8.25$\pm$0.05 &0.18$\pm$0.02 \\
$(R$--$I)$ &11.10$\pm$0.10 &1.66$\pm$0.08 &  - &8.25$\pm$0.05 &0.18$\pm$0.02 \\
$(V$--$I)$ &11.10$\pm$0.10 &1.66$\pm$0.08 &  - &8.25$\pm$0.05 &0.18$\pm$0.02 \\
$(V$--$R)$ &11.10$\pm$0.10 &1.66$\pm$0.08 &  - &8.25$\pm$0.05 &0.18$\pm$0.02 \\
$(B$--$R)$ &11.10$\pm$0.10 &1.66$\pm$0.08 &  - &8.25$\pm$0.05 &0.18$\pm$0.02 \\
      &               &              &    &        &            \\
Mean  &11.10$\pm$0.04&1.66$\pm$0.03&      &8.25$\pm$0.02 &0.18$\pm$0.01 \\
\hline
\multicolumn{5}{l}{NGC 6802: $E(B$--$V)=0.80\pm0.07$, ~~ $[Fe/H]=-0.30\pm0.13$, ~~ $Z=0.009\pm0.003$} \\
\hline
$(B$--$V)$ &11.20$\pm$0.10 &1.74$\pm$0.08 &9.00$-$9.05 &9.05$\pm$0.05 &1.12$\pm$0.14 \\
$(R$--$I)$ &11.10$\pm$0.15 &1.66$\pm$0.12 &9.00$-$9.05 &9.05$\pm$0.10 &1.12$\pm$0.29 \\
$(V$--$I)$ &11.20$\pm$0.15 &1.74$\pm$0.12 &9.00$-$9.05 &9.05$\pm$0.10 &1.12$\pm$0.29 \\
$(V$--$R)$ &11.20$\pm$0.10 &1.74$\pm$0.08 &9.00$-$9.05 &9.05$\pm$0.10 &1.12$\pm$0.29 \\
$(B$--$R)$ &11.20$\pm$0.10 &1.74$\pm$0.08 &9.00$-$9.05 &9.05$\pm$0.05 &1.12$\pm$0.14 \\
      &            &            &            &            &            \\
Mean  &11.19$\pm$0.05 &1.73$\pm$0.04 &       &9.05$\pm$0.03 &1.12$\pm$0.08 \\
\hline
\multicolumn{5}{l}{NGC6866: $E(B$--$V)=0.06\pm0.05$, ~~ $[Fe/H]= -0.10\pm0.05$, ~~ $Z= 0.015\pm0.002$} \\
\hline
$(B$--$V)$ &10.70$\pm$0.02 &1.38$\pm$0.01 &8.75$-$8.85 &8.85$\pm$0.05 &0.71$\pm$0.09 \\
$(R$--$I)$ &10.50$\pm$0.05 &1.26$\pm$0.03 &8.80$-$8.90 &8.90$\pm$0.05 &0.79$\pm$0.10 \\
$(V$--$I)$ &10.50$\pm$0.08 &1.26$\pm$0.05 &8.80$-$8.90 &8.90$\pm$0.05 &0.79$\pm$0.10 \\
$(V$--$R)$ &10.50$\pm$0.06 &1.26$\pm$0.04 &8.80$-$8.90 &8.90$\pm$0.05 &0.79$\pm$0.10 \\
$(B$--$R)$ &10.50$\pm$0.03 &1.26$\pm$0.02 &8.80$-$8.90 &8.90$\pm$0.05 &0.79$\pm$0.10 \\
      &            &            &            &            &            \\
Mean  & 10.61$\pm$0.02 &  1.32$\pm$0.01 &            &  8.89$\pm$0.02 &  0.75$\pm$0.04 \\
\hline
\multicolumn{5}{l}{NGC 7062: $E(B$--$V)=0.43\pm0.08$, ~~ $[Fe/H]=-0.31\pm0.09$, ~~ $Z=0.010\pm0.002$} \\
\hline

$(B$--$V)$ &11.40$\pm$0.01 &1.91$\pm$0.12 &8.75$-$8.85 &8.85$\pm$0.05 &0.71$\pm$0.09 \\
$(R$--$I)$ &11.40$\pm$0.07 &1.91$\pm$0.06 &8.75$-$8.85 &8.85$\pm$0.05 &0.71$\pm$0.09 \\
$(V$--$I)$ &11.40$\pm$0.06 &1.91$\pm$0.05 &8.75$-$8.85 &8.85$\pm$0.05 &0.71$\pm$0.09 \\
$(V$--$R)$ &11.40$\pm$0.03 &1.91$\pm$0.03 &8.75$-$8.85 &8.85$\pm$0.05 &0.71$\pm$0.09 \\
$(B$--$R)$ &11.40$\pm$0.03 &1.91$\pm$0.04 &8.75$-$8.85 &8.85$\pm$0.05 &0.71$\pm$0.09 \\
      &            &            &            &            &            \\
Mean& 11.40$\pm$0.02& 1.91$\pm$0.02 &                  & 8.84$\pm$0.02  & 0.71$\pm$0.04 \\
\hline
\multicolumn{5}{l}{NGC 436: $E(B$--$V)=0.40\pm0.07$, ~~ $[Fe/H]=-0.55\pm0.33$, ~~ $Z=0.005\pm0.004$}\\
\hline
$(B$--$V)$ & 11.90$\pm$0.09 &2.40$\pm$0.10 &   - &8.25$\pm$0.20 &0.18$\pm$0.10 \\
$(R$--$I)$ & 11.90$\pm$0.20 &2.40$\pm$0.22 &   - &8.25$\pm$0.20 &0.18$\pm$0.10 \\
$(V$--$I)$ & 11.90$\pm$0.10 &2.40$\pm$0.11 &   - &8.25$\pm$0.10 &0.18$\pm$0.05 \\
$(V$--$R)$ & 11.90$\pm$0.10 &2.40$\pm$0.11 &   - &8.25$\pm$0.15 &0.18$\pm$0.07 \\
$(B$--$R)$ & 11.90$\pm$0.10 &2.40$\pm$0.11 &   - &8.30$\pm$0.20 &0.20$\pm$0.12 \\
      &            &            &            &            &            \\
Mean  & 11.90$\pm$0.05 &  2.40$\pm$0.05 &            &  8.26$\pm$0.07 &  0.18$\pm$0.03 \\
\hline
\end{tabular}  
\end{table}

\clearpage

\begin{table}
\renewcommand\thetable{S2}
\tiny
\centering
\caption {The derived astrophysical parameters of the O\!Cs, NGC~1798, NGC~1857, NGC~7142, Be~73, and Haf~04.}
\begin{tabular}{lccccc}
\hline
Colour & $(V_{0}$--$M_{V}) $ & $d$ (kpc) & log($A$)-interval & log($A$) & $A$ (Gyr)\\
\hline
\multicolumn{5}{l}{NGC 1798: $E(B$--$V)=0.47\pm0.07$, ~~ $[Fe/H]=-0.50\pm0.28$, ~~ $Z=0.006\pm0.004$}  \\
\hline
$(B$--$V)$ &12.70$\pm$0.07 &3.47$\pm$0.12 &9.15$-$9.25 &9.25$\pm$0.05 &1.78$\pm$0.22 \\
$(R$--$I)$ &12.70$\pm$0.15 &3.47$\pm$0.24 &9.15$-$9.25 &9.25$\pm$0.10 &1.78$\pm$0.46 \\
$(V$--$I)$ &12.70$\pm$0.15 &3.47$\pm$0.24 &9.15$-$9.25 &9.25$\pm$0.10 &1.78$\pm$0.46 \\
$(V$--$R)$ &12.70$\pm$0.05 &3.47$\pm$0.08 &9.15$-$9.25 &9.25$\pm$0.05 &1.78$\pm$0.22 \\
$(B$--$R)$ &12.70$\pm$0.10 &3.47$\pm$0.16 &9.15$-$9.25 &9.25$\pm$0.05 &1.78$\pm$0.22 \\

           &            &            &            &            &            \\
  Mean & 12.70$\pm$0.04 &  3.47$\pm$0.06 &            &  9.25$\pm$0.03 &  1.78$\pm$0.22 \\
\hline
    
\multicolumn{5}{l}{NGC 1857: $E(B$--$V)=0.47\pm0.08$, ~~ $[Fe/H]=-0.36\pm0.19$, ~~ $Z=0.008\pm0.003$} \\
\hline
$(B$--$V)$ &12.00$\pm$0.07 &2.51$\pm$0.08 &8.40$-$8.50 &8.50$\pm$0.10 &0.32$\pm$0.08 \\
$(R$--$I)$ &12.00$\pm$0.20 &2.51$\pm$0.23 &8.40$-$8.50 &8.50$\pm$0.15 &0.32$\pm$0.13 \\
$(V$--$I)$ &12.00$\pm$0.10 &2.51$\pm$0.12 &8.40$-$8.50 &8.50$\pm$0.10 &0.32$\pm$0.08 \\
$(V$--$R)$ &12.00$\pm$0.10 &2.51$\pm$0.12 &8.40$-$8.50 &8.50$\pm$0.10 &0.32$\pm$0.08 \\
$(B$--$R)$ &11.90$\pm$0.10 &2.40$\pm$0.11 &8.45$-$8.55 &8.55$\pm$0.10 &0.35$\pm$0.09 \\
\hline
           &            &            &            &            &            \\
 Mean & 11.98$\pm$0.04 &2.49$\pm$0.05 &            &8.51$\pm$0.05 &0.32$\pm$0.04 \\
\hline
      
\multicolumn{5}{l}{NGC 7142: $E(B$--$V)=0.35\pm0.08$, ~~ $[Fe/H]=-0.16\pm0.12$, ~~ $Z=0.013\pm0.004$}  \\
\hline
$(B$--$V)$ &11.60$\pm$0.11 &2.10$\pm$0.11 &9.45$-$9.55 &9.55$\pm$0.15 &3.55$\pm$1.46 \\
$(R$--$I)$ &11.60$\pm$0.20 &2.10$\pm$0.19 &9.45$-$9.55 &9.55$\pm$0.15 &3.55$\pm$1.46 \\
$(V$--$I)$ &11.60$\pm$0.10 &2.10$\pm$0.10 &9.45$-$9.55 &9.55$\pm$0.15 &3.55$\pm$1.46 \\
$(V$--$R)$ &11.60$\pm$0.10 &2.10$\pm$0.10 &9.45$-$9.55 &9.55$\pm$0.15 &3.55$\pm$1.46 \\
$(B$--$R)$ &11.60$\pm$0.10 &2.10$\pm$0.10 &9.45$-$9.55 &9.55$\pm$0.10 &3.55$\pm$0.92 \\
      &            &            &            &            &            \\
    Mean & 11.60$\pm$0.05 &  2.10$\pm$0.05 &            &  9.55$\pm$0.06 &  3.55$\pm$0.57 \\
\hline
\multicolumn{5}{r}{Be 73: $E(B$--$V)=0.28\pm0.06$, ~~ $[Fe/H]=-0.21\pm0.06$, ~~ $Z = 0.012\pm0.002$} \\
\hline
$(B$--$V)$ &14.50$\pm$0.04 &7.94$\pm$0.15 &9.10$-$9.15 &9.15$\pm$0.05 &1.41$\pm$0.17 \\
$(R$--$I)$ &14.45$\pm$0.10 &7.76$\pm$0.36 &9.10$-$9.15 &9.15$\pm$0.10 &1.41$\pm$0.37 \\
$(V$--$I)$ &14.50$\pm$0.10 &7.94$\pm$0.37 &9.10$-$9.15 &9.15$\pm$0.05 &1.41$\pm$0.17 \\
$(V$--$R)$ &14.50$\pm$0.10 &7.94$\pm$0.37 &9.10$-$9.15 &9.15$\pm$0.05 &1.41$\pm$0.17 \\
$(B$--$R)$ &14.50$\pm$0.05 &7.94$\pm$0.18 &9.10$-$9.15 &9.15$\pm$0.05 &1.41$\pm$0.17 \\
      &            &            &            &            &            \\
      Mean & 14.50$\pm$0.03 &7.93$\pm$0.11 &       &9.15$\pm$0.02 &1.41$\pm$0.08 \\
\hline

\multicolumn{5}{r}{Haf04:  $E(B$--$V)=0.47\pm0.09$, ~~ $[Fe/H]=-0.33\pm0.19$, ~~ $Z= 0.009\pm0.008$}\\
\hline
$(B$--$V)$ &13.30$\pm$0.09 &4.57$\pm$0.19 &8.60$-$8.70 &8.60$\pm$0.10 &0.40$\pm$0.10 \\
$(R$--$I)$ &13.10$\pm$0.15 &4.17$\pm$0.29 &8.65$-$8.75 &8.65$\pm$0.10 &0.45$\pm$0.12 \\
$(V$--$I)$ &13.10$\pm$0.10 &4.17$\pm$0.29 &8.65$-$8.75 &8.65$\pm$0.10 &0.45$\pm$0.12 \\
$(V$--$R)$ &13.20$\pm$0.10 &4.37$\pm$0.20 &8.60$-$8.70 &8.60$\pm$0.10 &0.40$\pm$0.10 \\
$(B$--$R)$ &13.30$\pm$0.05 &4.57$\pm$0.21 &8.65$-$8.75 &8.65$\pm$0.10 &0.45$\pm$0.12 \\
      &            &            &            &            &            \\
   Mean & 13.22$\pm$0.05 &  4.39$\pm$0.10 &         &8.63$\pm$0.04 &0.42$\pm$0.05 \\
\hline
\end{tabular}  
\end{table} 

\clearpage

\begin{table}
\renewcommand\thetable{S3}
\tiny
\centering
\caption {The derived astrophysical parameters of the O\!Cs, NGC~2215, Rup~01, Be~35, Be~37, and Haf~08.}
\begin{tabular}{lccccc}
\hline
Colour & $(V_{0}$--$M_{V})$ & $d$ (kpc) & log($A$)-interval & log($A$) & $A$ (Gyr) \\
\hline
\multicolumn{5}{r}{NGC~2215: $E(B$--$V)=0.23\pm0.07$, ~~ $[Fe/H]=-0.40\pm0.27$, ~~ $Z=0.008\pm0.005$}  \\
\hline
$(B$--$V)$ &9.60$\pm$0.07 &0.83$\pm$0.03 &8.80$-$8.85 &8.80$\pm$0.05 &0.63$\pm$0.08 \\
$(R$--$I)$ &9.60$\pm$0.10 &0.83$\pm$0.04 &8.85$-$8.90 &8.85$\pm$0.10 &0.71$\pm$0.18 \\
$(V$--$I)$ &9.60$\pm$0.10 &0.83$\pm$0.04 &8.85$-$8.90 &8.85$\pm$0.10 &0.71$\pm$0.18 \\
$(V$--$R)$ &9.60$\pm$0.05 &0.83$\pm$0.02 &8.80$-$8.85 &8.80$\pm$0.05 &0.63$\pm$0.08 \\
$(B$--$R)$ &9.60$\pm$0.05 &0.83$\pm$0.02 &8.80$-$8.85 &8.80$\pm$0.10 &0.63$\pm$0.16 \\
      &            &            &            &            &            \\

 Mean &9.60$\pm$0.03 &0.83$\pm$0.01 &         &8.81$\pm$0.03 &0.64$\pm$0.05 \\
\hline
\multicolumn{5}{r}{Rup~01: $E(B$--$V)=0.17\pm0.06$, ~~ $[Fe/H]=-0.25\pm0.18$, ~~ $Z=0.011\pm0.005$} \\
\hline
$(B$--$V)$ &10.90$\pm$0.05 &1.51$\pm$0.04 &8.65$-$8.70 &8.65$\pm$0.05 &0.45$\pm$0.05 \\
$(R$--$I)$ &10.80$\pm$0.15 &1.45$\pm$0.10 &8.70$-$8.75 &8.70$\pm$0.10 &0.50$\pm$0.13 \\
$(V$--$I)$ &10.80$\pm$0.15 &1.45$\pm$0.10 &8.70$-$8.75 &8.70$\pm$0.10 &0.50$\pm$0.13 \\
$(V$--$R)$ &10.80$\pm$0.10 &1.45$\pm$0.07 &8.70$-$8.75 &8.70$\pm$0.05 &0.50$\pm$0.06 \\
$(B$--$R)$ &10.80$\pm$0.10 &1.45$\pm$0.07 &8.70$-$8.75 &8.70$\pm$0.10 &0.50$\pm$0.13 \\
      &            &            &            &            &            \\
 Mean &10.85$\pm$0.04 &1.48$\pm$0.03 &          &8.68$\pm$0.03 &0.48$\pm$0.04 \\
\hline
\multicolumn{5}{r}{Be~35: $E(B$--$V)=0.11\pm0.07$, ~~ $[Fe/H]=-0.13\pm0.18$, ~~ $Z=0.014\pm0.006$} \\
\hline
$(B$--$V)$ &13.50$\pm$0.07 &5.01$\pm$0.15 &8.85$-$8.95 &8.95$\pm$0.10 &0.89$\pm$0.23 \\
$(R$--$I)$ &13.50$\pm$0.15 &5.01$\pm$0.35 &8.85$-$8.95 &8.95$\pm$0.15 &0.89$\pm$0.37 \\
$(V$--$I)$ &13.50$\pm$0.10 &5.01$\pm$0.23 &8.85$-$8.95 &8.95$\pm$0.10 &0.89$\pm$0.23 \\
$(V$--$R)$ &13.50$\pm$0.10 &5.01$\pm$0.23 &8.85$-$8.95 &8.95$\pm$0.05 &0.89$\pm$0.11 \\
$(B$--$R)$ &13.50$\pm$0.10 &5.01$\pm$0.23 &8.85$-$8.95 &8.95$\pm$0.05 &0.89$\pm$0.11 \\
      &            &            &            &            &            \\
Mean  & 13.50$\pm$0.04 &  5.01$\pm$0.10 &            &8.95$\pm$0.03 &0.89$\pm$0.06 \\
\hline
\multicolumn{5}{r}{Be~37: $E(B$--$V)=0.05\pm0.05$, ~~ $[Fe/H]=-0.05\pm0.08$, ~~ $Z=0.017\pm0.003$}  \\
\hline
$(B$--$V)$ &13.60$\pm$0.03 &5.25$\pm$0.07 &8.80$-$8.85 &8.80$\pm$0.05 &0.63$\pm$0.07 \\
$(R$--$I)$ &13.60$\pm$0.10 &5.25$\pm$0.24 &8.80$-$8.85 &8.80$\pm$0.05 &0.63$\pm$0.08 \\
$(V$--$I)$ &13.60$\pm$0.10 &5.25$\pm$0.24 &8.80$-$8.85 &8.80$\pm$0.10 &0.63$\pm$0.16 \\
$(V$--$R)$ &13.65$\pm$0.10 &5.37$\pm$0.25 &8.80$-$8.85 &8.80$\pm$0.05 &0.63$\pm$0.08 \\
$(B$--$R)$ &13.60$\pm$0.05 &5.25$\pm$0.12 &8.80$-$8.85 &8.80$\pm$0.05 &0.63$\pm$0.08 \\
      &            &            &            &            &            \\
 Mean &13.60$\pm$0.02 &5.25$\pm$0.06 &            &8.80$\pm$0.02 &0.63$\pm$0.06 \\
\hline
\multicolumn{5}{r}{Haf~08: $E(B$--$V)=0.32\pm0.07$, ~~ $[Fe/H]=-0.39\pm0.26$, ~~ $Z=0.008\pm0.005$}  \\
\hline
$(B$--$V)$ &11.90$\pm$0.06 &2.40$\pm$0.07 &8.60$-$8.75 &8.75$\pm$0.10 &0.56$\pm$0.15 \\
$(R$--$I)$ &11.80$\pm$0.10 &2.30$\pm$0.11 &8.60$-$8.75 &8.75$\pm$0.15 &0.56$\pm$0.23 \\
$(V$--$I)$ &11.90$\pm$0.10 &2.40$\pm$0.11 &8.60$-$8.75 &8.75$\pm$0.10 &0.56$\pm$0.15 \\
$(V$--$R)$ &11.90$\pm$0.10 &2.40$\pm$0.11 &8.60$-$8.75 &8.75$\pm$0.10 &0.56$\pm$0.15 \\
$(B$--$R)$ &11.90$\pm$0.10 &2.40$\pm$0.11 &8.60$-$8.75 &8.75$\pm$0.10 &0.56$\pm$0.15 \\
      &            &            &            &            &            \\
Mean  &11.88$\pm$0.04 &  2.38$\pm$0.04 &          &8.75$\pm$0.05 &0.56$\pm$0.07 \\
\hline
\end{tabular}  
\end{table} 

\clearpage

\begin{table}
\renewcommand\thetable{S4}
\tiny
\centering
\caption {The derived astrophysical parameters of the O\!Cs, Ki~23, NGC~2186, NGC~2304, and NGC~2360.}
\begin{tabular}{lccccc}
\hline 
Colour & $(V_{0}$--$M_{V})$ & $d$ (kpc) & log($A$)-interval & log($A$) & $A$ (Gyr) \\
\hline   
\multicolumn{5}{r}{Ki~23: $E(B$--$V)=0.03\pm0.02$, ~~ $[Fe/H]=-0.11\pm0.11$, ~~ $Z=0.015\pm0.004$}  \\
\hline
$(B$--$V)$ &12.40$\pm$0.04 &3.02$\pm$0.05 & $-$  &9.25$\pm$0.03 &1.78$\pm$0.13 \\
$(R$--$I)$ &12.40$\pm$0.10 &3.02$\pm$0.14 & $-$  &9.25$\pm$0.05 &1.78$\pm$0.22 \\
$(V$--$I)$ &12.40$\pm$0.10 &3.02$\pm$0.14 & $-$  &9.25$\pm$0.05 &1.78$\pm$0.22 \\
$(V$--$R)$ &12.40$\pm$0.05 &3.02$\pm$0.07 & $-$  &9.25$\pm$0.03 &1.78$\pm$0.13 \\
$(B$--$R)$ &12.40$\pm$0.05 &3.02$\pm$0.07 & $-$  &9.25$\pm$0.03 &1.78$\pm$0.13 \\
      &            &            &            &            &            \\
 Mean & 12.40$\pm$0.02 &  3.02$\pm$0.03 &  9.25$\pm$0.02 & 1.78 $\pm$0.07   \\
\hline
\multicolumn{5}{r}{NGC~2186: $E(B$--$V)=0.26\pm0.07$, ~~ $[Fe/H]=-0.39\pm0.26$, ~~ $Z=0.008\pm0.005$} \\
\hline
$(B$--$V)$ &11.40$\pm$0.03&1.91$\pm$0.03 &8.50$-$8.60 &8.50$\pm$0.10 &0.32$\pm$0.08 \\
$(R$--$I)$ &11.40$\pm$0.15&1.91$\pm$0.13 &8.50$-$8.60 &8.50$\pm$0.15 &0.32$\pm$0.13 \\
$(V$--$I)$ &11.40$\pm$0.15&1.91$\pm$0.13 &8.50$-$8.60 &8.50$\pm$0.15 &0.32$\pm$0.13 \\
$(V$--$R)$ &11.40$\pm$0.10&1.91$\pm$0.09 &8.55$-$8.65 &8.55$\pm$0.10 &0.35$\pm$0.09 \\
$(B$--$R)$ &11.40$\pm$0.10&1.91$\pm$0.09 &8.50$-$8.60 &8.50$\pm$0.10 &0.32$\pm$0.08 \\
      &            &            &            &            &            \\
Mean  &11.40$\pm$0.03 &1.91$\pm$0.03 &            &8.51$\pm$0.05 &0.32$\pm$0.04 \\
\hline
\multicolumn{5}{r}{NGC~2304: $E(B$--$V)=0.03\pm0.03$, ~~ $[Fe/H]=-0.20\pm0.18$, ~~ $Z=0.012\pm0.005$}\\
\hline
$(B$--$V)$ &12.80$\pm$0.02 &3.63$\pm$0.03 &8.90$-$8.95 &8.95$\pm$0.03 &0.89$\pm$0.06 \\
$(R$--$I)$ &12.70$\pm$0.10 &3.47$\pm$0.16 &8.95$-$9.00 &9.00$\pm$0.05 &1.00$\pm$0.12 \\
$(V$--$I)$ &12.70$\pm$0.05 &3.47$\pm$0.08 &8.95$-$9.00 &9.00$\pm$0.03 &1.00$\pm$0.07 \\
$(V$--$R)$ &12.80$\pm$0.10 &3.63$\pm$0.17 &8.90$-$8.95 &8.95$\pm$0.05 &0.89$\pm$0.11 \\
$(B$--$R)$ &12.80$\pm$0.05 &3.63$\pm$0.08 &8.90$-$8.95 &8.95$\pm$0.03 &0.89$\pm$0.06 \\
      &            &            &            &            &            \\
Mean  & 12.79$\pm$0.02 &  3.61$\pm$0.03 &            &  8.97$\pm$0.02 & 0.93$\pm$0.03 \\
\hline
 \multicolumn{5}{r}{NGC~2360: $E(B$--$V)=0.09\pm0.06$, ~~ $[Fe/H]=-0.11\pm0.11$, ~~ $Z=0.015\pm0.004$}\\
\hline
$(B$--$V)$ &10.30$\pm$0.05 &1.15$\pm$0.03 &9.00$-$9.05 &9.05$\pm$0.05 &1.12$\pm$0.14 \\
$(R$--$I)$ &10.30$\pm$0.10 &1.15$\pm$0.05 &9.00$-$9.05 &9.05$\pm$0.08 &1.12$\pm$0.23 \\
$(V$--$I)$ &10.20$\pm$0.05 &1.10$\pm$0.03 &9.00$-$9.05 &9.05$\pm$0.05 &1.12$\pm$0.14 \\
$(V$--$R)$ &10.20$\pm$0.05 &1.10$\pm$0.03 &9.00$-$9.05 &9.05$\pm$0.05 &1.12$\pm$0.14 \\
$(B$--$R)$ &10.30$\pm$0.05 &1.15$\pm$0.03 &9.00$-$9.05 &9.05$\pm$0.05 &1.12$\pm$0.14 \\
      &            &            &       &            &            \\
Mean &10.25$\pm$0.02 &1.12$\pm$0.01 &            &9.05$\pm$0.02 &1.12$\pm$0.07 \\
\hline
\end{tabular}  
\end{table}

\begin{table*}
\renewcommand\thetable{S5}
\centering
\tiny
\caption {Possible RC candidates, which are classified in the CM diagrams of 12 O\!Cs from our sample.
The $(\alpha,~\delta)$ coordinates, the $I$ and $V$ magnitudes, and the the distances, $d_{\rm I}$
and $d_{\rm V}$, (in kpc) from the RC calibrations, are listed in Columns~1--2, 3--4, and 5--6,
respectively.  The distances $d_{\rm I}$ and $d_{\rm V}$, derived from CM plots, have also been
given for each cluster in its header together with its name.  The symbols ``M'' and ``NM'', of
Columns 5--6, indicate possible members and non-members according to distance comparisons.}
\begin{tabular}{cccccc}
\hline
$\alpha(^{\circ})$& $\delta(^{\circ})$ & $I$ & $V$ &$d_{\rm I}$&$d_{\rm V}$ \\
\hline
NGC 6802 & $d_{\rm I}=1.74\pm0.12$   &  $d_{\rm V}=1.74\pm0.08$          &            &            &            \\
\hline
292.6643 &   20.2566 & 12.554$\pm$0.003 & 14.619$\pm$0.005 & 1.73$\pm$0.11, M  & 2.03$\pm$0.22, M \\
292.6472 &   20.2642 & 12.657$\pm$0.002 & 14.692$\pm$0.004 & 1.81$\pm$0.12, M  & 2.10$\pm$0.23, M \\
292.6505 &   20.2758 & 12.734$\pm$0.002 & 14.735$\pm$0.004 & 1.88$\pm$0.12, M  & 2.14$\pm$0.24, NM \\
292.6286 &   20.2743 & 12.704$\pm$0.002 & 14.774$\pm$0.004 & 1.85$\pm$0.12, M  & 2.18$\pm$0.24, NM \\
292.6619 &   20.2335 & 12.852$\pm$0.002 & 14.879$\pm$0.003 & 1.98$\pm$0.13, M  & 2.29$\pm$0.25, NM \\
292.6327 &   20.2461 & 12.851$\pm$0.002 & 14.967$\pm$0.003 & 1.98$\pm$0.13, M  & 2.39$\pm$0.26, NM \\
292.6354 &   20.2632 & 12.898$\pm$0.002 & 14.980$\pm$0.005 & 2.03$\pm$0.13, NM & 2.40$\pm$0.26, NM \\
292.6738 &   20.2562 & 12.971$\pm$0.002 & 14.996$\pm$0.004 & 2.10$\pm$0.14, NM & 2.42$\pm$0.27, NM \\
292.6019 &   20.2584 & 12.996$\pm$0.001 & 15.037$\pm$0.005 & 2.12$\pm$0.14, NM & 2.46$\pm$0.27, NM \\
292.6936 &   20.2544 & 13.054$\pm$0.001 & 15.094$\pm$0.004 & 2.18$\pm$0.14, NM & 2.53$\pm$0.28, NM \\
\hline
           &            &            &            &            &            \\
NGC 6866 & $d_{\rm I}=1.26\pm0.05$   &  $d_{\rm V}=1.38\pm0.01$          &            &            &            \\
\hline
300.9904 &   44.1077 & 10.667$\pm$0.003 & 11.706$\pm$0.003 & 1.42$\pm$0.09, NM & 1.53$\pm$0.12, NM \\
300.9799 &   44.1401 & 10.580$\pm$0.001 & 11.625$\pm$0.002 & 1.37$\pm$0.09, M  & 1.47$\pm$0.12, M \\
\hline
         &            &            &            &            &            \\
NGC 7062 & $d_{\rm I}=1.91\pm0.05$   &  $d_{\rm V}=1.91\pm0.12$          &            &            &            \\
\hline
320.8357 &   46.3748 & 10.971$\pm$0.008 & 12.417$\pm$0.004 & 1.17$\pm$0.12, NM & 1.25$\pm$0.14, NM \\
320.8369 &   46.3736 & 11.521$\pm$0.004 & 12.940$\pm$0.007 & 1.51$\pm$0.16, NM & 1.59$\pm$0.18, NM \\
320.8638 &   46.3705 & 11.509$\pm$0.002 & 12.981$\pm$0.003 & 1.50$\pm$0.16, NM & 1.62$\pm$0.18, M \\
320.9146 &   46.3702 & 11.574$\pm$0.001 & 13.095$\pm$0.002 & 1.54$\pm$0.16, NM & 1.71$\pm$0.19, M \\
320.8914 &   46.3852 & 11.531$\pm$0.001 & 13.102$\pm$0.003 & 1.51$\pm$0.16, NM & 1.71$\pm$0.19, M \\
\hline
         &            &            &            &            &            \\
 Ki05       & $d_{\rm I}=1.74\pm0.08$  & $d_{\rm V}=1.74\pm0.09$ &            &            &            \\
\hline
48.6470 &   52.7242 & 13.213$\pm$0.006 & 14.988$\pm$0.004 &   2.57$\pm$0.27, NM &   2.78$\pm$0.31, NM \\
\hline
48.6601 &   52.6974 & 13.203$\pm$0.006 & 14.947$\pm$0.003 & 2.56$\pm$0.27, NM& 2.73$\pm$0.31, NM \\
48.6924 &   52.6893 & 12.918$\pm$0.006 & 14.712$\pm$0.003 & 2.24$\pm$0.23, M & 2.45$\pm$0.28, NM \\
48.5764 &   52.7114 & 12.810$\pm$0.007 & 14.609$\pm$0.005 & 2.13$\pm$0.22, M & 2.33$\pm$0.26, M \\
48.6160 &   52.7113 & 12.717$\pm$0.006 & 14.541$\pm$0.002 & 2.04$\pm$0.21, M & 2.26$\pm$0.26, M \\
48.6462 &   52.7332 & 12.601$\pm$0.006 & 14.385$\pm$0.002 & 1.94$\pm$0.20, M & 2.10$\pm$0.24, M \\
48.7148 &   52.7269 & 12.467$\pm$0.007 & 14.210$\pm$0.003 & 1.82$\pm$0.19, M & 1.94$\pm$0.22, M \\
48.6806 &   52.7040 & 12.401$\pm$0.005 & 14.197$\pm$0.003 & 1.77$\pm$0.18, M & 1.93$\pm$0.22, M \\
\hline
         &            &            &            &            &            \\
NGC 1798 & $d_{\rm I}=3.47\pm0.24$   &  $d_{\rm V}=3.47\pm0.12$          &            &            &            \\
\hline
77.8858 &   47.6875 & 13.708$\pm$0.006 & 15.123$\pm$0.003 & 3.98$\pm$0.35, M  & 4.10$\pm$0.40, NM \\
77.8938 &   47.6883 & 13.808$\pm$0.005 & 15.242$\pm$0.003 & 4.16$\pm$0.36, NM & 4.34$\pm$0.42, NM \\
77.9343 &   47.6977 & 13.899$\pm$0.008 & 15.299$\pm$0.003 & 4.34$\pm$0.38, NM & 4.45$\pm$0.44, NM \\
77.9175 &   47.6885 & 13.873$\pm$0.007 & 15.330$\pm$0.003 & 4.29$\pm$0.38, NM & 4.51$\pm$0.44, NM \\
77.9600 &   47.6997 & 13.995$\pm$0.006 & 15.397$\pm$0.003 & 4.54$\pm$0.40, NM & 4.66$\pm$0.46, NM \\
77.9427 &   47.7011 & 14.099$\pm$0.006 & 15.490$\pm$0.003 & 4.76$\pm$0.42, NM & 4.86$\pm$0.48, NM \\
77.8680 &   47.7047 & 14.066$\pm$0.005 & 15.497$\pm$0.004 & 4.69$\pm$0.41, NM & 4.88$\pm$0.48, NM \\
77.8698 &   47.6774 & 14.086$\pm$0.008 & 15.525$\pm$0.004 & 4.73$\pm$0.41, NM & 4.94$\pm$0.48, NM \\
77.9460 &   47.6833 & 14.136$\pm$0.006 & 15.542$\pm$0.004 & 4.84$\pm$0.42, NM & 4.98$\pm$0.49, NM \\
77.9427 &   47.6869 & 14.190$\pm$0.007 & 15.614$\pm$0.004 & 4.96$\pm$0.43, NM & 5.15$\pm$0.50, NM \\
77.8992 &   47.6949 & 14.153$\pm$0.008 & 15.624$\pm$0.004 & 4.88$\pm$0.43, NM & 5.17$\pm$0.51, NM \\
77.9163 &   47.7094 & 14.268$\pm$0.006 & 15.644$\pm$0.003 & 5.15$\pm$0.45, NM & 5.22$\pm$0.51, NM \\
77.9197 &   47.6774 & 14.291$\pm$0.006 & 15.707$\pm$0.004 & 5.20$\pm$0.45, NM & 5.37$\pm$0.53, NM \\
77.9542 &   47.6659 & 14.302$\pm$0.008 & 15.721$\pm$0.004 & 5.23$\pm$0.46, NM & 5.41$\pm$0.53, NM \\
77.8936 &   47.6955 & 14.315$\pm$0.007 & 15.754$\pm$0.006 & 5.26$\pm$0.46, NM & 5.49$\pm$0.54, NM \\

\hline
\end{tabular} 
\end{table*}

\clearpage

\begin{table*}
\renewcommand\thetable{S5}
\centering
\tiny
\begin{tabular}{cccccc}
\hline
$\alpha(^{\circ})$& $\delta(^{\circ})$ & $I$ & $V$ &$d_{\rm I}$&$d_{\rm V}$ \\
      &            &            &            &            &            \\
NGC 7142 & $d_{\rm I}=2.10\pm0.10$   &  $d_{\rm V}=2.10\pm0.11$          &            &            &            \\
\hline
326.1734 &   65.7971 & 12.840$\pm$0.003 & 14.251$\pm$0.003 & 2.97$\pm$0.31, NM & 3.26$\pm$0.37, NM \\
326.3668 &   65.7518 & 12.722$\pm$0.003 & 14.236$\pm$0.003 & 2.82$\pm$0.30, NM & 3.24$\pm$0.36, NM \\
326.3041 &   65.8276 & 12.747$\pm$0.003 & 14.185$\pm$0.003 & 2.85$\pm$0.30, NM & 3.16$\pm$0.36, NM \\
326.1826 &   65.7785 & 12.616$\pm$0.000 & 14.135$\pm$0.003 & 2.67$\pm$0.28, NM & 3.10$\pm$0.35, NM \\
326.3872 &   65.7920 & 12.382$\pm$0.003 & 13.829$\pm$0.003 & 2.50$\pm$0.25, NM & 2.68$\pm$0.30, NM \\
326.3351 &   65.8086 & 12.233$\pm$0.002 & 13.763$\pm$0.003 & 2.25$\pm$0.23, M & 2.60$\pm$0.30, NM \\
326.2400 &   65.8102 & 12.319$\pm$0.002 & 13.738$\pm$0.002 & 2.33$\pm$0.24, M & 2.57$\pm$0.29, NM \\
326.2530 &   65.7657 & 12.023$\pm$0.002 & 13.4966$\pm$0.002 & 2.04$\pm$0.21, M & 2.30$\pm$0.26, M \\

\hline
        &            &            &            &            &            \\
Ru 01 & $d_{\rm I}=1.45\pm0.10$   &  $d_{\rm V} =1.51\pm0.04$          &            &            &            \\
\hline
99.0931  &   -14.1485 & 10.842$\pm$0.002 & 12.055$\pm$0.002 & 1.40$\pm$0.11, M & 1.53$\pm$0.14 M \\
\hline
         &            &            &            &            &            \\
 Be 35   & $d_{\rm I}=5.01\pm0.23$   & $d_{\rm V} =5.01\pm0.15$   &            &            &            \\
\hline
107.4635 &   2.7182 & 13.095$\pm$0.003 & 14.168$\pm$0.003 & 4.16$\pm$0.38, NM& 4.42$\pm$0.45, M \\
107.4976 &   2.6956 & 13.295$\pm$0.003 & 14.359$\pm$0.003 & 4.56$\pm$0.42, M & 4.83$\pm$0.49, M \\
107.4861 &   2.7698 & 13.455$\pm$0.003 & 14.501$\pm$0.003 & 4.91$\pm$0.45, M & 5.15$\pm$0.52, M \\
107.4932 &   2.7685 & 13.564$\pm$0.003 & 14.599$\pm$0.004 & 5.17$\pm$0.47, M & 5.39$\pm$0.55, M \\
107.4923 &   2.7378 & 13.612$\pm$0.000 & 14.652$\pm$0.004 & 5.28$\pm$0.48, M & 5.52$\pm$0.56, M \\
107.4935 &   2.7336 & 13.796$\pm$0.003 & 14.832$\pm$0.004 & 5.75$\pm$0.52, M & 6.00$\pm$0.61, NM \\
107.4684 &   2.7372 & 13.784$\pm$0.004 & 14.811$\pm$0.004 & 5.72$\pm$0.52, M & 5.94$\pm$0.60, NM \\
107.4698 &   2.7234 & 13.831$\pm$0.005 & 14.816$\pm$0.008 & 5.84$\pm$0.53, NM& 5.96$\pm$0.60, NM \\
107.4899 &   2.7355 & 13.974$\pm$0.004 & 15.001$\pm$0.004 & 6.24$\pm$0.57  NM& 6.49$\pm$0.66, NM \\
\hline
         &            &            &            &            &            \\
 Be 37  & $d_{\rm I}=5.25\pm0.24$    &  $d_{\rm V}=5.25\pm0.07$  &            &            &            \\
\hline
110.1164&   -1.0266 & 13.360$\pm$0.003 & 14.396$\pm$0.003 & 4.97$\pm$0.33, M & 5.35$\pm$0.42, NM \\
110.0793&   -1.0450 & 13.250$\pm$0.002 & 14.251$\pm$0.003 & 4.72$\pm$0.31, NM & 5.00$\pm$0.40, M \\
110.0558&   -0.9965 & 13.286$\pm$0.002 & 14.260$\pm$0.003 & 4.80$\pm$0.32, M & 5.02$\pm$0.40, M \\
110.0808&   -0.9972 & 13.528$\pm$0.005 & 14.432$\pm$0.004 & 5.37$\pm$0.35, M & 5.44$\pm$0.43, M \\
110.0442&   -0.9862 & 13.333$\pm$0.009 & 14.243$\pm$0.004 & 4.91$\pm$0.32, M & 4.98$\pm$0.39, M \\
110.0810&   -1.0268 & 13.337$\pm$0.002 & 14.242$\pm$0.003 & 4.92$\pm$0.32, M & 4.98$\pm$0.39, M \\
110.0678&   -1.0113 & 13.154$\pm$0.003 & 14.017$\pm$0.004 & 4.52$\pm$0.30, M &4.49$\pm$0.36, NM \\
110.0768&   -1.0240 & 12.940$\pm$0.002 & 13.850$\pm$0.002 & 4.10$\pm$0.27, NM &4.16$\pm$0.33, NM \\
\hline
        &            &            &            &            &            \\
Ki 23   &  $d_{\rm I}=3.02\pm0.14$   &  $d_{\rm V}=3.02\pm0.05$    &            &            &            \\
\hline
110.444 &   -0.9787 & 12.585$\pm$0.002 & 13.598$\pm$0.002 & 3.54$\pm$0.10, NM & 3.81$\pm$0.20, NM \\
110.4522&   -0.9922 & 12.375$\pm$0.002 & 13.428$\pm$0.002 & 3.21$\pm$0.09, M  & 3.52$\pm$0.19, NM \\
110.4692&   -0.9875 & 12.477$\pm$0.001 & 13.572$\pm$0.002 & 3.37$\pm$0.10, NM & 3.77$\pm$0.20, NM \\

\hline
        &            &            &            &            &            \\
NGC 2304  & $d_{\rm I}=3.47\pm0.08$     & $d_{\rm V}=3.63\pm0.03$     &            &            &            \\
\hline
         &           &          I &          V &    $d_{I}$   &         dV \\
103.8089 &   17.9764 & 12.136$\pm$0.002 & 13.190$\pm$0.002 & 2.88$\pm$0.12, NM & 3.16$\pm$0.19, NM \\
103.8109 &   17.9954 & 12.758$\pm$0.002 & 13.678$\pm$0.003 & 3.83$\pm$0.16, NM & 3.95$\pm$0.24, NM \\
103.8096 &   17.9854 & 13.053$\pm$0.002 & 14.008$\pm$0.003 & 4.39$\pm$0.18, NM & 4.60$\pm$0.28, NM \\
103.7918 &   17.9802 & 13.144$\pm$0.002 & 14.125$\pm$0.003 & 4.58$\pm$0.19, NM & 4.86$\pm$0.29, NM \\
\hline
         &            &            &            &            &            \\
NGC 2360 & $d_{\rm I}=1.10\pm0.03$        & $d_{\rm V}=1.15\pm0.03$   &            &            &            \\
\hline
109.4799 &   -15.6987 & 10.462$\pm$0.007 & 11.506$\pm$0.002 & 1.26$\pm$0.10, NM & 1.33$\pm$0.12, NM \\
109.3857 &   -15.6498 & 10.457$\pm$0.005 & 11.485$\pm$0.003 & 1.26$\pm$0.10, NM & 1.32$\pm$0.12, NM \\
109.4300 &   -15.6185 & 10.078$\pm$0.005 & 11.204$\pm$0.003 & 1.06$\pm$0.08, M  & 1.16$\pm$0.10, M \\
109.4528 &   -15.5971 & 10.301$\pm$0.007 & 11.354$\pm$0.003 & 1.17$\pm$0.09, M  & 1.24$\pm$0.11, M \\
109.4140 &   -15.6100 & 10.256$\pm$0.006 & 11.279$\pm$0.003 & 1.15$\pm$0.09, M  & 1.20$\pm$0.11, M \\
109.4293 &   -15.6256 & 10.062$\pm$0.006 & 11.089$\pm$0.002 & 1.05$\pm$0.08, M  & 1.10$\pm$0.10, M \\
109.4529 &   -15.6614 &  9.972$\pm$0.006 & 11.038$\pm$0.003 & 1.00$\pm$0.08, M  & 1.08$\pm$0.10, M \\
109.3985 &   -15.6252 & 10.061$\pm$0.005 & 11.058$\pm$0.002 & 1.05$\pm$0.08, M  & 1.09$\pm$0.10, M \\
109.3961 &   -15.6432 &  9.720$\pm$0.006 & 10.797$\pm$0.003 & 0.90$\pm$0.07, NM & 0.96$\pm$0.09, NM \\
109.4474 &   -15.6358 &  9.578$\pm$0.007 & 10.732$\pm$0.003 & 0.84$\pm$0.07, NM & 0.94$\pm$0.08, NM \\
\hline
\end{tabular}  
\end{table*} 

\clearpage

\begin{table*}
\renewcommand\thetable{S6}
\centering
\tiny
\caption{Possible BS candidates in NGC~7142. The coordinates, $(\alpha,~\delta)$, of the candidates are
given in Columns~1--2, while $V$ magnitudes and colour indices for $VBRI$ photometry, together with their
uncertainties, have been listed in Columns~3--8.}
\begin{tabular}{lccccccc}
\hline
 $\alpha(^{\circ})$ & $\delta(^{\circ})$ & $V$ & $B-V$ & $R-I$ & $V-I$ & $V-R$ & $B-R$ \\
\hline
326.3132 &65.8234 & 15.443$\pm$0.006 & 0.676$\pm$0.010 & 0.499$\pm$0.009 & 0.894$\pm$0.008 & 0.395$\pm$0.009 & 1.071$\pm$0.011 \\
326.2488 &65.7197 & 15.436$\pm$0.006 & 0.732$\pm$0.011 & 0.498$\pm$0.009 & 0.945$\pm$0.008 & 0.447$\pm$0.009 & 1.179$\pm$0.011 \\
326.2103 &65.8170 & 15.406$\pm$0.004 & 0.764$\pm$0.008 & 0.522$\pm$0.005 & 0.979$\pm$0.005 & 0.457$\pm$0.006 & 1.221$\pm$0.008 \\
326.2602 &65.7404 & 15.344$\pm$0.004 & 0.744$\pm$0.008 & 0.530$\pm$0.005 & 1.002$\pm$0.005 & 0.472$\pm$0.006 & 1.216$\pm$0.008 \\
326.2210 &65.7472 & 15.201$\pm$0.004 & 0.712$\pm$0.007 & 0.499$\pm$0.006 & 0.888$\pm$0.006 & 0.389$\pm$0.006 & 1.101$\pm$0.007 \\
326.3679 &65.8245 & 15.097$\pm$0.004 & 0.550$\pm$0.007 & 0.417$\pm$0.006 & 0.739$\pm$0.006 & 0.322$\pm$0.006 & 0.872$\pm$0.007 \\
326.1821 &65.7889 & 14.913$\pm$0.004 & 0.572$\pm$0.006 & 0.365$\pm$0.004 & 0.702$\pm$0.005 & 0.337$\pm$0.005 & 0.909$\pm$0.005 \\
326.3591 &65.7573 & 14.900$\pm$0.004 & 0.760$\pm$0.007 & 0.500$\pm$0.004 & 0.945$\pm$0.005 & 0.445$\pm$0.005 & 1.205$\pm$0.007 \\
326.4176 &65.7372 & 14.859$\pm$0.004 & 0.660$\pm$0.007 & 0.467$\pm$0.004 & 0.855$\pm$0.005 & 0.388$\pm$0.005 & 1.048$\pm$0.007 \\
326.1879 &65.8063 & 14.630$\pm$0.003 & 0.417$\pm$0.005 & 0.322$\pm$0.006 & 0.569$\pm$0.005 & 0.247$\pm$0.005 & 0.664$\pm$0.006 \\
326.2186 &65.7718 & 14.476$\pm$0.003 & 0.401$\pm$0.005 & 0.305$\pm$0.003 & 0.507$\pm$0.005 & 0.202$\pm$0.005 & 0.603$\pm$0.006 \\
326.2166 &65.7727 & 14.223$\pm$0.003 & 0.442$\pm$0.005 & 0.323$\pm$0.004 & 0.555$\pm$0.004 & 0.232$\pm$0.004 & 0.674$\pm$0.005 \\
\hline
\end{tabular}  
\end{table*}

\clearpage

\begin{figure*}
\renewcommand\thefigure{S1}
\centering
\epsfig{file=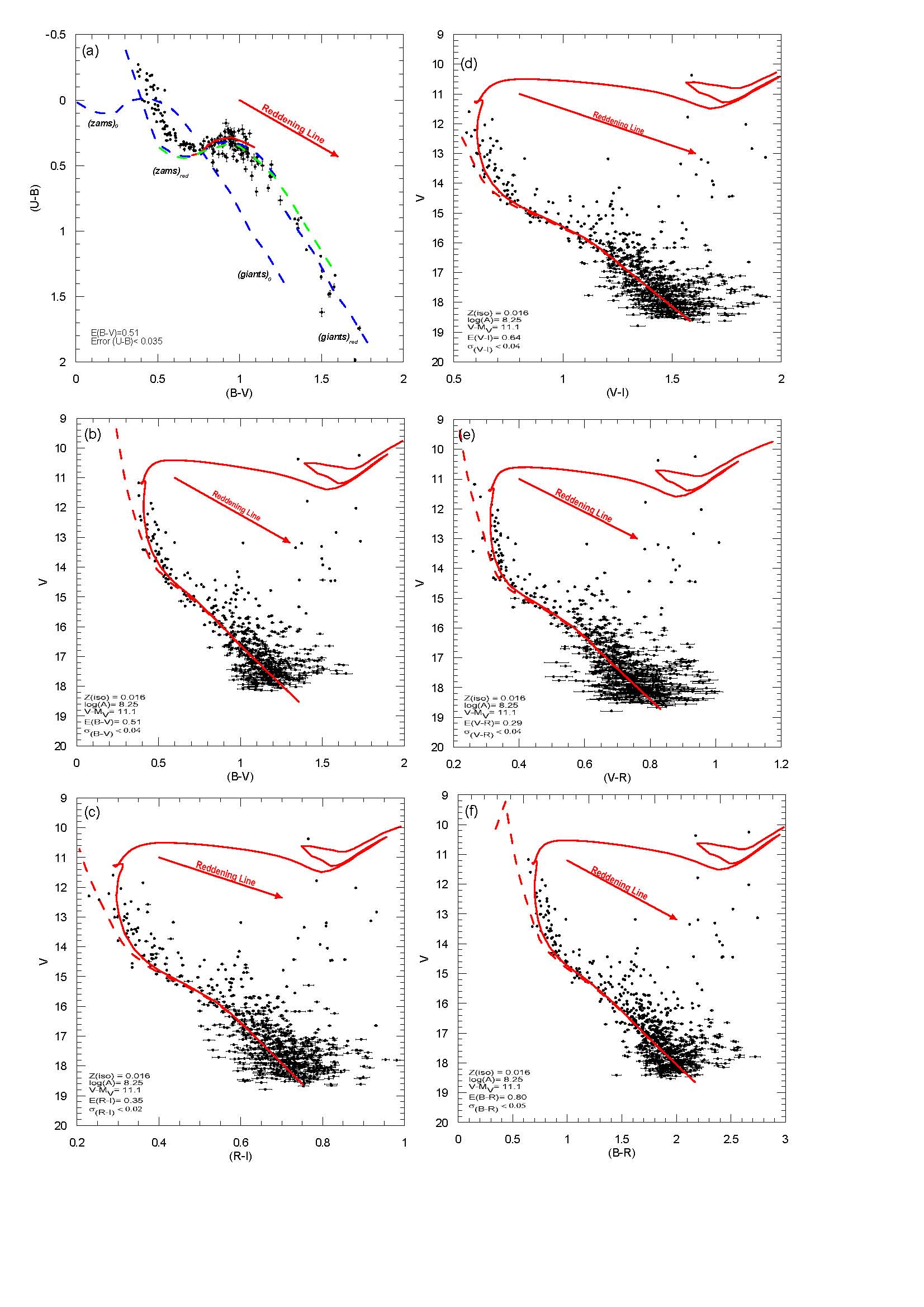, width=12cm, height=12.4cm}
\caption {Panel (a): the $(U$--$B, B$--$V)$ diagram of NGC~6694.  
Panels~(b)$-$(f): CM diagrams of $(V,B$--$V)$, $(V,R$--$I)$,  
$(V,V$--$I)$,  $(V,V$--$R)$, and  $(V,B$--$R)$. 
The symbols are the same as Fig.~1 and Figs.~2--3.}
\end{figure*}

\begin{figure*}
\renewcommand\thefigure{S2}
\centering
\epsfig{file=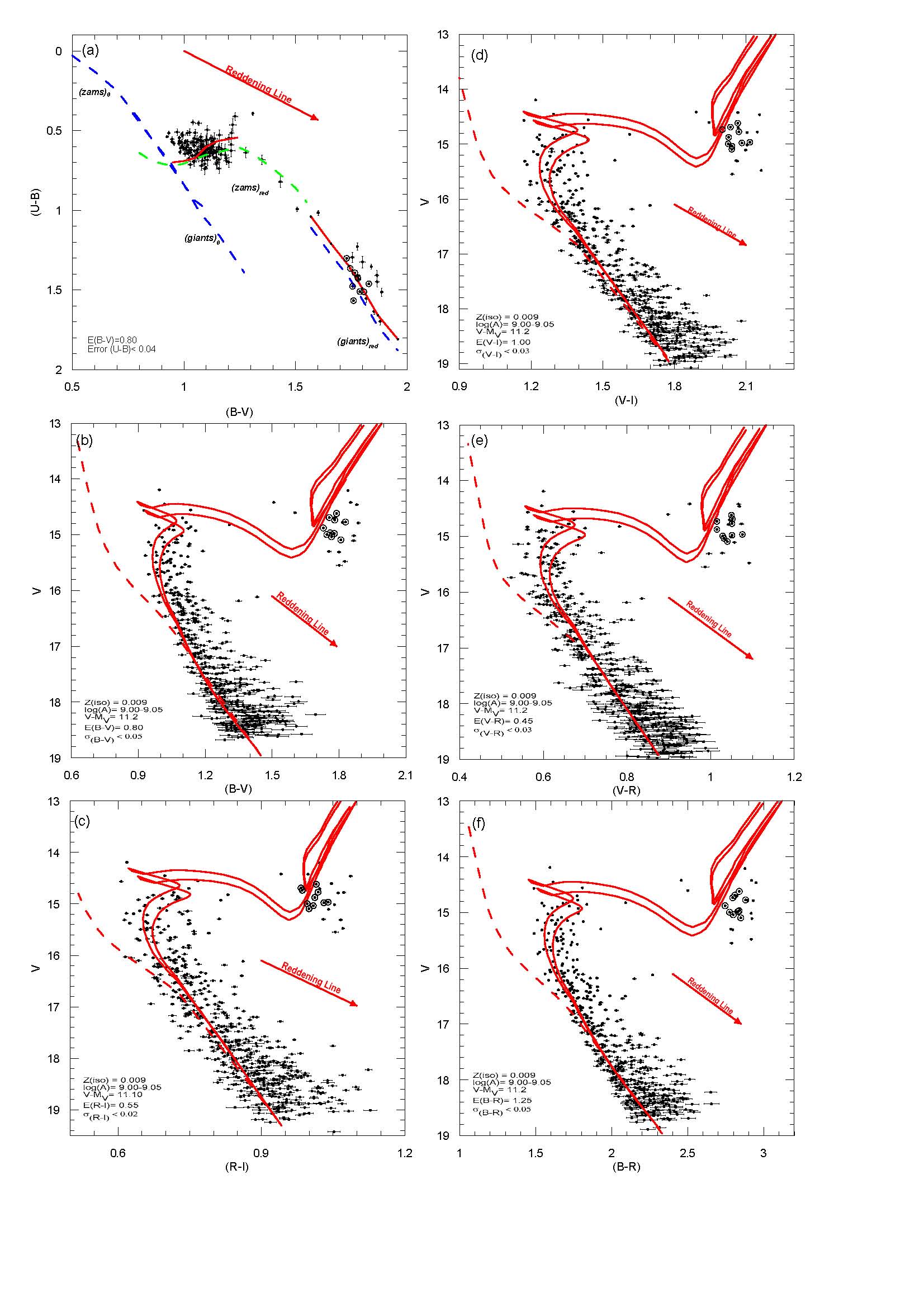, width=12cm, height=12.4cm}
\caption {Panel (a): the $(U$--$B, B$--$V)$ diagram of NGC~6802.  
Panels~(b)--(f): CM diagrams of $(V,B$--$V)$, $(V,R$--$I)$,  
$(V,V$--$I)$,  $(V,V$--$R)$, and  $(V,B$--$R)$. 
The symbols are the same as Fig.~1 and Figs.~2--3.}
\end{figure*}

\begin{figure*}
\renewcommand\thefigure{S3}
\centering
\epsfig{file=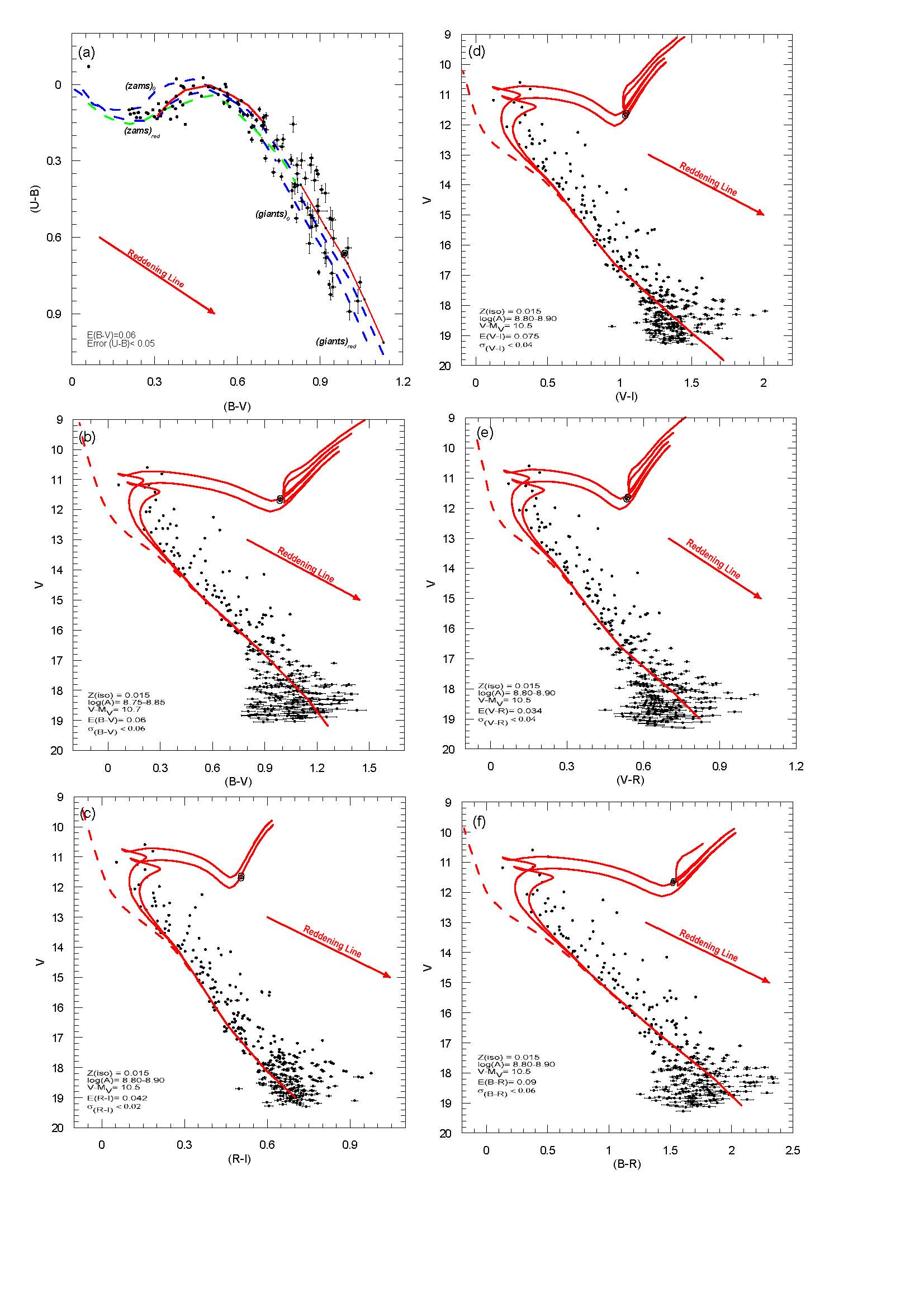, width=12cm, height=12.4cm}
\caption {Panel (a): the $(U$--$B, B$--$V)$ diagram of NGC~6866.  
Panels~(b)--(f): CM diagrams of $(V,B$--$V)$, $(V,R$--$I)$,  
$(V,V$--$I)$,  $(V,V$--$R)$, and  $(V,B$--$R)$. 
The symbols are the same as Fig.~1 and Figs.~2--3.}
\end{figure*}

\begin{figure*}
\renewcommand\thefigure{S4}
\centering
\epsfig{file=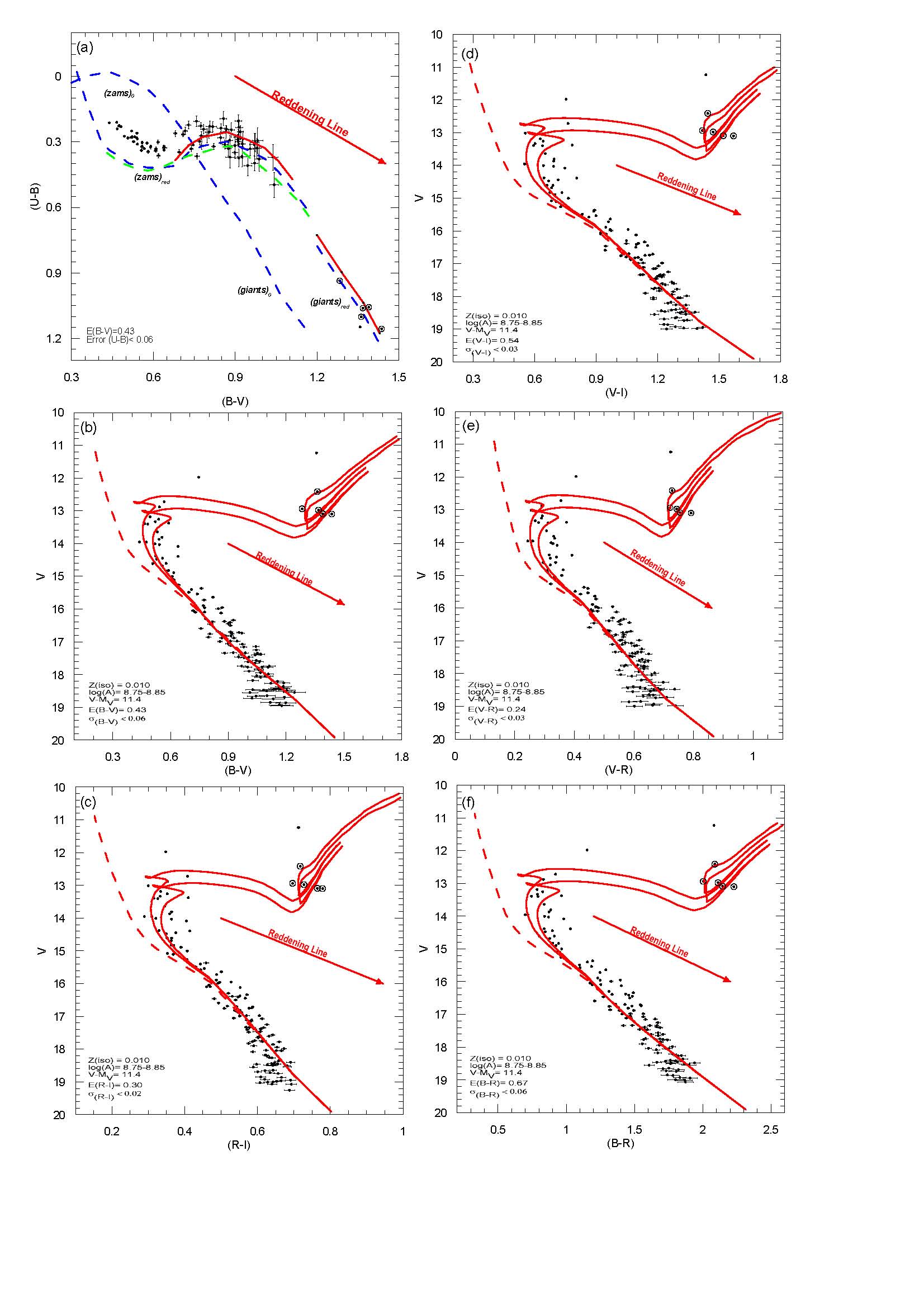, width=12cm, height=12.4cm}
\caption {Panel (a): the $(U$--$B, B$--$V)$ diagram of NGC~7062.  
Panels~(b)--(f): CM diagrams of $(V,B$--$V)$, $(V,R$--$I)$,  
$(V,V$--$I)$,  $(V,V$--$R)$, and  $(V,B$--$R)$. 
The symbols are the same as Fig.~1 and Figs.~2--3.}
\end{figure*}

\begin{figure*}
\renewcommand\thefigure{S5}
\centering
\epsfig{file=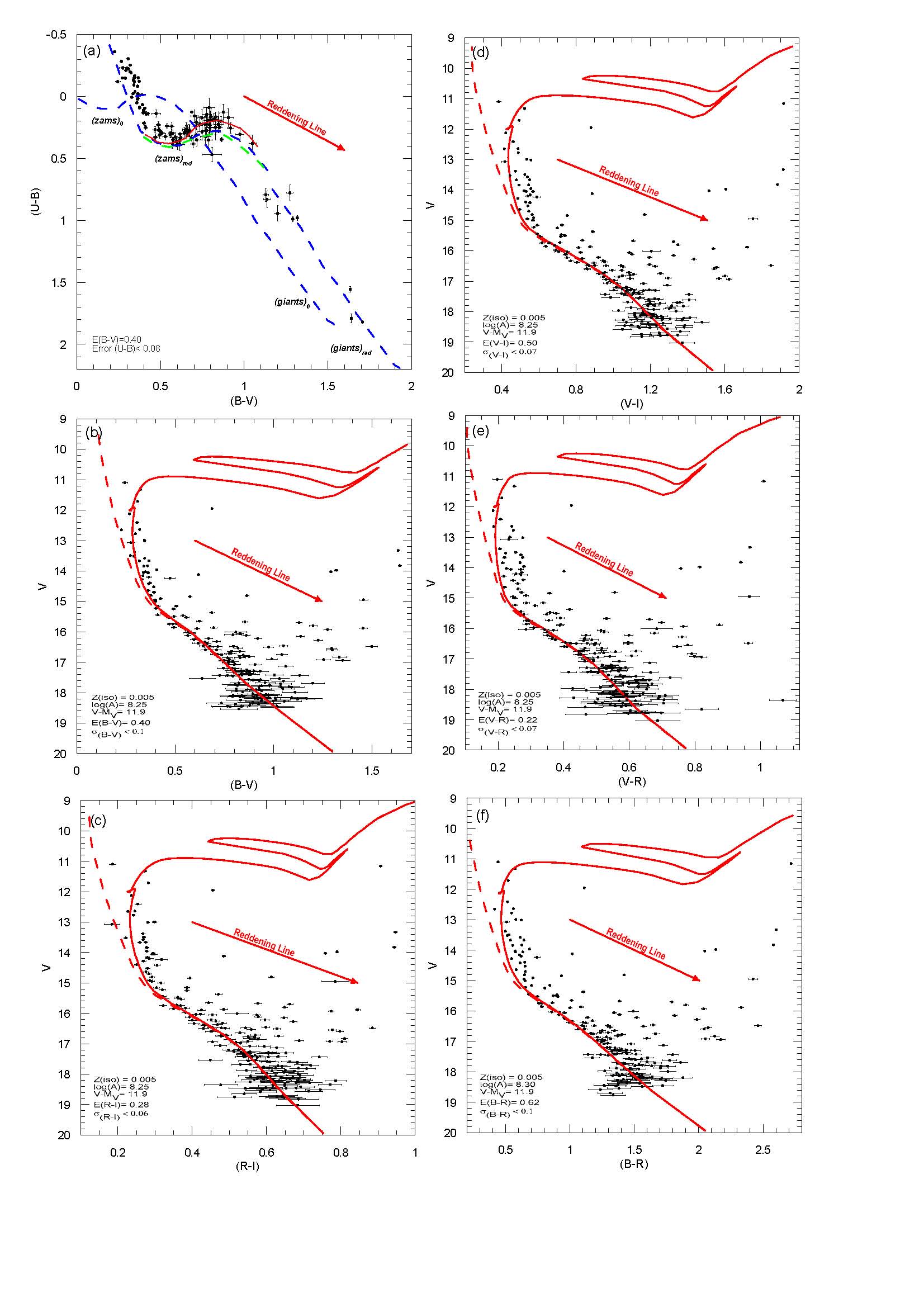, width=12cm, height=12.4cm}
\caption {Panel (a): the $(U$--$B, B$--$V)$ diagram of NGC~436.  
Panels~(b)--(f): CM diagrams of $(V,B$--$V)$, $(V,R$--$I)$,  
$(V,V$--$I)$,  $(V,V$--$R)$, and  $(V,B$--$R)$. 
The symbols are the same as Fig.~1 and Figs.~2--3.}
\end{figure*}

\begin{figure*}
\renewcommand\thefigure{S6}
\centering
\epsfig{file=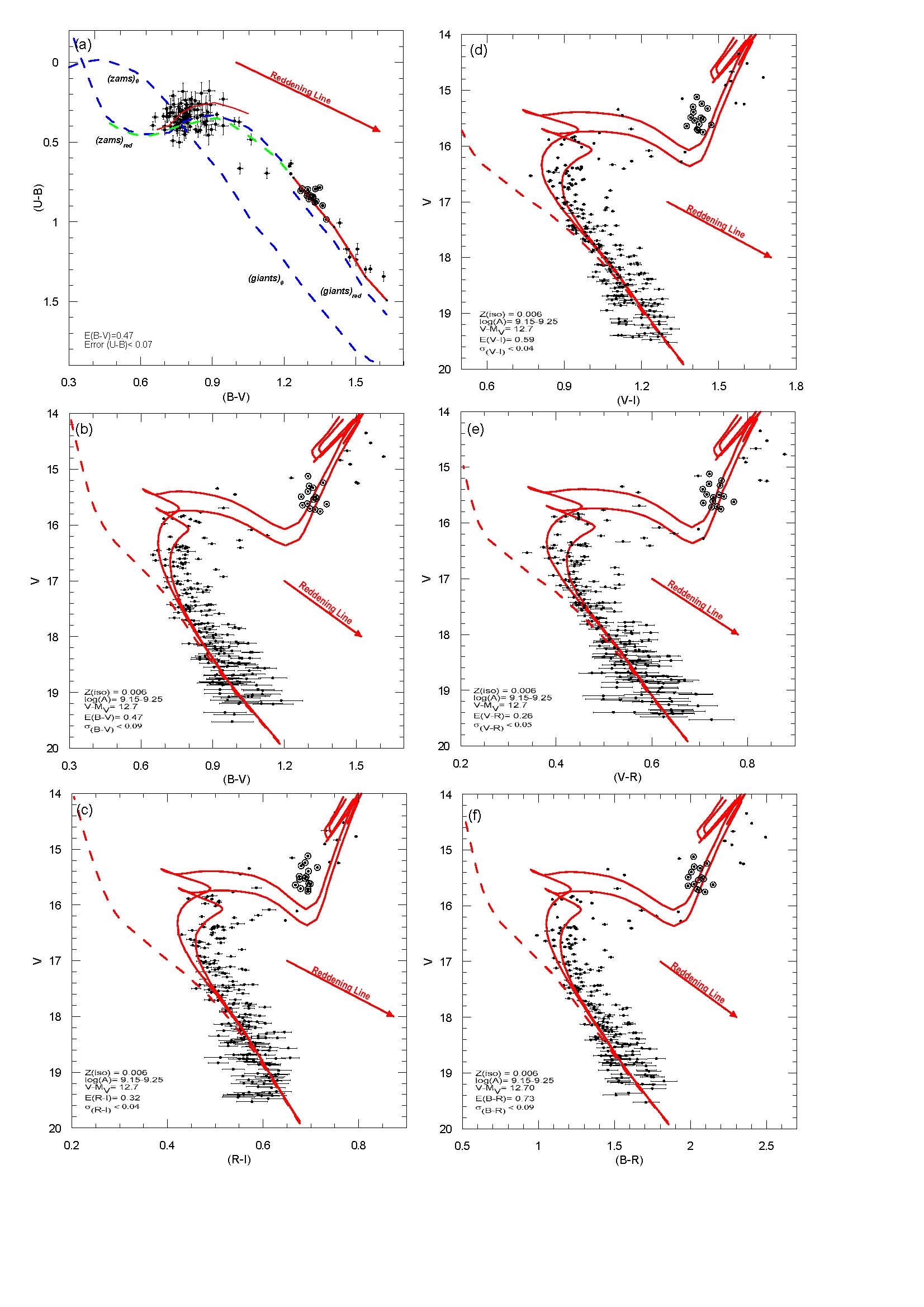, width=12cm, height=12.4cm}
\caption {Panel (a): the $(U$--$B, B$--$V)$ diagram of NGC~1798.  
Panels~(b)--(f): CM diagrams of $(V,B$--$V)$, $(V,R$--$I)$,  
$(V,V$--$I)$,  $(V,V$--$R)$, and  $(V,B$--$R)$. 
The symbols are the same as Fig.~1 and Figs.~2--3.}
\end{figure*}

\begin{figure*}
\renewcommand\thefigure{S7}
\centering
\epsfig{file=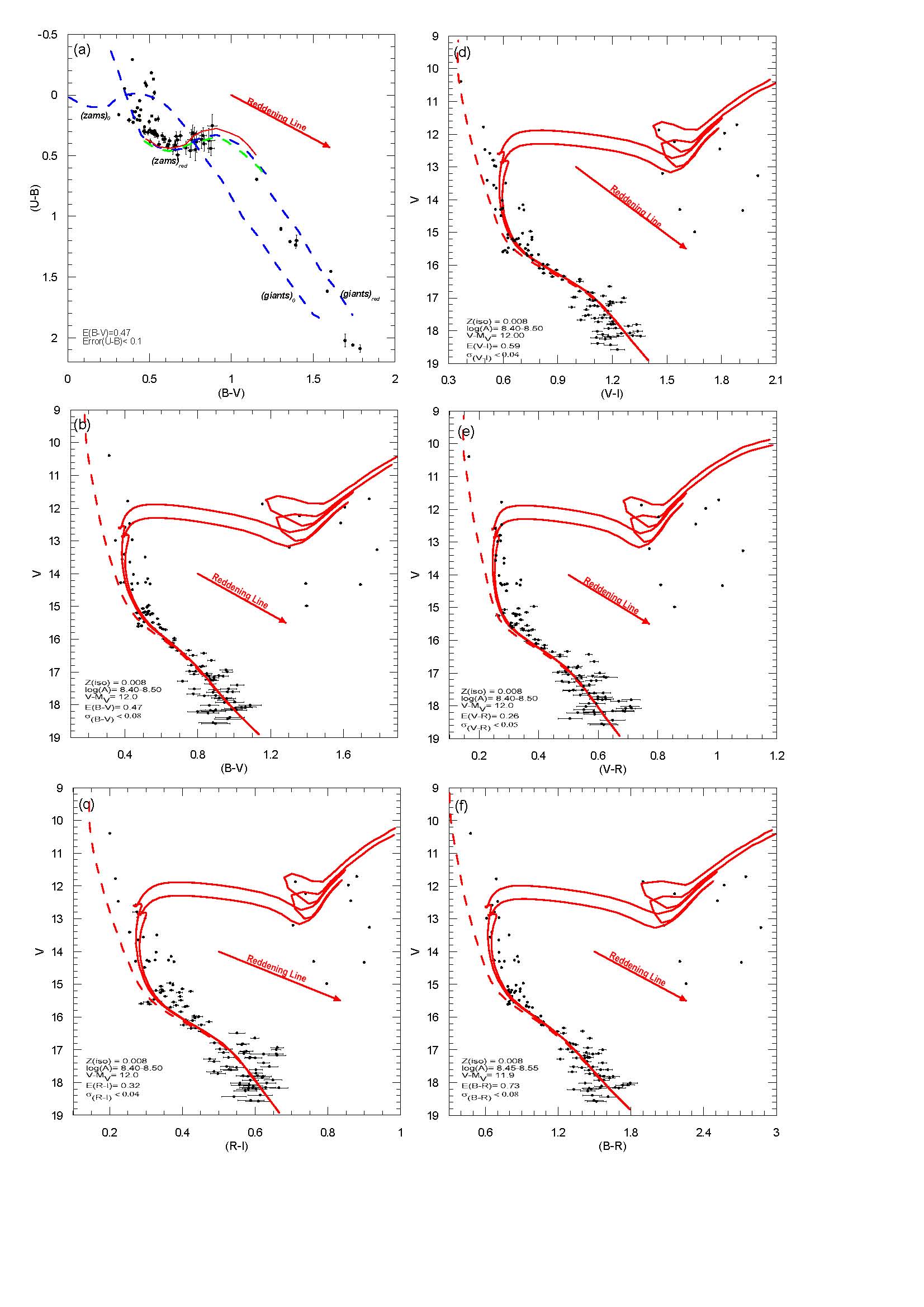, width=12cm, height=12.4cm}
\caption {Panel (a): the $(U$--$B, B$--$V)$ diagram of NGC~1857.  
Panels~(b)--(f): CM diagrams of $(V,B$--$V)$, $(V,R$--$I)$,  
$(V,V$--$I)$,  $(V,V$--$R)$, and  $(V,B$--$R)$. 
The symbols are the same as Fig.~1 and Figs.~2--3.}
\end{figure*}

\begin{figure*}
\renewcommand\thefigure{S8}
\centering
\epsfig{file=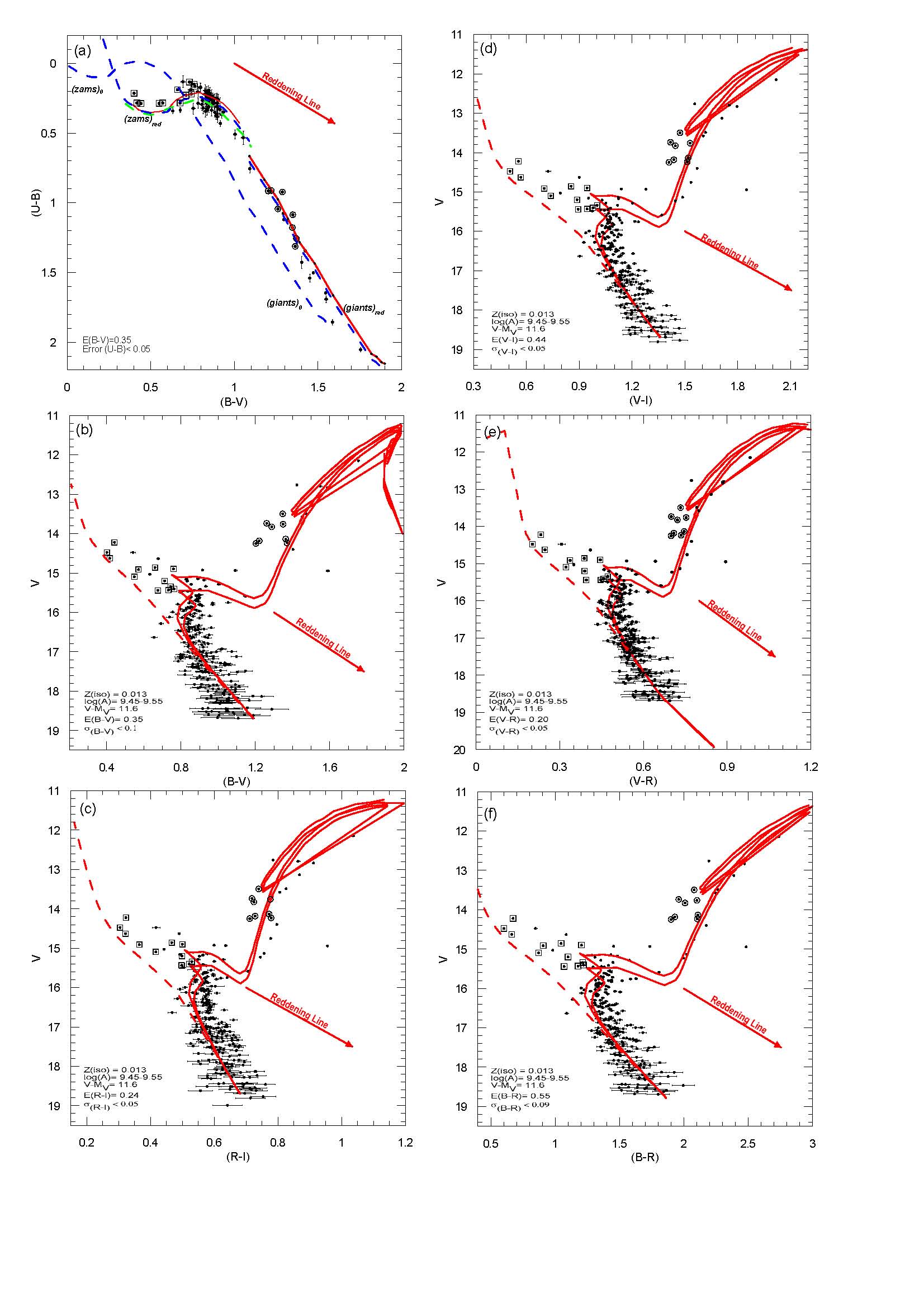, width=12cm, height=12.4cm}
\caption {Panel (a): the $(U$--$B, B$--$V)$ diagram of NGC~7142.  
Panels~(b)--(f): CM diagrams of $(V,B$--$V)$, $(V,R$--$I)$,  
$(V,V$--$I)$,  $(V,V$--$R)$, and  $(V,B$--$R)$. 
The symbols are the same as Fig.~1 and Figs.~2--3; the B\!S candidates
are marked with open squares.}
\end{figure*}

\begin{figure*}
\renewcommand\thefigure{S9}
\centering
\epsfig{file=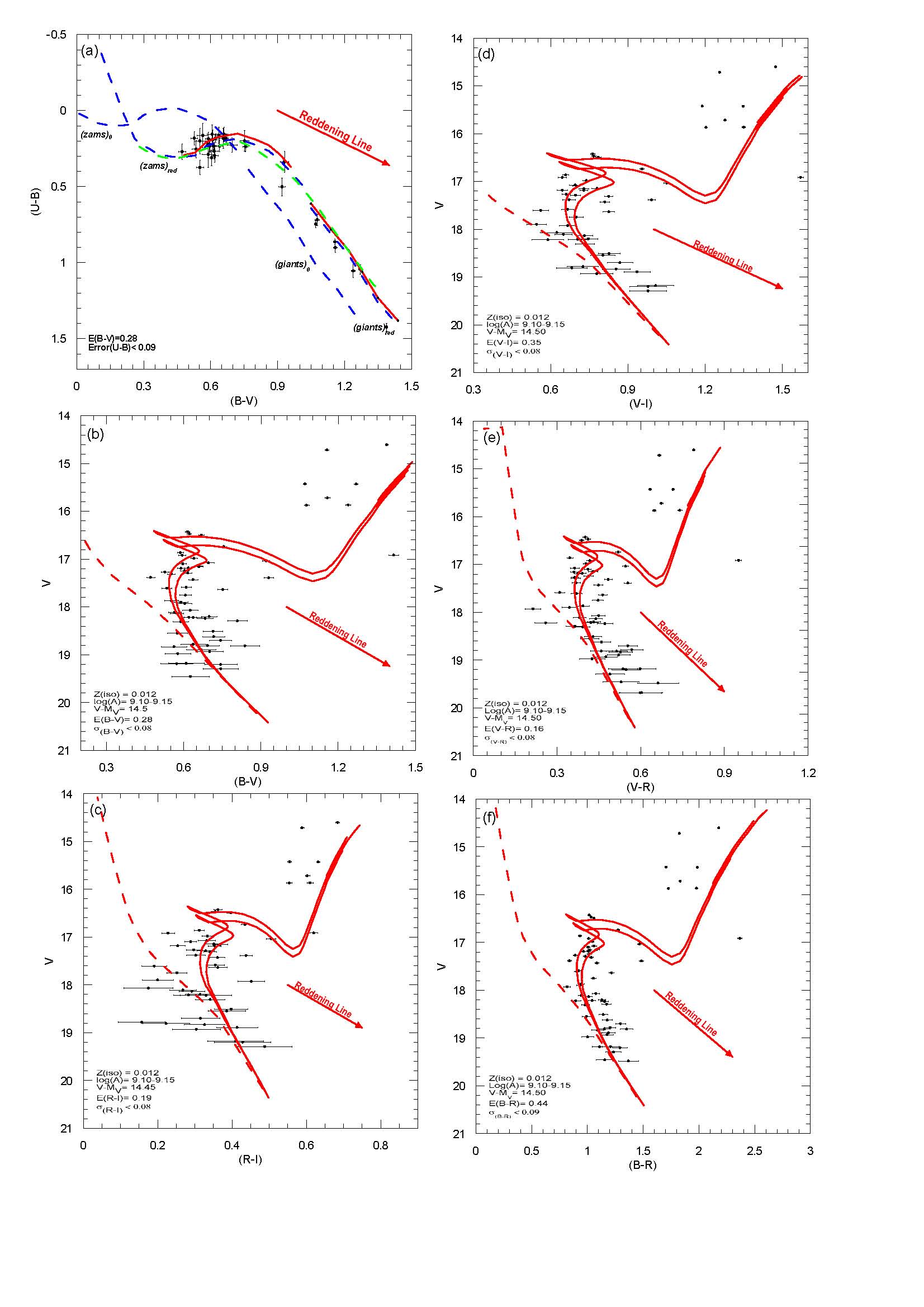, width=12cm, height=12.4cm}
\caption {Panel (a): the $(U$--$B, B$--$V)$ diagram of Be~73.  
Panels~(b)--(f): CM diagrams of $(V,B$--$V)$, $(V,R$--$I)$,  
$(V,V$--$I)$,  $(V,V$--$R)$, and  $(V,B$--$R)$. 
The symbols are the same as Fig.~1 and Figs.~2--3.}
\end{figure*}

\begin{figure*}
\renewcommand\thefigure{S10}
\centering
\epsfig{file=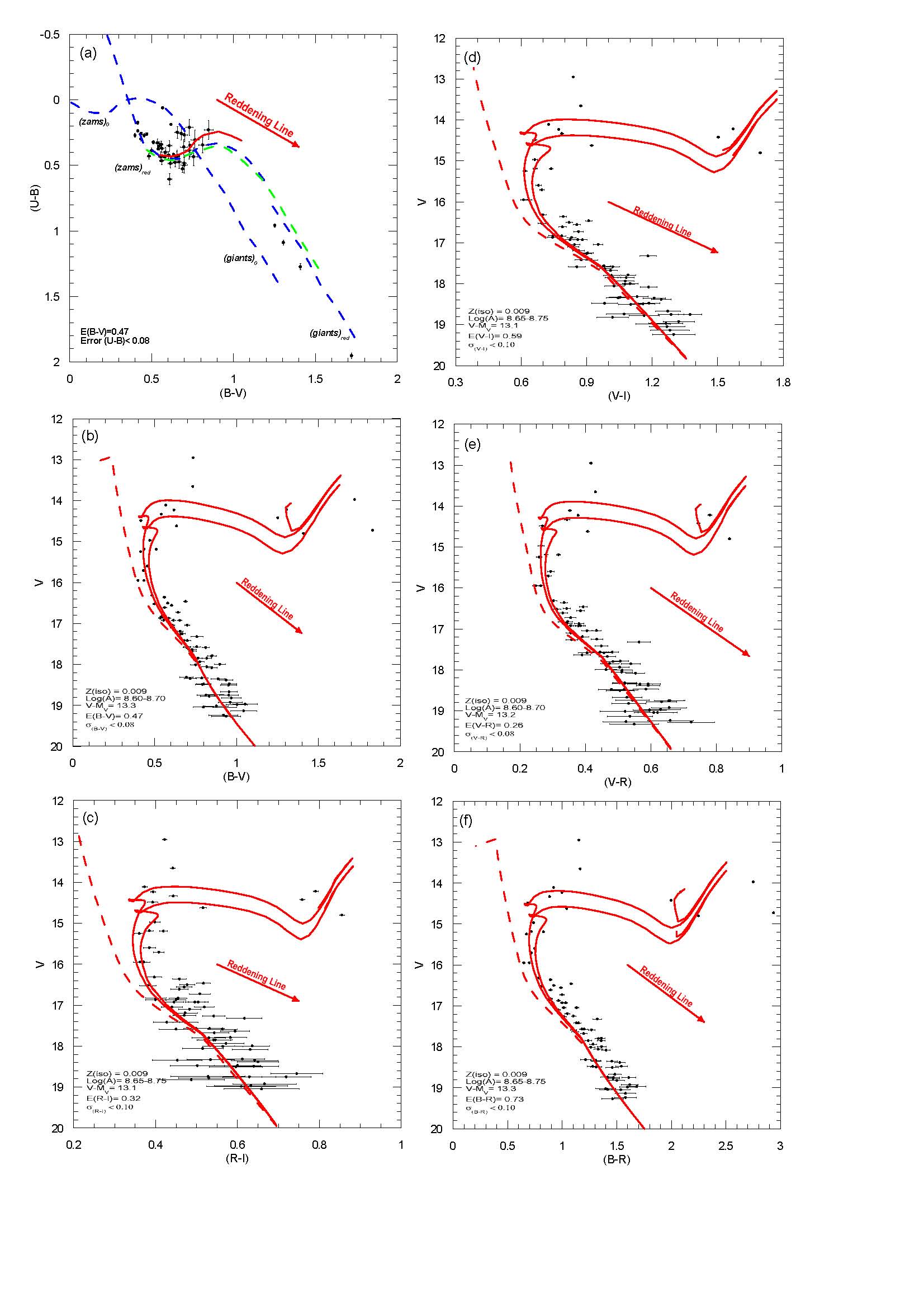, width=12cm, height=12.4cm}
\caption {Panel (a): the $(U$--$B, B$--$V)$ diagram of Haf~04.  
Panels~(b)--(f): CM diagrams of $(V,B$--$V)$, $(V,R$--$I)$,  
$(V,V$--$I)$,  $(V,V$--$R)$, and  $(V,B$--$R)$. 
The symbols are the same as Fig.~1 and Figs.~2--3.}
\end{figure*}

\begin{figure*}
\renewcommand\thefigure{S11}
\centering
\epsfig{file=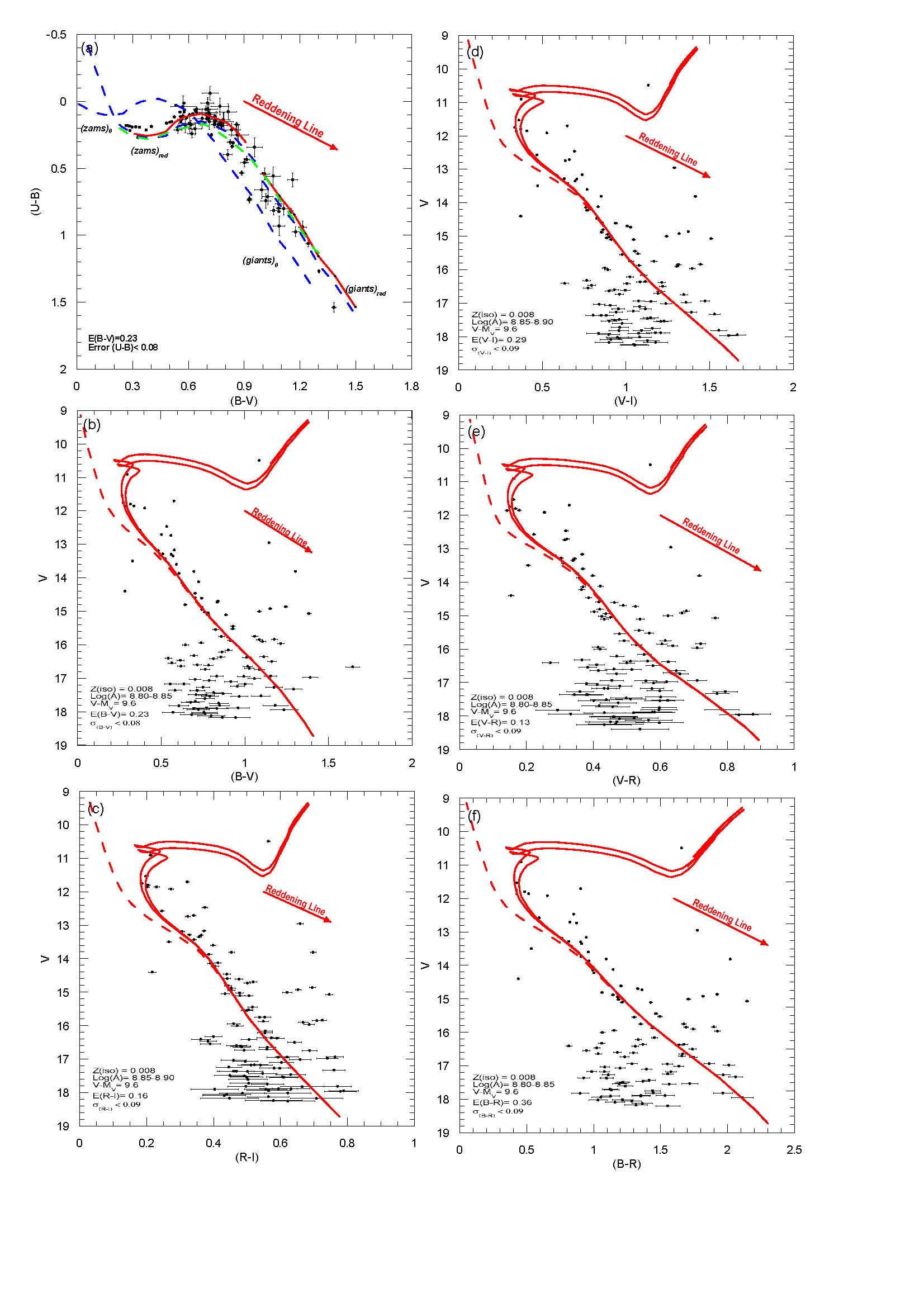, width=12cm, height=12.4cm}
\caption {Panel (a): the $(U$--$B, B$--$V)$ diagram of NGC~2215.  
Panels~(b)--(f): CM diagrams of $(V,B$--$V)$, $(V,R$--$I)$,  
$(V,V$--$I)$,  $(V,V$--$R)$, and  $(V,B$--$R)$. 
The symbols are the same as Fig.~1 and Figs.~2--3.}
\end{figure*}

\begin{figure*}
\renewcommand\thefigure{S12}
\centering
\epsfig{file=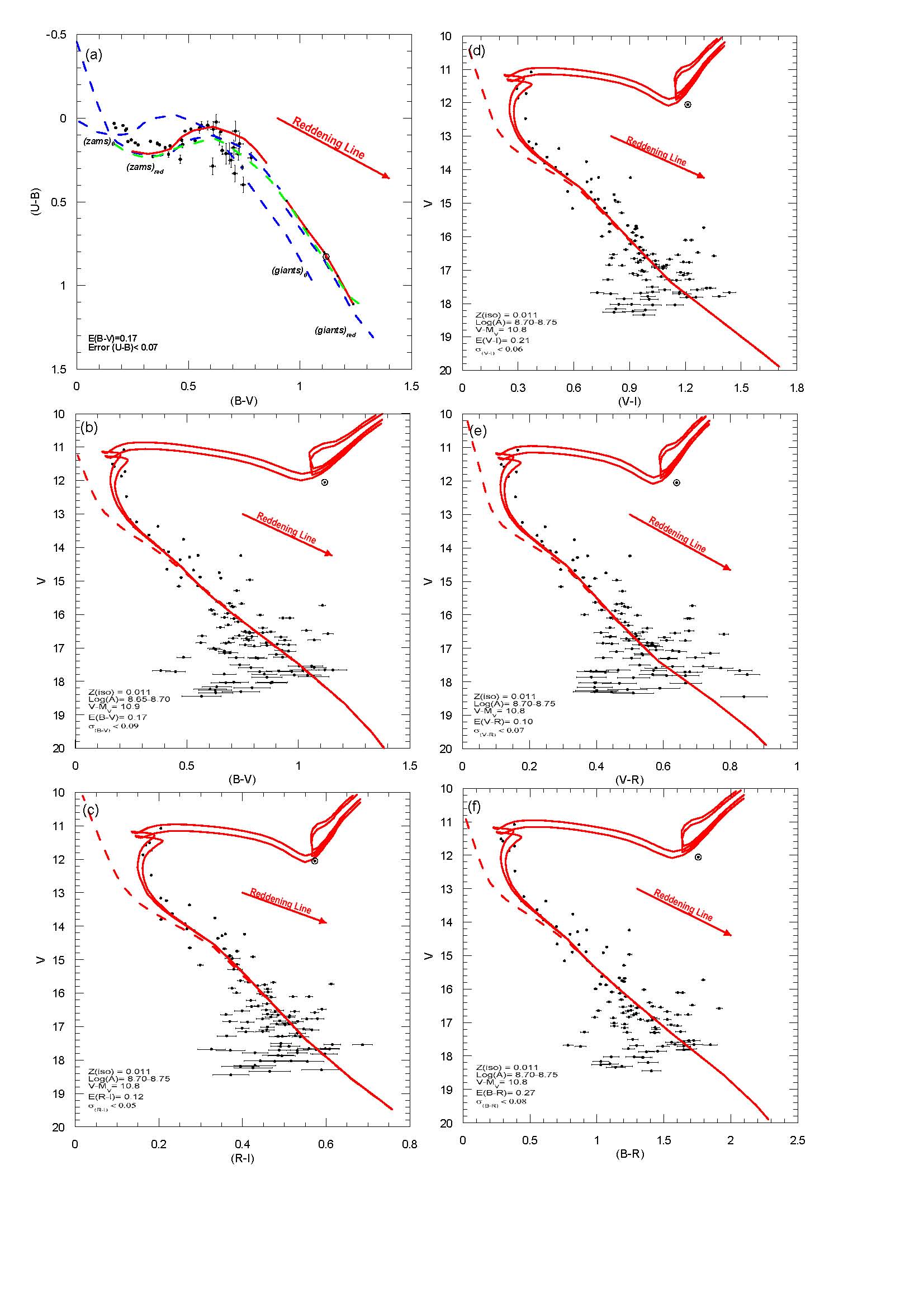, width=12cm, height=12.4cm}
\caption {Panel (a): the $(U$--$B, B$--$V)$ diagram of Rup~01.  
Panels~(b)--(f): CM diagrams of $(V,B$--$V)$, $(V,R$--$I)$,  
$(V,V$--$I)$,  $(V,V$--$R)$, and  $(V,B$--$R)$. 
The symbols are the same as Fig.~1 and Figs.~2--3.}
\end{figure*}

\begin{figure*}
\renewcommand\thefigure{S13}
\centering
\epsfig{file=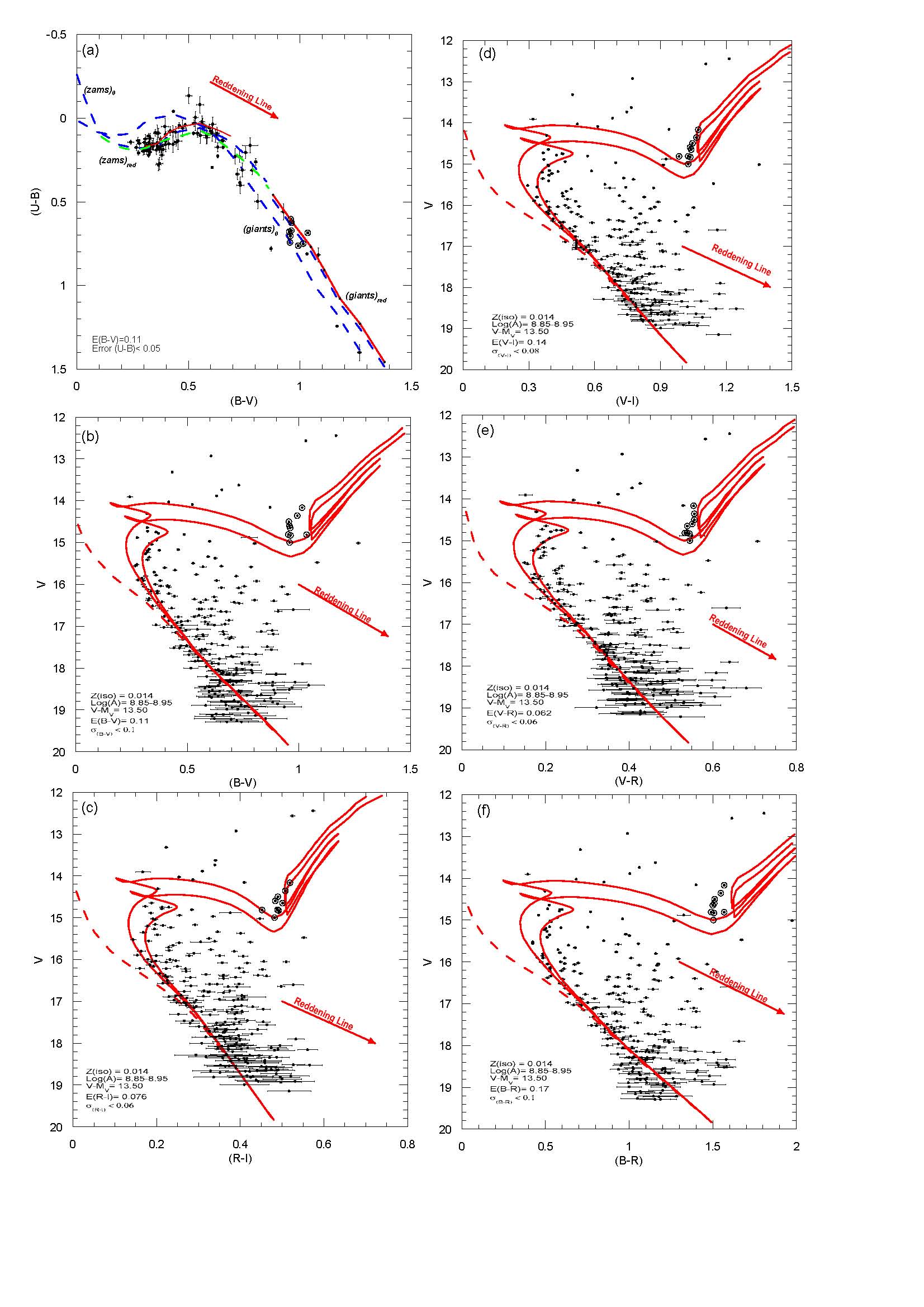, width=12cm, height=12.4cm}
\caption {Panel (a): the $(U$--$B, B$--$V)$ diagram of Be~35.  
Panels~(b)--(f): CM diagrams of $(V,B$--$V)$, $(V,R$--$I)$,  
$(V,V$--$I)$,  $(V,V$--$R)$, and  $(V,B$--$R)$. 
The symbols are the same as Fig.~1 and Figs.~2--3.}
\end{figure*}

\begin{figure*}
\renewcommand\thefigure{S14}
\centering
\epsfig{file=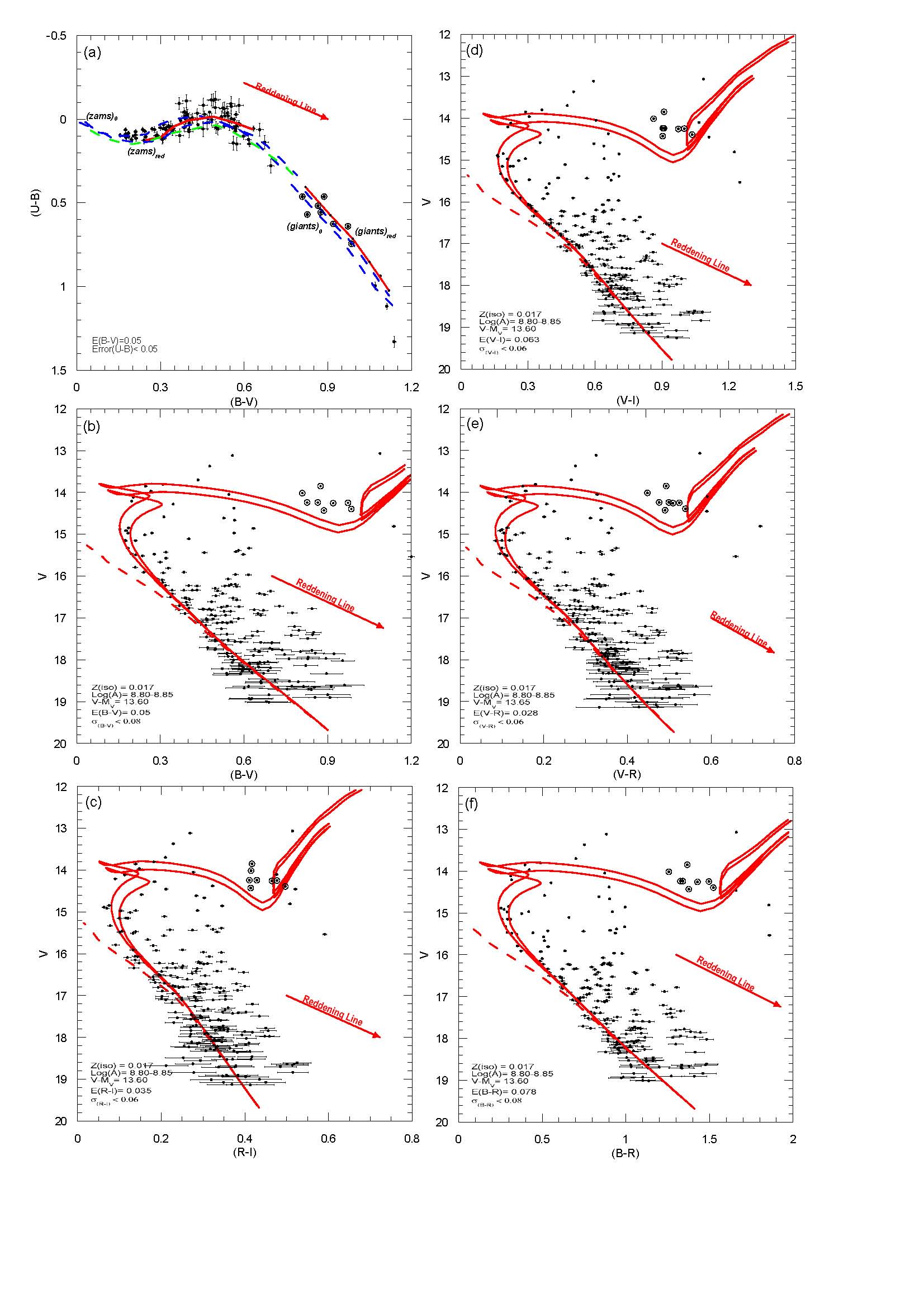, width=12cm, height=12.4cm}
\caption {Panel (a): the $(U$--$B, B$--$V)$ diagram of Be~37.  
Panels~(b)--(f): CM diagrams of $(V,B$--$V)$, $(V,R$--$I)$,  
$(V,V$--$I)$,  $(V,V$--$R)$, and  $(V,B$--$R)$. 
The symbols are the same as Fig.~1 and Figs.~2--3.}
\end{figure*}

\begin{figure*}
\renewcommand\thefigure{S15}
\centering
\epsfig{file=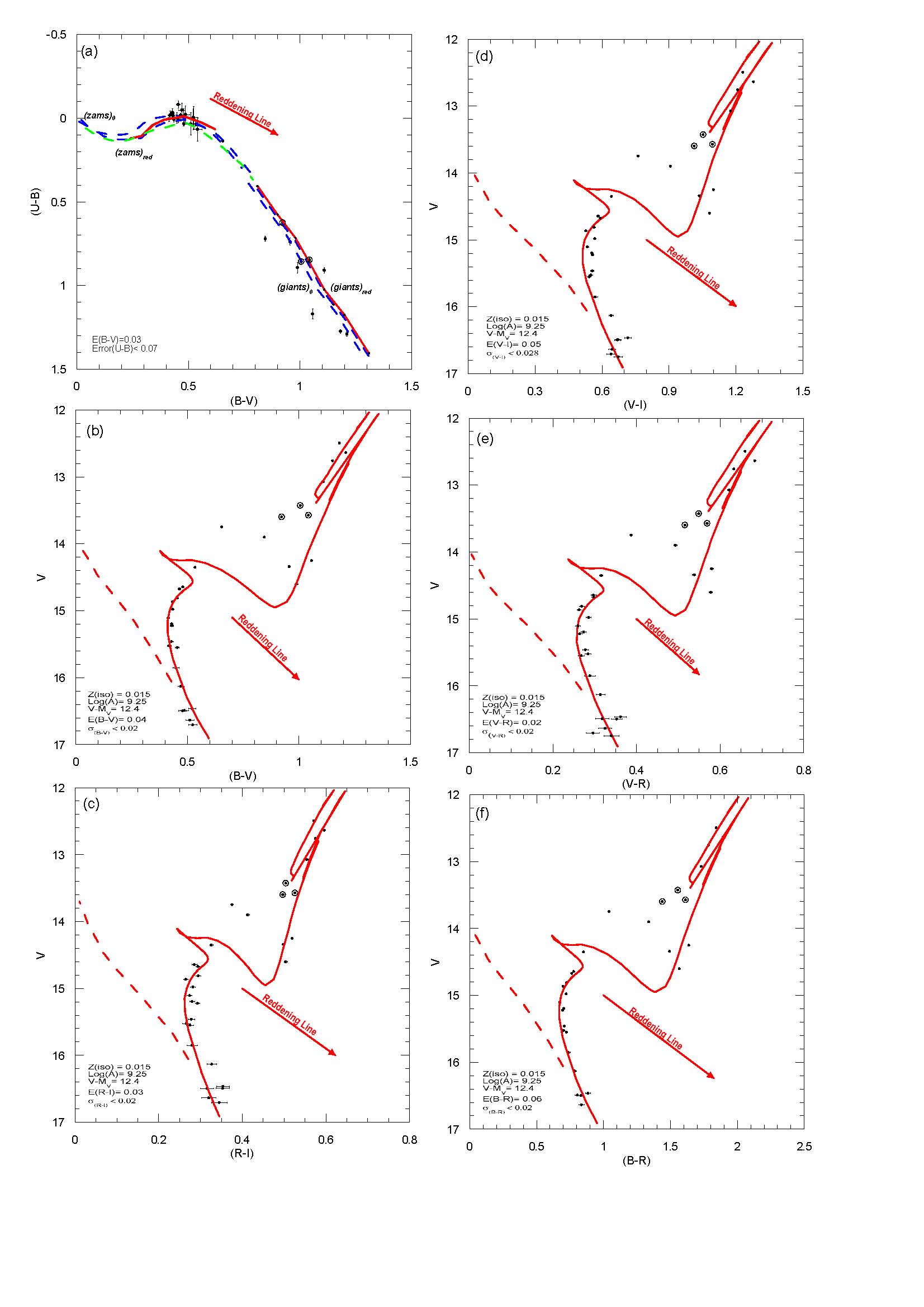, width=12cm, height=12.4cm}
\caption {Panel (a): the $(U$--$B, B$--$V)$ diagram of Ki~23.  
Panels~(b)--(f): CM diagrams of $(V,B$--$V)$, $(V,R$--$I)$,  
$(V,V$--$I)$,  $(V,V$--$R)$, and  $(V,B$--$R)$. 
The symbols are the same as Fig.~1 and Figs.~2--3.}
\end{figure*}

\begin{figure*}
\renewcommand\thefigure{S16}
\centering
\epsfig{file=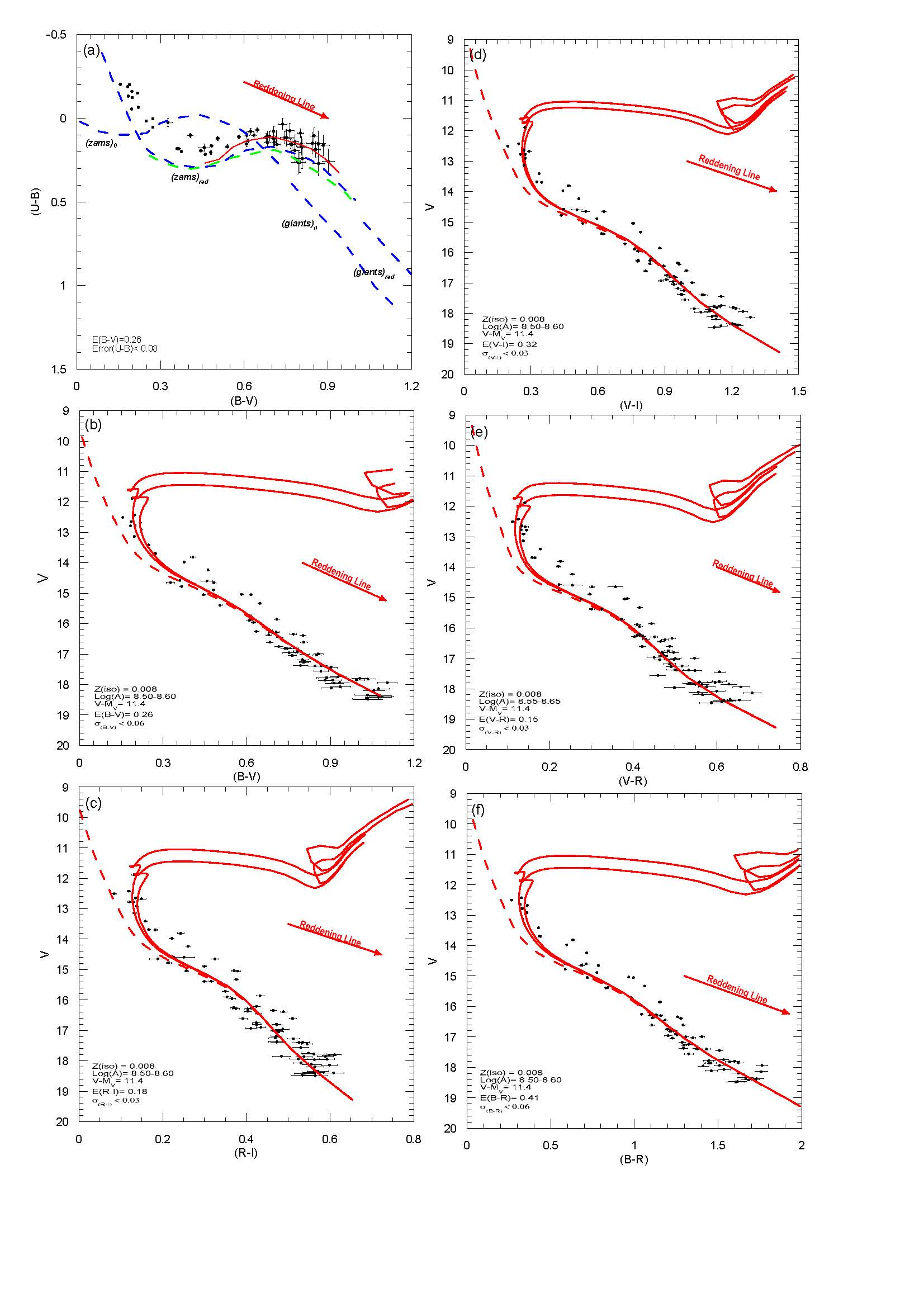, width=12cm, height=12.4cm}
\caption {Panel (a): the $(U$--$B, B$--$V)$ diagram of NGC~2186.  
Panels~(b)--(f): CM diagrams of $(V,B$--$V)$, $(V,R$--$I)$,  
$(V,V$--$I)$,  $(V,V$--$R)$, and  $(V,B$--$R)$. 
The symbols are the same as Fig.~1 and Figs.~2--3.}
\end{figure*}

\clearpage

\begin{figure*}
\renewcommand\thefigure{S17}
\centering
\epsfig{file=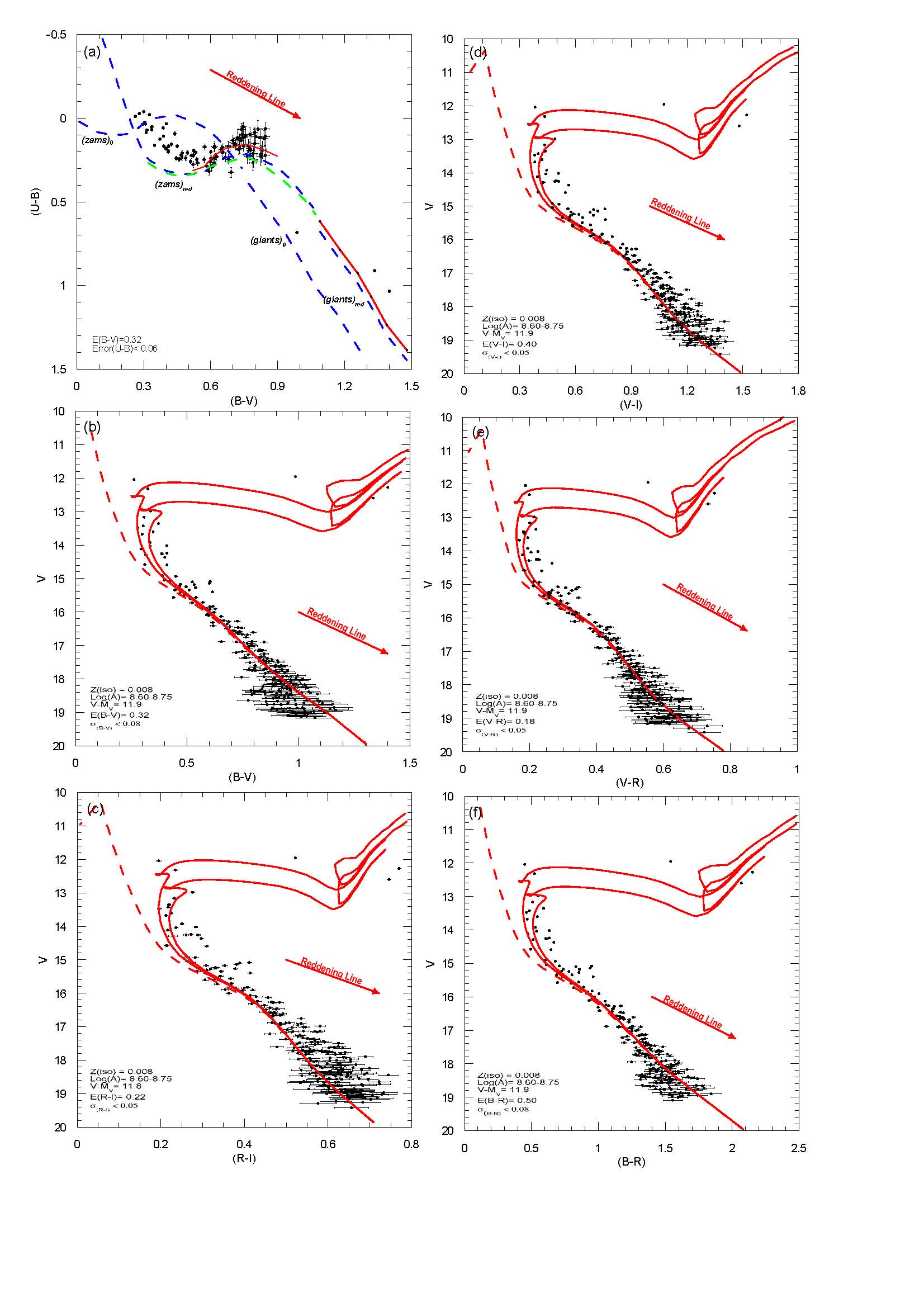, width=12cm, height=12.4cm}
\caption {Panel (a): the $(U$--$B, B$--$V)$ diagram of Haf~08.  
Panels~(b)--(f): CM diagrams of $(V,B$--$V)$, $(V,R$--$I)$,  
$(V,V$--$I)$,  $(V,V$--$R)$, and  $(V,B$--$R)$. 
The symbols are the same as Fig.~1 and Figs.~2--3.}
\end{figure*}

\begin{figure*}
\renewcommand\thefigure{S18}
\centering
\epsfig{file=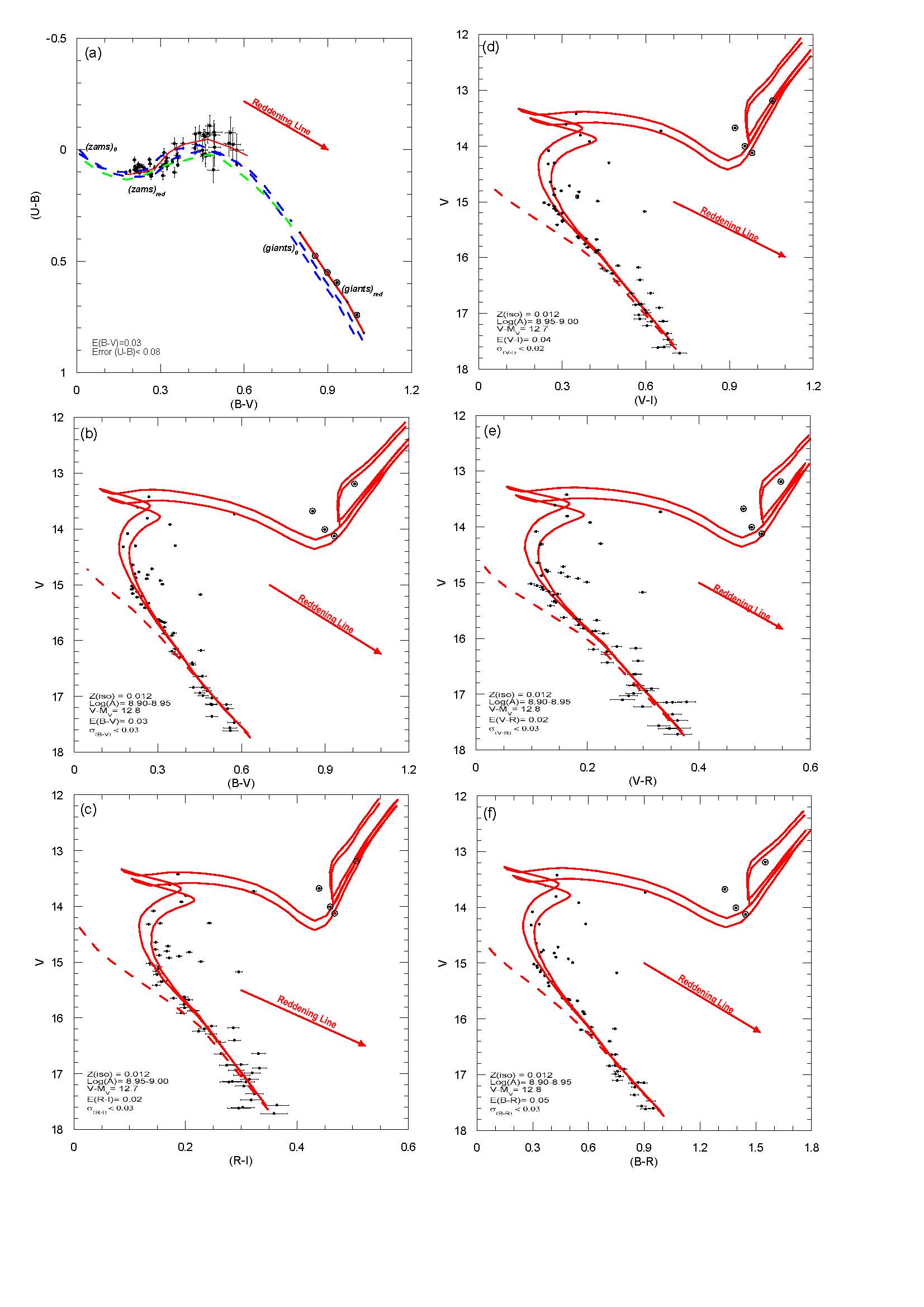, width=12cm, height=12.4cm}
\caption {Panel (a): the $(U$--$B, B$--$V)$ diagram of NGC~2304.  
Panels~(b)--(f): CM diagrams of $(V,B$--$V)$, $(V,R$--$I)$,  
$(V,V$--$I)$,  $(V,V$--$R)$, and  $(V,B$--$R)$. 
The symbols are the same as Fig.~1 and Figs.~2--3.}
\end{figure*}

\begin{figure*}
\renewcommand\thefigure{S19}
\centering
\epsfig{file=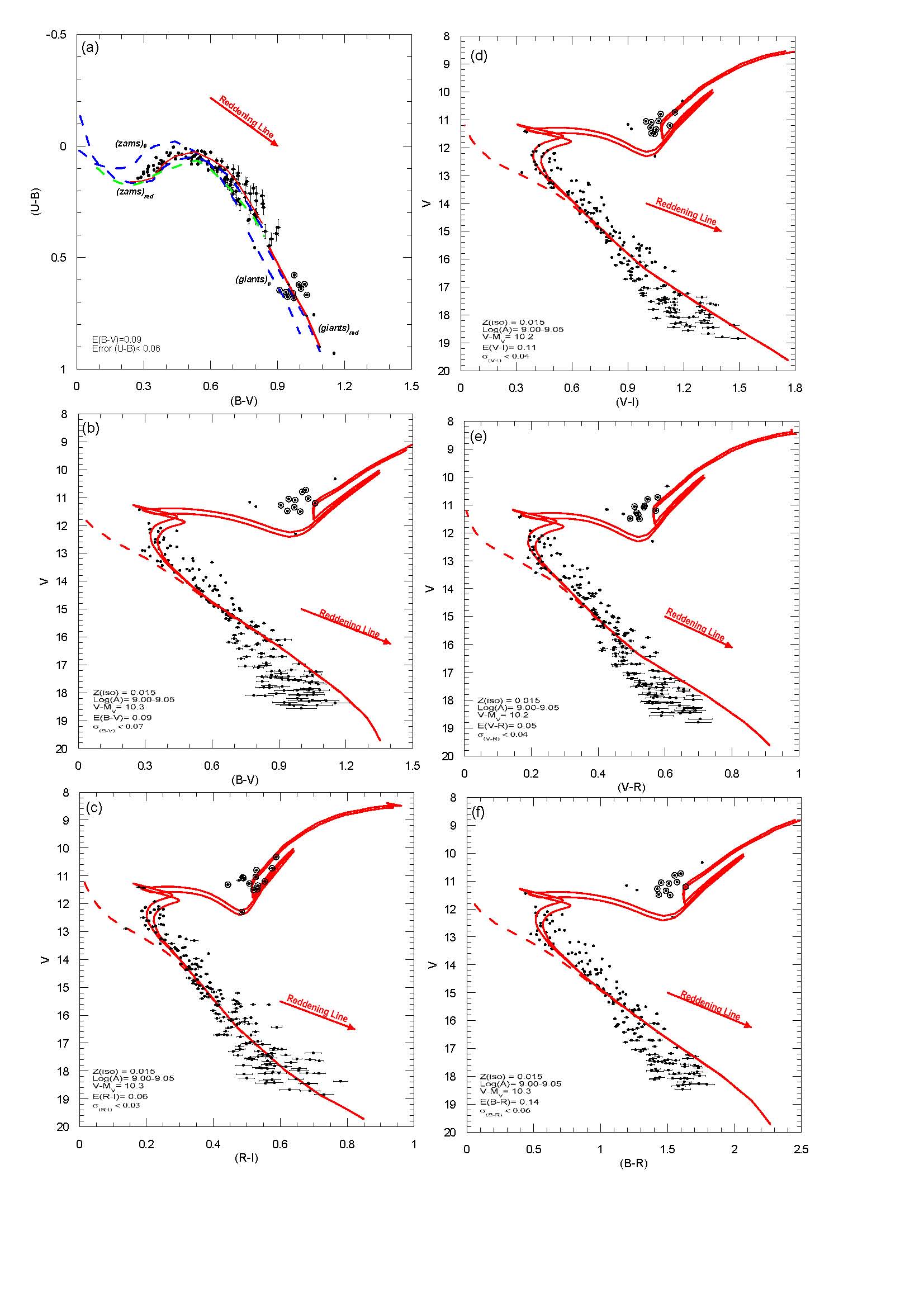, width=12cm, height=12.4cm}
\caption {Panel (a): the $(U$--$B, B$--$V)$ diagram of NGC~2360.  
Panels~(b)--(f): CM diagrams of $(V,B$--$V)$, $(V,R$--$I)$,  
$(V,V$--$I)$,  $(V,V$--$R)$, and  $(V,B$--$R)$. 
The symbols are the same as Fig.~1 and Figs.~2--3.}
\end{figure*}

\clearpage


\begin{thebibliography}{}

\bibitem[Akkaya et al.\ (2010)]{akk10}
Akkaya, \.I., Schuster, W.~J., Michel, R., Chavarr\'ia-K, C.,  
Moitinho, A., V\'azquez, R. \& Karata\c{s}, Y. 2010, RMxAA, 46, 385 (A10)

\bibitem[Ann et al.\ (2002)]{ann02}
Ann, H.B., Leo, S.H., Sung, H., Lee, M.G.,  
Kim, S.L., Chun, M.~Y., Jeon, Y.~B., Park, B.~G., Yuk, I.~S., 2002, \aj, 123, 905 

\bibitem[Arce \& Goodman (1999)]{arc99}
Arce, H.~G., \& Goodman, A.~A. 1999., \apj, 512, L135

\bibitem[Asplund et al.\ (2009)]{asp09}
Asplund, M., Sauval, A.J., Scott, P., 2009, \araa, 47, 481

\bibitem[Babu (1989)]{babu89}
Babu, G.S.D., 1989, JApA, 10, 295

\bibitem[Bahcall \& Soneira (1980)]{bs80}
Bahcall, J.N., \& Soneira, R.M., 1980, \apj, 238, 17
 
\bibitem[Barbaro et al.\ (1969)]{barb69}
Barbaro, G., Dallaporta, N., Fabris, G., 1969, Ap\&SS, 3, 123

\bibitem[Becker et al.\ (1976)]{bec76}
Becker, W., Svolopoulos, S.N., Fang, Ch., 1976, Kataloge photographischer und photoelektrischer 
Helligkeiten von 25 galaktischen Sternhaufen im RGU- und UcBV-System

\bibitem[Bellazzini et al.\ (2004)]{bel04}
Bellazzini, M., Ibata, R.A, Monaco, L., Martin, N.F., Irwin, M.J., Lewis, G.F., 2004, \mnras, 354, 1263

\bibitem[Bertelli et al.\ (1994)]{ber94}
Bertelli, G., Bressan, A., Chiosi, C., Fagotto, F., \& Nasi, E. 1994, \aaps, 106, 275

\bibitem[Bonatto \& Bica (2007)]{bon07}
Bonatto, Ch., Bica, E., 2007, \aap, 473, 445

\bibitem[Bonifacio et al.\ (2000)]{bon}
Bonifacio, P., Monai, S., \& Beers, T.~C. 2000, \aj, 120, 2065

\bibitem[Brunthaler et al.\ (2011)]{bru11}
Brunthaler, A., Reid, M.J., Menten, K.M., et.al. 2011, AN, 332, No.5, 461


\bibitem[Caffau et al.\ (2009)]{caf09}
Caffau, E., Maiorca, E., Bonifacio, P., Faraggiana, R., Steffen, M., Ludwig, H.G., Kamp, I., Busso, M., 2009, \aap, 498, 877

\bibitem[Cameron (1985)]{cam85}
Cameron, L.~M. 1985, \aap, 147, 47

\bibitem[Camargo et al.\ (2009)]{cam09}
Camargo, D., Bonatto, C., Bica, E., 2009, \aap, 508, 211

\bibitem[Cambresy et al.\ (2005)]{camb05}
Cambresy, L., Jarett, T.H., Beichman, C.A., 2005, \aap, 435, 131

\bibitem[Canterna et al.\ (1986)]{can86}
Canterna, R., Geisler, D., Harris, H.C., Olszewski, E., Schommer, R., 1986, \aj, 92, 79
 
\bibitem[Carney (2001)]{carn01}
Carney, B. 2001, Star Clusters, Saas-Fee Advanced Course 28, Lecture Notes 1998,
Swiss Society for Astrophysics and Astronomy, eds.\ L.~Labhardt and B.~Binggeli
(Berlin:  Springer-Verlag) pp.~1-222

\bibitem[Carraro \& Chiosi (1994)]{car94} 
Carraro, G., Chiosi, C., 1994, \aap, 287, 761

\bibitem[Carraro et al.\ (1998)]{car98}
Carraro, G., Ng, Y.K., Portinari, L., 1998, \mnras, 296, 1045

\bibitem[Carraro \& Vallenari (2000)]{car00}
Carraro, G., Vallenari, A., 2000, \aaps, 142, 59

\bibitem[Carraro et al.\ (2005)]{car05}
Carraro, G., Geisler, D., Moitinho, A., Baume, G., Vazquez, R.A., 2005, \aap, 442, 917

\bibitem[Carraro et al.\ (2007)]{car07}
Carraro, G., Geisler, D., Villanova, S., Frinchaboy, P.M., Majewski, S.R., 2007, \aap, 476

\bibitem[Carraro et al.\ (2008)]{car08}
Carraro, G., Moitinho, A., Vazquez, R.A., 2008, \mnras, 385, 1597

\bibitem[Carraro et al.\ (2010a)]{car10a}
Carraro, G., Vazquez, R.A., Costa, E., Perren, G., Moitinho, A., 2010a, \aj, 718, 683

\bibitem[Carraro et al.\ (2010b)]{car10b}
Carraro, G., Costa, E., Ahumada, J.A., 2010b, \aj, 140, 954

\bibitem[Chen et al.\ (1999)]{chen99}
Chen, B., Figueras, F., Torra, J., Jordi, C, Luri, X., Galadi-Enríquez, D., 1999, \aap, 352, 459

\bibitem[Chen et al.\ (2003)]{chen03}
Chen, L., Hou, J.L., Wang, J.J., 2003, \aj, 125, 1397


\bibitem[Claret et al.\ (2003)]{clar03}
Claret, A., Paunzen, E., Maitzen, H.M., 2003, \aap, 412, 91

\bibitem[Claria et al.\ (1989)]{cla89}
Claria, J.J., Lapasset, E., Minniti, D., 1989, \aaps, 78, 363

\bibitem[Claria et al.\ (1999)]{cla99}
Claria, J.J., Mermillod, J.~C., Piatti, A.E., 1999, \aaps, 134, 301

\bibitem[Claria et al.\ (2008)]{cla08}
Claria, J.J., Piatti, A.E., Mermillod, J.~C., Palma, T., 2008, AN, 329, 609

\bibitem[Clem \& Landoldt (2013)]{clem13} 
Clem, J.~L., Landoldt, A., 2013, \aj, 146, 88

\bibitem[Crinklaw \& Talbert (1991)]{cri91}
Crinklaw, G., Talbert, F.D., 1991, \pasp, 103, 536

\bibitem[Cuffey (1940)]{cuf40}
Cuffey, J., 1940, \apj, 92,303

\bibitem[Dias et al.\ (2002)]{dia02}
Dias, W.~S., Alessi, B.~S., Moitinho, A., \& L\'epine, J.~R.~D., 2002, \aap, 389, 871

\bibitem[Dias et al.\ (2012)]{dia12}
Dias, W.~S., Alessi, B.~S., Moitinho, A., \& L\'epine, J.~R.~D., 2012, 2010yCat, VizieR On-line Data Catalog: B/ocl. 

\bibitem[Durgapal et al.\ (2001)]{dur01}
Durgapal, A.K., Pandey, A.K., Mohan, V., 2001, \aap, 372, 71

\bibitem[Eggen (1968)]{egg68}
Eggen, O.J., 1968, \apj, 152, 83

\bibitem[Fenkart et al.\ (1972)]{fen72}
Fenkart, R.P., Buser, R., Ritter, H., Schmitt, H., Steppe, H., Wagner, R., Wiedemann, D.,  1972, \aaps, 7, 487

\bibitem[Friel \& Janes (1993)]{fri93}
Friel, E.D. \& Janes, K.A., 1993, \aap, 267, 75

\bibitem[Friel et al.\ (1995)]{fri95}
Friel, E.D., 1995, \araa, 33, 381

\bibitem[Friel et al.\ (2002)]{fri02}
Friel, E.D., Janes, K.A., Tavarez, M., Scott, J., Katsanis, R., Lotz, J., Hong, L., Miller, N., 2002, \aj, 124, 2693

\bibitem[Geisler  et al.\ (1991)]{gei91}
Geisler, D., Claria, J.J., Dante, M., 1991, \aj, 102, 1836

\bibitem[Geisler  et al.\ (1992)]{gei92}
Geisler, D., Claria, J.J., Dante, M., 1992, \aj, 104, 1892

\bibitem[Gieles  et al.\ (2006)]{gie06}
Gieles, M., Portegies-Zwart, S., Athanassoula, E., Baumgardt, H., Lamers, H.J.G.L.M., Sipior, M, Leenaarts, J., 2006, \mnras, 371, 793

\bibitem[Gim et al.\ (1998)]{gim98}
Gim, M., Vandenberg, D.A.,  Stetson, P., Hesser, J., Zurek, D.R., 1998, \pasp, 110, 1318

\bibitem[Girardi et al.\ (2000)]{gir00}
Girardi, L., Bressan, A., Bertelli, G., \& Chiosi, C., 2000, \aaps, 141, 371
(http://pleiadi.pd.astro.it)

\bibitem[Groenewegen (2008)]{gro08}
Groenewegen, M.A.T., 2008, \aap, 488, 935

\bibitem[Hamdani et al.\ (2000)]{ham00}
Hamdani, S., North, P., Mowlari, N., Raboud, D., Mermillod, J.~C., 2000, \aap, 360, 509

\bibitem[Hardie (1962)]{har62} 
Hardie, R. H., 1962, in "Photometric Reductions", chapter 8 of "Astronomical Techniques", edited by W. A. Hiltner, 
from the series "Stars and Stellar Systems", vol III, general editor G. P. Kuiper and associated general editor B. M. Middlehurst,
University of Chicago Press, \mbox{ISBN 0-226-45963-2}

\bibitem[Hasegawa et al.\ (2008)]{has08}
Hasegawa, T., Sakamoto, T., Malasan, H.L., 2008, PASJ, 60, 1267

\bibitem[Haywood (2008)]{hay08}
Haywood, M., 2008, \mnras, 388, 1175

\bibitem[Hoag et al.\ (1961)]{hoa61}
Hoag, A.A., Johnson, H. L., Iriarfe, B., Mitchell, R., Hallam, K.L., Sharpless, S., 1961, Publication of the U.S. Naval Obeservatory 2nd Ser., 17, 1

\bibitem[Houdek \& Gough (2011)]{hou11}
Houdek, G., Gough, D.O., 2011, \mnras, 418, 1217

\bibitem[Howell (1989)]{how89} 
Howell, S.B., 1989, \pasp, 101, 616

\bibitem[Howell (1990)]{how90} 
Howell, S.B., 1990, in ASP Conf. Ser.~8, CCDs in Astronomy,
ed. G.H. Jacoby (San Francisco:  ASP), 312 

\bibitem[Huestamendia et al.\ (1991)]{hue91} 
Huestamendia, G., del Rio, G., Mermillod J.~C., 1991, \aaps, 87, 153 

\bibitem[Jacobson et al.\ (2008)]{jac08} 
Jacobson, H., Friel, E.D., Pilachowski C.A., 2008, \aj, 135, 2341 

\bibitem[Jacobson et al.\ (2011)]{jac11} 
Jacobson, H., Pilachowski, C.A., Friel E.D.,  2011, \aj, 142, 59 

\bibitem [Johnson et al.\ (1966)]{john66}
Johnson, H.L., Mitchell, R.I., Iriart, B., and Wisniewski, W.Z., 1966, Comm. Lunar and Planet. Labor., University of Arizona 6, No:92,85 

\bibitem[Joshii (2005)]{josh05}
Joshii, Y.C., 2005, \mnras, 362, 1259

\bibitem[Karata\c{s} \& Schuster (2006)]{ks06}
Karata\c{s}, Y., Schuster W.J., 2006, \mnras, 371, 1793

\bibitem[King (1952a)]{kin52a} 
King, I., 1952a, ApJ, 115, 580

\bibitem[King (1952b)]{kin52b} 
King, I., 1952b, \aj, 57, 253 

\bibitem[Landolt (1983)]{lan83}
Landolt, A.~U., 1983, \aj, 88, 439

\bibitem[Landolt (1992)]{lan92}
Landolt, A.~U., 1992, \aj, 104, 340

\bibitem[Lata et al.\ (2010)]{lat10}
Lata, S., Pandey, A.K., Kumar, B., Bhatt, H., Pace, G., Sharma, S., 2010, \aj, 139, 378


\bibitem[Lepine et al.\ (2011)]{lep11}
Lepine, J.R.D., Cruz, P. Scarano, S., Barros, D.A., Dias, W.S., Pompeia, L., Andrievsky S.M., Garraro G., Famaey B., 2011, \mnras, 417, 698

\bibitem[Lynga{\aa} (1987)]{lyn87}
Lynga, G.,  Computer Based Catalogue of Open Cluster Data, Observatoire de Strasbourg, Centre de Donn\'ees
Stellaires, Strasbourg,1987, {\aa}.

\bibitem[Maciejewski \& Niedzielski (2007)]{mac07}
Maciejewski, G., Niedzielski, A., 2007, \aap, 467, 1065

\bibitem[Magrini et al.\ (2009)]{mag09}
Magrini, L., Sestio, P., Randich, S., Galli, D., 2009, \aap, 494, 95

\bibitem[Marigo et al.\ (2008)]{mar08}
Marigo, P., Girardi, L., Bressan, A., Groenewegen, M.~A.~T., Silva, L., \& Granato, G.~L., 2008, \aap, 482, 883 (M08)

\bibitem[Martin et al.\ (2004)]{mar04}
Martin, N.F., Ibata, R.A., Bellazzini, M., Irwin, M.J., Lewis, G.F., Dehnen, W., 2004, \mnras, 348, 12

\bibitem[McFarland (2010)]{mc10}
McFarland, J., 2010, BSc Thesis, Universidad Aut\'onoma de Baja California,
M\'exico (e-mail to  bibens@astrosen.unam.mx)

\bibitem[Melbourne (1959)]{mel59}
Melbourne, W.G., 1959, PhD. thesis, Calif. Inst. of Tech.

\bibitem[Melbourne (1960)]{mel60}
Melbourne, W.G., 1960, \apj, 132, 101

\bibitem[Mermilliod (1992)]{mer92}
Mermilliod,, J.C., 1992, Bull. Inform, CDS, 40, 115

\bibitem[Michel (2014)]{mich14} 
Michel, R., 2014, in preparation 

\bibitem[Mitchell (1960)]{mit60} 
Mitchell, R. I., 1960, ApJ, 132,68

\bibitem[Moffat \& Vogt (1975)]{mof75}
Moffat, A.F.J., Vogt N., 1975, \aaps, 20, 85

\bibitem[Moitinho (2010)]{moi10}
Moitinho, A., 2010, Star clusters: basic galactic building blocks, Proceedings IAU Symposium No.266, eds. R.de Grijs \& J.R.D. Lepine

\bibitem[Momany et al.\ (2006)]{mom06}
Momany, Y., Zaggia, S., Gilmore, G., Piotto, G., Carraro, G., Bedin, L. R., de Angeli, F., 2006, \aap, 451, 515

\bibitem[Netopil et al.\ (2007)]{net07}
Netopil, M., Paunzen, E., Maitzen, H.M., Pintado, O.I., Claret, A., Miranda, L.F., Iliev, Kh. I., Casanova, V., 2007, \aap, 462, 591

\bibitem[Newberg  et al.\ (2002)]{new02}
Newberg, H.J. et al., 2002, \apj, 569, 245

\bibitem[Ortolani et al.\ (2005)]{ort05}
Ortolani, S., Bica, E., Barbuy, B., Zoccali, M., 2005, \aap, 429, 607

\bibitem[Pandey et al.\ (2003)]{pan03}
Pandey, A.K., Upadhyay, K., Nakada, Y., Ogura, K., 2003, \aap, 397, 191

\bibitem[Park \& Lee (1999)]{par99}
Park, H.S., Lee, M.G., 1999, \mnras, 304, 883

\bibitem[Paunzen \& Netopil (2006)]{pau06}
Paunzen, E., \& Netopil, M. 2006, \mnras, 371, 1641

\bibitem[Paunzen et al.\ (2010)]{pau10}
Paunzen, E., Heiter, U., Netopil, M., \& Soubiran, C., 2010, \aap, 517, 32

\bibitem[Peniche et al.\ (1990)]{pen90}
Peniche, R., Pe\~na, J.H., D\'iaz Martinez, S.H., G\'omez, T., 1990, RMxAA, 20, 127

\bibitem[Phelps et al.\ (1994a)]{phel94a}
Phelps, R.L., Jane,s K.A., Montgomery, K.A., 1994a, \aj, 107, 1079

\bibitem[Phelps \& Janes (1994b)]{phel94b}
Phelps, R.L., Janes, K.A., 1994b, \apjs, 90, 31

\bibitem[Piatti et al.\ (1995)]{pia95}
Piatti, A.E., Claria, J.J., Abadi, M.G., 1995, \aj, 110, 2813

\bibitem[Piatti et al.\ (2008)]{pia08}
Piatti, A.E., Claria, J.J., Parisi, M.C., Ahumada, A.V., 2008, Balt A, 17, 67


\bibitem[Piskunov et al.\ (2008)]{pis08}
Piskunov, A.E., Kharchenko, N.V., Schilbach, E., Roser, S., Scholz, R.-D., Zinnecker, H., 2008, \aap, 487, 557


\bibitem[Salaris et al.\ (2004)]{sal04}
Salaris, M., Weiss, A., Percival, S.M., 2004, \aap, 414, 163

\bibitem[Sandage (1969)]{san69}
Sandage, A. 1969, \apj, 158, 1115

\bibitem[Scarano \& Lepine (2013)]{sl13}
Scarano, S., Lepine, J.R.D., 2013, \mnras, 428, 625


\bibitem[Schlegel et al.\ (1998)]{sch98}
Schlegel, D.J., Finkbeiner, D. P., \& Davis, M., 1998, \apj, 500, 525

\bibitem[Schmidt-Kaler (1982)]{schm82}
Schmidt-Kaler, Th.\ 1982, in Landolt-Bornstein, Numerical Data and 
Functional Relationships in Science and Technology, New Series, 
Group VI, Vol.2b, eds.\ K.~Schaifers \& H.~H. Voigt (Berlin:  Springer), p.~14
(SK82)

\bibitem[Sch{\"o}nrich \& Binney (2009)]{sch09}
Sch{\"o}nrich, R., Binney, J., 2009, \mnras, 396, 203

\bibitem[Schuster et al.\ (2004)]{sch04}
Schuster, W.~J., Beers, T.~C., Michel, R., Nissen, P.~E., \& Garc\'{\i}a, G.,
2004, \aap, 422, 527

\bibitem[Schuster et al.\ (2007)]{sch07}
Schuster,, W.~J., Michel, R., Dias, W., Tapia-Peralta, T., V\'azquez, R.,
Macfarland J., Chavarr\'{\i}a, C., Santos, C., \& Moitinho, A. 2007,
Galaxy Evolution Across the Hubble Time, eds.\ F.~Combes and J.~Palou$\breve{\rm s}$, 
Proceedings of the International Astronomical Union, IAU Symposium No.~235,
(Cambridge, United Kingdom:  Cambridge University Press), p.~331


\bibitem[Schuster \& Parrao (2001)]{sp01}
Schuster, W.~J., Parrao, L., 2001, RevMexAA, 37, 187


\bibitem[Stetson (1987)]{ste87}
Stetson, P.~B., 1987, \pasp, 99, 191

\bibitem[Stetson (1990)]{ste90}
Stetson, P.~B., 1990, \pasp, 102, 932

\bibitem[Tadross (2003)]{tad03}
Tadross, A.L., 2003, NewAst, 8, 737

\bibitem[Tadross (2008)]{tad08}
Tadross, A.L., 2008, \mnras, 389, 285

\bibitem[Tapia et al.\ (2010)]{tap10}
Tapia, M.~T., Schuster, W.~J., Michel, R., Chavarr\'ia-K, C., Dias, W.~S., V\'azquez, R., \&
Moitinho, A. 2010, \mnras, 401, 621 (T10)


\bibitem[Twarog et al.\ (1997)]{twa97}
Twarog, B.A., Ashman, K.M., Anthony-Twarog, B.J., 1997, \aj, 114, 2556


\bibitem[Yi et al.\ (2003)]{yi03}
Yi, S.K., Kim, Y.C., Demarque, P., 2003, ApJS, 144, 259

\bibitem[Yong et al.\ (2005)]{yong05}
Yong, D., Carney, B., Teixera de Almeida, M.L., 2005, \aj, 130, 597


\bibitem[Yong et al.\ (2012)]{yong12}
Yong, D., Carney, B., Friel, E., 2012, \aj, 144, 95


\bibitem[Zwitter et al.\ (2008)]{zwi08}
Zwitter, T. et al., 2008, \aj, 136, 421


\end{thebibliography}
\end{document}